%% file: main.tex
\documentclass[pdftex,bookmarks,hidelinks]{article}
\pdfoutput=1
\usepackage{jheppub}
\usepackage{graphicx,epstopdf,amsmath,amsfonts,amssymb,appendix,comment}
\usepackage{color,slashed,braket,multirow}
\usepackage{epsfig,mathrsfs,latexsym,color,url,etoolbox}
\usepackage{bbm}
\usepackage{ulem}
\usepackage{mathtools}
\usepackage{footnote}
\usepackage[usenames,dvipsnames,table]{xcolor}
\usepackage{wrapfig}
\usepackage{pbox}
\usepackage[utf8]{inputenc}
\usepackage[T1]{fontenc}
\usepackage{caption}
\usepackage{subcaption}

\definecolor{nicered}{rgb}{0.7,0.1,0.1}
\definecolor{nicegreen}{rgb}{0.1,0.5,0.1}
\DeclareMathAlphabet{\mathbbold}{U}{bbold}{m}{n}

\newcommand{\beq}{\begin{equation}}
\newcommand{\eeq}{\end{equation}}
\newcommand\bout{\bgroup\markoverwith{\textcolor{blue}{\rule[0.5ex]{4pt}{0.8pt}}}\ULon}

\usepackage{hyperref}
\hypersetup{colorlinks,citecolor= nicegreen,linkcolor= nicered}

\newcommand\snowmass{\begin{center}\rule[-0.2in]{\hsize}{0.01in}\\\rule{\hsize}{0.01in}\\
\vskip 0.1in Submitted to the  Proceedings of the US Community Study\\ 
on the Future of Particle Physics (Snowmass 2021)\\ 
\rule{\hsize}{0.01in}\\\rule[+0.2in]{\hsize}{0.01in} \end{center}}

\begin{document}
\input{common/FrontMatter}

\setcounter{tocdepth}{2}
\flushbottom

\input{Sections/Introduction}

\input{Sections/Theory}


\input{Sections/Experiment}


\input{Sections/Conclusions}

\input{common/EndMatter}

\end{document}

%% file: common/FrontMatter.tex
\pagestyle{titlepage}

\date{\today}

\title{{\huge Searches for Baryon Number Violation in Neutrino Experiments: A White Paper}}

\snowmass{}

\author[1]{P. S. B.  Dev\footnote{Editor},}
\author[2]{L. W. Koerner$^*$,}
\author[3]{S. Saad$^*$,}
\author[3]{S. Antusch,}
\author[4,5]{M. Askins,}
\author[6]{K. S. Babu,}
\author[7,8]{J. L. Barrow,}
\author[9]{J. Chakrabortty,}
\author[10]{A. de Gouv\^{e}a,}
\author[11]{Z. Djurcic,}
\author[12]{S. Girmohanta,}
\author[13]{I. Gogoladze,}
\author[11]{M. C. Goodman,}
\author[14]{A. Higuera,}
\author[15]{D. Kalra,}
\author[15]{G. Karagiorgi,}
\author[16]{E. Kearns,}
\author[17]{V. A. Kudryavtsev,}
\author[18]{T. Kutter,}
\author[19]{M. Malinsk\'{y},}
\author[20]{D. A. Martinez Caicedo,}
\author[21]{R. N. Mohapatra,}
\author[22]{P. Nath,}
\author[8]{S. Nussinov,}
\author[23]{J. P. Ochoa-Ricoux,}
\author[24]{V. Pec,}
\author[11]{A. Rafique,}
\author[20]{J. Rodriguez Rondon,}
\author[12]{R. Shrock,}
\author[23]{H. W. Sobel,}
\author[18]{T. Stokes,}
\author[25]{M. Strait,}
\author[26]{R. Svoboda,}
\author[12,27]{S. Syritsyn,}
\author[28]{V. Takhistov,}
\author[29]{Y.-T. Tsai,}
\author[28,30]{R. A. Wendell,}
\author[31,32]{Y.-L. Zhou}

\affiliation[1]{Department of Physics and McDonnell Center for the Space Sciences, Washington University, St. Louis, MO 63130, USA}
\affiliation[2]{Department of Physics, University of Houston, Houston, TX 77204, USA}
\affiliation[3]{Department of Physics, University of Basel, Klingelbergstrasse\ 82, CH-4056 Basel, Switzerland}
\affiliation[4]{Department of Physics, University of California, Berkeley, CA 94720, USA}
\affiliation[5]{Lawrence Berkeley National Laboratory, 1 Cyclotron Road, Berkeley, CA 94720-8153, USA}
\affiliation[6]{Department of Physics, Oklahoma State University, Stillwater, OK 74078, USA}
\affiliation[7]{The Massachusetts Institute of Technology, Department of Physics, 77 Massachusetts Avenue, Building 4, Room 304, Cambridge, MA 02139, USA}
\affiliation[8]{Tel Aviv University, Tel Aviv 6997801, Israel}
\affiliation[9]{Department of Physics, Indian Institute of Technology Kanpur, Kanpur 208016, 
India}
\affiliation[10]{Northwestern University, Department of Physics \& Astronomy, 2145 Sheridan Road, Evanston, IL 60208, USA}
\affiliation[11]{Argonne National Laboratory, 
Argonne, IL 60439, 
USA}
\affiliation[12]{C. N. Yang Institute for Theoretical Physics and Department of Physics and Astronomy, Stony Brook University, Stony Brook, NY 11794, USA}
\affiliation[13]{Department of Physics \& Astronomy, 
224 Sharp Laboratory, 
University of Delaware,
Newark,  DE 19716, USA}
\affiliation[14]{Department of Physics and Astronomy, Rice University, Houston, TX 77005, USA}
\affiliation[15]{Department of Physics, Columbia University, New York, NY 10025, USA}
\affiliation[16]{Department of Physics, Boston University, Boston, MA 02215, USA}
\affiliation[17]{University of Sheffield, Sheffield S3 7RH, United Kingdom}
\affiliation[18]{Department of Physics and Astronomy, Louisiana State University, Baton Rouge, LA 70803, USA}
\affiliation[19]{Institute of Particle and Nuclear Physics, Charles University, V Hole\v{s}ovi\v{c}k\'ach 2, 180 00 Praha 8, Czech Republic}
\affiliation[20]{Department of Physics, South Dakota School of Mines and Technology, Rapid City, SD 57701, USA}
\affiliation[21]{Maryland Center for Fundamental Physics, Department of Physics, University of Maryland, College Park, MD 20742,
USA}
\affiliation[22]{Department of Physics, Northeastern University, Boston, MA 02115-5000, USA}
\affiliation[23]{Department of Physics and Astronomy, University of California, Irvine, CA 92697, USA}
\affiliation[24]{Institute of Physics, The Czech Academy of Sciences, Prague, Czech Republic}
\affiliation[25]{School of Physics and Astronomy, University of Minnesota Twin Cities, Minneapolis, MN 55455, USA}
\affiliation[26] {Department of Physics and Astronomy, University of California Davis, Davis, CA 95616, USA}
\affiliation[27]{RIKEN-BNL Research Center, Brookhaven National Lab, Upton, NY 11973, USA}
\affiliation[28]{Kavli Institute for the Physics and Mathematics of the Universe (WPI), UTIAS, The University of Tokyo, Kashiwa, Chiba 277-8583, Japan}
\affiliation[29]{SLAC National Accelerator Laboratory, Menlo Park, CA 94025, USA}
\affiliation[30]{Department of Physics, Kyoto University, Kyoto, Kyoto 606-8502, Japan}
\affiliation[31]{School of Fundamental Physics and Mathematical Sciences, Hangzhou Institute for Advanced Study, UCAS, Hangzhou, China}
\affiliation[32]{International Centre for Theoretical Physics Asia-Pacific, Beijing/Hangzhou, China}

\emailAdd{bdev@wustl.edu}
\emailAdd{lkoerner@central.uh.edu}
\emailAdd{shaikh.saad@unibas.ch}

\abstract{Baryon number conservation is not guaranteed by any fundamental symmetry within the Standard Model, and therefore has been a subject of experimental and theoretical scrutiny for decades.   So far, no evidence for baryon number violation has been observed.  Large underground detectors have long been used for both neutrino detection and searches for baryon number violating processes.  The next generation of large neutrino detectors will seek to improve upon the limits set by past and current experiments and will cover a range of lifetimes predicted by several Grand Unified Theories. In this White Paper, we summarize theoretical motivations and experimental aspects of searches for baryon number violation in neutrino experiments.
}

\maketitle


%% file: Sections/Introduction.tex
\section{Introduction and Executive Summary}\label{sec:intro}

\subsection{Theory Summary }
The stability of ordinary matter has long been a subject of both theoretical and experimental interests.  The electron is stable because of electric charge conservation. On the other hand, in the Standard Model (SM), the stability of the proton is guaranteed by the accidental global symmetry of baryon number at the renormalizable level. In models of quark-lepton unification, such as the Grand Unified Theories (GUTs), baryon number is necessarily violated. As a result, the proton is not stable, and decays dominantly into $e^+\pi^0$ (in non-supersymmetric theories) or $K^+\bar{\nu}$ (in supersymmetric theories). Another compelling reason for baryon non-conservation is that understanding the origin of matter in the universe requires it as one of the three fundamental Sakharov conditions in addition to CP violation and thermal non-equilibrium.

This striking prediction of GUTs on proton decay, which are otherwise inaccessible to laboratory experiments, motivated the construction of large-scale water Cherenkov detectors like IMB, Kamiokande, and its subsequent upgrade, Super-Kamiokande. Although there is no direct evidence of proton decay so far, but only stringent lower limits on the proton lifetime, it is important to continue the searches for proton (and bound neutron) decay and other baryon number violating (BNV) processes in general. 
A large class of GUTs predict proton lifetime to be within an order of magnitude above the current experimental limit.  
It is also important to keep in mind that the same experiments originally constructed to search for proton decay have now become truly multi-purpose experiments. In particular, they have played a major role in neutrino physics, starting with the serendipitous detection of SN1987A neutrinos, as well as  the discovery of neutrino oscillations in atmospheric and solar neutrinos. 

Therefore, the significance of current and next-generation neutrino experiments {\it simultaneously} searching for baryon number violation and studying neutrino properties cannot be overemphasized. While the main focus of the BNV experiments is on proton decay searches, there also exist other equally important baryon and/or lepton number violating processes, such as dinucleon decays and neutron-antineutron oscillations which must be studied as well along with their experimental detection prospects. Possible connections of BNV observables to other beyond the Standard Model physics, such as neutrino mass, baryogenesis, dark matter, flavor physics, and gravitational waves are also being explored. The recent lattice developments for the relevant nucleon and nuclear matrix elements of effective BNV operators are also crucial for reducing the theoretical uncertainties in the BNV predictions.  

\subsection{Experimental Summary }

Experiments have long sought evidence of the decay of the proton as proof of physics beyond the SM (BSM).  The lower limit on the proton's lifetime is currently of order $10^{34}$~years.  Experimental searches that seek to probe beyond this limit therefore need a huge source of protons and years of exposure.
The large mass and long operation times of detectors used for observation of neutrino oscillations make them well-suited for searches for baryon number violation, including nucleon decay and neutron-antineutron oscillations.  The Super-Kamiokande neutrino experiment, which uses a water Cherenkov detector with a fiducial mass of 22.5~ktons and began operation in 1996, has published leading limits on 30 baryon number violating processes.  Next-generation neutrino detectors, such as DUNE (40~kton fiducial mass liquid argon TPC), Hyper-Kamiokande (190~kton fiducial mass water Cherenkov), and JUNO (20~kton fiducial mass liquid scintillator), all include searches for baryon number violation as a major component of their physics programs and hope to improve upon the limits set by Super-Kamiokande, if not observe baryon number violation for the first time.  

Detector mass is a crucial characteristic in next-generation baryon number violation searches.  For small detectors, the exposure required to improve upon limits already set by Super-Kamiokande can exceed the likely lifetime of the experiment. Clearly Hyper-Kamiokande has the advantage in this respect. That being said, detector technology is also extremely important; DUNE's excellent imaging capabilities and JUNO's superb timing resolution offer advantages in some channels over Hyper-Kamiokande's larger mass.  NOvA, a currently-running neutrino experiment with a 14~kton segmented liquid scintillator detector, is developing a search for neutron-antineutron oscillations that could potentially have sensitivity comparable to current limits. THEIA is a proposed water-based liquid scintillator detector that would combine the advantages of the large mass of a water Cherenkov detector with the good resolution of a liquid scintillator detector.  With this worldwide program, should a baryon number violating signal be observed by any of the detectors in the next generation, confirmation from other detectors using different technologies would provide powerful evidence of physics beyond the Standard Model.

In addition to detector mass and technology, simulation and analysis techniques can also affect the potential of these searches.  As with neutrino interactions, the experimental community has come to understand how important nuclear effects are in predicting the characteristics of final-state particles.  Final-state interactions in the nucleus alter the multiplicity and momenta of final-state particles.  Uncertainties in modeling final state interactions therefore introduce uncertainties into the signal efficiency estimates and lifetime limits.  Furthermore, analysis techniques are continually improving.  For example, Super-Kamiokande made improvements to the search for proton decay via $p \to e^+\pi^0$ by reducing backgrounds via neutron tagging.  Potential improvements to searches in a liquid argon TPC could come from tagging of nuclear de-excitations.

The experimental neutrino physics community has long been conducting searches for baryon number violation using neutrino detectors.  The next generation of neutrino detectors, situated in underground laboratories with good shielding to reduce cosmic ray backgrounds, will allow the continued pursuit of this goal, with massive detectors and continually improving analysis techniques.

%% file: Sections/Theory.tex
\section{Theoretical Motivations}\label{sec:theory}
\subsection{Historical Context} 
``Is ordinary matter stable?'' This question has been a subject of experimental and theoretical interest over many decades. Ordinary matter made up of electrons and nucleons (protons and neutrons) is stable first because electrons are stable because of electric charge conservation; but what about nucleons? So far, experimental searches for the decay of protons and bound neutrons have not found evidence for nucleon decay and have only led to decay mode dependent upper limits on the nucleon lifetime. From a theoretical point of view, the stability of matter can be guaranteed by assigning a baryon number $B=+1$ to the proton (the lightest baryon), which is known as the principle of conservation of baryon number formulated by Weyl in 1929~\cite{Weyl:1929fm} (see also Refs.~\cite{stuckelberg1938wechselwirkungskrafte,stueckelberg1938forces,wigner1949invariance}). Baryon number conservation, however, is not guaranteed by any fundamental symmetry within the SM. Instead, baryon number is an accidental classical (global) symmetry of the SM Lagrangian, which is violated by small amounts via non-perturbative effects (namely, SU(2)$_L$ sphalerons) ~\cite{tHooft:1976rip,tHooft:1976snw,Manton:1983nd,Kuzmin:1985mm}. These are however negligibly small at temperatures low compared with the electroweak scale, but are important in the early universe.

After the recent discovery~\cite{CMS:2012qbp,ATLAS:2012yve} of the Higgs boson, the SM seems to be remarkably complete; furthermore, as modified to include nonzero neutrino masses and lepton mixing (which were absent in the original SM), it is generally consistent with the current experimental data. (Here, we note that there are indications of possible violation of lepton flavor universality in the $B$-meson decays reported by the LHCb experiment \cite{LHCb:2021trn}. Furthermore, the  $(g-2)_\mu$ measured value shows a significant deviation compared to the SM prediction recently confirmed by the Fermilab experiment \cite{Muong-2:2021ojo}. There is an intensive experimental program to check these indications.)  However, it is clear that the SM has several deficiencies~\cite{ParticleDataGroup:2020ssz}. For example, even apart from its prediction of (i) zero neutrino masses, the original SM accommodated, but did not explain, the quark and charged lepton masses or the observed Cabibbo-Kobayashi-Maskawa (CKM) quark mixing. Other deficiencies include the fact that (ii) there is no cold Dark Matter (DM) candidate in the SM (aside from primordial black holes, whose possible contributions to DM have long been studied), (iii) the observed matter-antimatter asymmetry of the Universe cannot be explained, (iv) the fundamental reason for electric charge quantization is not understood, and (v) the Higgs hierarchy (fine-tuning) problem is not solved. 
These are clear indications that the SM must be extended. A natural pathway towards building an ultraviolet (UV) complete model follows directly from the unification of the electromagnetic and weak forces already realized within the SM. These two interactions are of quite different strengths at low energy scales, owing to the electroweak symmetry breaking (EWSB), which produces masses of $O(10^2)$ GeV for the $W$ and $Z$ vector bosons, while keeping the photon massless. However, at energy scales above the EWSB scale, the SU(2)$_L$ and U(1)$_Y$ gauge interactions have coupling that are of comparable strength. 

Therefore, a natural expectation is that all forces might unify at some higher scale. Historically, the first such attempts at unification were the Pati-Salam (PS) model~\cite{Pati:1973rp,Pati:1974yy} that unifies quarks and leptons and the  Georgi-Glashow (GG)~\cite{Georgi:1974sy} model that unifies all three forces (electromagnetic, weak, and strong) within a fundamental symmetry group $SU(5)$, which also unifies particles and antiparticles. Spontaneous breaking of these symmetries at some ultrahigh-energy scale leads to the SM with three distinct unbroken symmetries that result in separate electroweak and strong forces. This idea of embedding the entire SM into a single unified group is called Grand Unification~\cite{Pati:1974yy, Georgi:1974sy, Georgi:1974yf, Georgi:1974my, Fritzsch:1974nn,Langacker:1980js}.

An immediate consequence of Grand Unified Theories (GUTs) is that baryon number is necessarily violated and the proton decays. These processes are mediated by the new superheavy gauge bosons associated with the GUT group.  Assuming the unified gauge coupling to take a value similar to the fine-structure constant, and using the best experimental lower limit of the proton lifetime  known around the time unified theories were proposed (1974)~\cite{Reines:1974pb}: $\tau_p\gtrsim 10^{30}$ yrs (using 20 tons of liquid scintillator),  one obtains a limit on the superheavy gauge boson mass $M_X\gtrsim 10^{14}$ GeV. This  corresponds to a scale at which SM gauge couplings are expected to be unified. Being twelve orders of magnitude larger than the electroweak scale, there is no hope to directly test GUT scale physics at colliders (see however Ref.~\cite{Croon:2019kpe} for indirect tests). On the other hand, proton decay (and the decay of neutrons that would otherwise be stably bound in nuclei) provides a clear signature of GUTs, and therefore, it was the prime target of early experiments using water Cherenkov and other detectors. Since 1 kiloton of water contains about $6\times 10^{32}$ nucleons, it was thought that proton decay predicted by GUTs should be within reach
of experimental searches.

The search for nucleon  decay has a long history. The first useful limits on  nucleon decay rates were obtained using radiochemical methods. Alternatively, nucleon  decays were also searched for using geochemical techniques. Both of these methods are discussed in Ref.~\cite{Rosen:1975ch}, and  a strong bound on the nucleon lifetime of $\tau_p> 2\times 10^{27}$ yrs was obtained  using the second method~\cite{Bennett:1981yz}.  The advantage of these indirect methods was that limits on nucleon lifetime were obtained independent of the decay modes.   On the other hand, the direct detection method is based on detecting particles emitted by the decay of a nucleon. The primary advantage of detectors of this type is that much larger
quantities of matter can be used as nucleon sources.  Furthermore, large backgrounds encountered in the nuclear experiments can be significantly reduced in the direct detection experiments. The CERN group~\cite{Backenstoss} carried out the first deep underground experiment hunting for proton decay in a railway tunnel (in Switzerland) that provided an upper bound on the nucleon lifetime of $3\times 10^{26}$ yrs (1960).   By analyzing results from earlier deep underground experiments, the proton lifetime reached about $\tau_p>3\times 10^{30}$ yrs by 1981~\cite{Learned:1979gp, Cherry:1981uq}.  For an overview of early proton decay experiments, see Ref.~\cite{Perkins:1984rg}.

During the 1980's, five underground detectors started searching for proton decay. Three of them were based on calorimeter-type detectors: (i) NUSEX (in Europe)~\cite{NUSEX:1988jvu}, (ii)  Frejus (in Europe)~\cite{Frejus:1990cot}, (iii)  Soudan-2 (in Minnesota)~\cite{Soudan-2:1993jwe}, and two of them were water Cerenkov detectors: (iv) IMB (in Ohio)~\cite{Irvine-Michigan-Brookhaven:1983iap}, (v) Kamiokande (in Japan)~\cite{Hirata:1988ad}.  After several years of operation, these experiments observed no clear indication of nucleon decay; however, some of these experiments led to unexpected groundbreaking discoveries in neutrino physics.  Kamiokande (Kamioka Nucleon Decay Experiment), funded in 1982, listed in its proposal the possibility of detecting neutrino bursts from supernovae as well as  studying neutrino oscillations through the measurements of atmospheric neutrinos. Shortly after Kamiokande started taking data in 1983, it was realized that this experiment could be upgraded to measure solar neutrinos.  The upgraded experiment Kamiokande-II started taking data in 1987 and, immediately after,  it detected a neutrino burst from supernova SN1987A~\cite{Kamiokande-II:1987idp}. Neutrinos from this supernova explosion were also detected by the IMB experiment~\cite{Bionta:1987qt}.   This historical observation demonstrates the excellent capability of water Cherenkov detectors to measure low-energy neutrinos. Two years later, Kamiokande-II also successfully detected solar neutrinos and confirmed the deficit of neutrinos from the sun~\cite{Hirata:1988ad}.

Subsequently, the construction of Super-K (Super-Kamiokande)
was approved in 1991, which had proton decay and neutrino astronomy (solar neutrinos and supernova neutrinos) in its top-listed search agenda.  The making of the Super-K detector was completed in 1996, and within two years of data taking,  the discovery of neutrino oscillation (of the atmospheric neutrinos) was announced in 1998 ~\cite{Super-Kamiokande:1998uiq}.    Remarkably, the data analysis exhibited that a muon neutrino produced in the atmosphere was converting to another neutrino flavor.  
The Super-K experiment also played a significant role in discovering neutrino oscillation in solar neutrinos.  Measurements of the solar neutrino flux at Super-K showed that solar electron neutrinos were transforming to different neutrino flavors~\cite{Super-Kamiokande:2001ljr,Super-Kamiokande:2001bfk}. These two major milestones clearly depict the importance of  experiments that are simultaneously searching for baryon number violating (BNV) processes and studying neutrino properties. At the time of writing this White Paper, no convincing evidence of proton decay has been reported, and the current lower limit on the proton lifetime exceeds $10^{34}$ yrs for some channels.

\subsection{Proton Decay in Grand Unified Theories} 
The basic idea of GUTs is the embedding of the entire SM gauge group $SU(3)_c\times SU(2)_L\times U(1)_Y\equiv {G}_{\textrm{SM}}$ in a larger (non-Abelian) group ${G}_{\textrm{GUT}}$, which thus involves a single gauge coupling. Importantly, electric charge quantization is guaranteed in this framework, since the electric charge operator is a generator of  ${G}_{\textrm{GUT}}$. Even though the values of three gauge couplings associated with ${G}_{\textrm{SM}}$ are different at the low energy scales, they are expected to get unified at the GUT scale. However, within the SM, these gauge couplings do not quite combine into a single coupling.  Interestingly, if TeV scale supersymmetry (SUSY)~\cite{Wess:1974tw, Witten:1981nf,Dimopoulos:1981yj,Dimopoulos:1981zb,Sakai:1981gr} is assumed, the Minimal Supersymmetric SM (MSSM) leads to unification at about $M_{\textrm{GUT}}=2\times 10^{16}$ GeV.  Since SUSY can also provide a solution to the hierarchy problem~\cite{Veltman:1980mj,Dimopoulos:1981au,Witten:1981nf,Dine:1981za} and since a DM candidate naturally arises in an R-parity-conserving SUSY theory if the lightest SUSY particle (LSP) does not carry electric and/or color charge,  SUSY GUTs~\cite{Nilles:1983ge} can be considered an appealing extensions of the SM. In order for SUSY to provide a solution to the fine-tuning problem with the Higgs mass, it was necessary that the SUSY breaking scale should not be much higher than the electroweak symmetry breaking scale of 250 GeV.  Thus, it was widely expected that supersymmetric partners of SM particles would be observed at the LHC.  However, no evidence of these superpartners (in particular, the squarks and gluinos, which interact strongly) has been seen at the LHC running at 14 TeV \cite{ParticleDataGroup:2020ssz}. This subsection summarizes the state of theoretical knowledge of both non-SUSY and SUSY GUTs.
In our discussion, we mostly focus on minimal models; however, nucleon decay predictions of a wide range of models are summarized in Table~\ref{tab:MODELS}. 
For reviews on this subject, see Refs.~\cite{Langacker:1980js,Nath:2006ut,Babu:2013jba,Hewett:2014qja}.

\begin{figure}[t!]\centering
\includegraphics[width=0.85\textwidth]{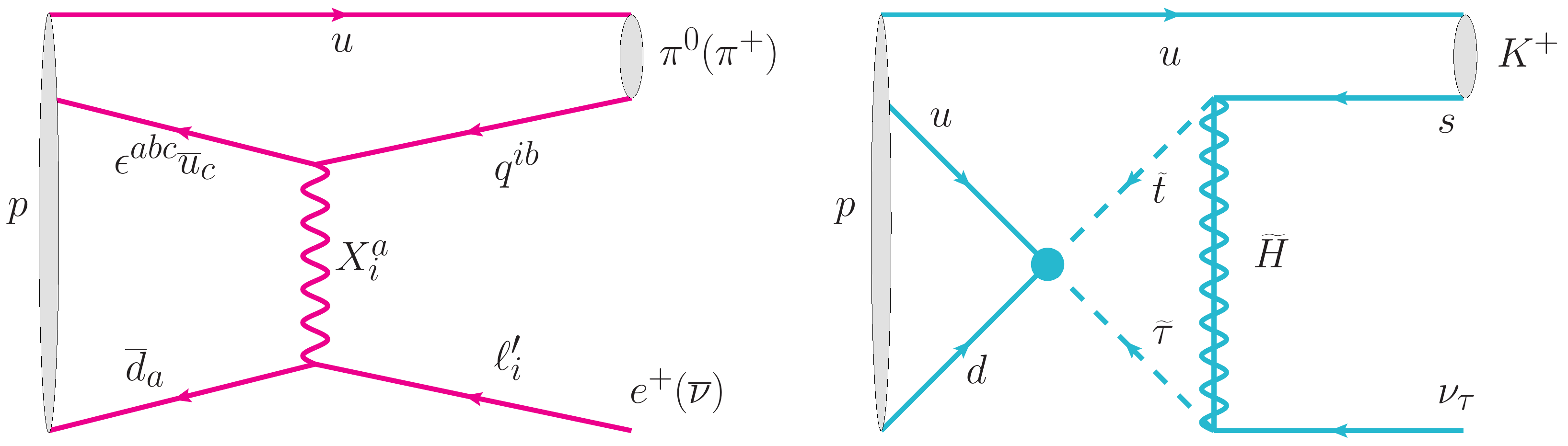}
\caption{Left-diagram: Dominant proton decay mode $p\to e^+ \pi^0$ in non-SUSY GUTs (here $(X_\mu)^a_i=(X^a_\mu, Y^a_\mu)^T$ is the iso-doublet gauge field). Right-diagram: Dominant proton decay mode $p\to \overline{\nu}K^+$ in SUSY GUTs (example diagram with Higgsino dressing, see text for details). The blob here represents the dimension-5 operator induced by colored Higgs exchange.}\label{fig:ProtonDecay}
\end{figure}

\textbf{\textbullet\; \underline{$SU(5)$ GUTs:\;}} The simplest GUT model is the GG model \cite{Georgi:1974sy} with the gauge group  ${G}_{\textrm{GUT}}=SU(5)$. In this model, the GUT group is spontaneously broken to the  ${G}_{\textrm{SM}}$ in a single step by a Higgs field in the adjoint ($\bf{24}$) representation. Finally, SM symmetry is broken down to $SU(3)_c\times U(1)_{\textrm{em}}$ when a scalar field in the fundamental ($\bf{5}$) representation acquires vacuum expectation value. The latter field contains the SM Higgs doublet that remains light, whereas its color-triplet partner needs to reside roughly above $10^{11}$\,GeV  
since it leads to proton decay predominantly through $p \rightarrow  \bar{\nu} K^+$ channel~\cite{Dorsner:2012uz}. The SM fermions of each generation are contained in a $\bf{\overline{5}}$- and a $\bf{10}$-dimensional representations. Notably, the $\bf{\overline{5}}$ contains the lepton doublet and the $d^c$ quark field, while the ${\bf 10}$ contains the quark doublet, and the $u^c$ as well as the $e^c$ fields. The gauge multiplet belonging to the adjoint representation contains twelve SM and twelve BSM vector bosons (labeled by $X$ and $Y$ with electric charges $4/3$ and $1/3$, respectively). These new gauge boson interactions   involve both quarks and leptons; consequently, the new interactions violate baryon number $B$, and lead to proton decay via dimension-6 operators of the form such as  $\overline{u^c}\gamma^\mu Q\overline{e^c}\gamma_\mu Q$ etc ~\cite{Weinberg:1981wj,Wilczek:1979hc,FileviezPerez:2004hn,Nath:2006ut}. An example diagram is presented in Fig.~\ref{fig:ProtonDecay} (diagram on the left).

Non-observation of proton decay requires these gauge bosons to be superheavy, and this bound can be  computed easily by approximating the left diagram in Fig.~\ref{fig:ProtonDecay} by a four-fermion interaction; by doing so, one obtains,  
\begin{align}
\tau_p\sim
\frac{16\pi^2  M^4_X}{g^4_{\textrm{GUT}}  m^5_p},
\end{align}
where $g_{\textrm{GUT}}$ is the unified gauge coupling and $m_p$ and $M_X$ are the proton and superheavy gauge boson masses, respectively. Then the current experimental bound of $\tau_p(p\to e^+\pi^0)> 2.4\times 10^{34}$ yrs from the Super-Kamiokande Collaboration  ~\cite{Super-Kamiokande:2020wjk} typically implies $M_X\sim M_{\textrm{GUT}}\gtrsim 5\times 10^{15}$ GeV. In addition to $p\to e^+ \pi^0$,  current (as well as future) experimental bounds (sensitivities) on other important proton decay modes are summarized in Table~\ref{tab:proton}.

\begin{table}[b!]
\centering
\footnotesize
\resizebox{0.75\textwidth}{!}{
\begin{tabular}{|c|c|c|}
\hline
\pbox{10cm}{Modes\\(partial lifetime)}  &\pbox{10cm}{Current limit [$90\%$ CL] \\ ($10^{34}$ years)} 
&\pbox{10cm}{\vspace{3pt}Future Sensitivity [$90\%$ CL]\\($10^{34}$ years)}\\ [1ex] \hline

$\tau_p \left( p\to e^+\pi^0  \right)$  &Super-K: $2.4$ ~\cite{Super-Kamiokande:2020wjk} &\pbox{10cm}{\vspace{3pt} Hyper-K (1900~kton-yrs): $7.8$~\cite{Hyper-Kamiokande:2018ofw} \\ DUNE (400~kton-yrs): $\sim$1.0~\cite{DUNE:2020ypp} \\ THEIA (800 kton-yrs): 4.1 \vspace{3pt}} \\ \hline

$\tau_p \left( p\to \mu^+\pi^0  \right)$  &Super-K: $1.6$~\cite{Super-Kamiokande:2020wjk} &Hyper-K (1900~kton-yrs): $7.7$~\cite{Hyper-Kamiokande:2018ofw}  \\ \hline

$\tau_p \left( p\to \overline{\nu}K^+ \right)$  &Super-K: $0.66$~\cite{Mine:2016mxy} &\pbox{10cm}{\vspace{3pt} Hyper-K (1900~kton-yrs): $3.2$~\cite{Hyper-Kamiokande:2018ofw}\\DUNE (400~kton-yrs): 1.3~\cite{DUNE:2020fgq}\\JUNO (200 kton-yrs): 1.9~\cite{JUNO:2015zny} \\ THEIA (800 kton-yrs) 3.8\vspace{3pt}}  \\ \hline

$\tau_p \left( p\to \overline{\nu}\pi^+\right)$  &Super-K: $0.039$~\cite{Super-Kamiokande:2013rwg} &$-$  \\ \hline

\end{tabular}
}
\caption{Current lower limits from Super-Kamiokande on partial lifetime for different proton decay modes are presented in the second column. The third column shows future sensitivities at $90\%$ confidence level (CL) of the Hyper-K, DUNE, JUNO, and THEIA detectors. 
}
\label{tab:proton}
\end{table}

In fact, the minimal $SU(5)$ GUT (GG model) in combination with imprecise unification of gauge couplings predicts the proton lifetime of order $\tau_p\approx 10^{28}-10^{32}$ yrs and was already ruled out by early proton decay experiments. Moreover, there are additional flaws of the GG model: (i) gauge couplings do not unify, (ii) it predicts an incorrect fermion mass relation, $m_e/m_\mu = m_d/m_s$, (iii) there is a doublet-triplet splitting problem~\cite{Randall:1995sh,Yamashita:2011an} (a generic problem in most of the GUT models), i.e., the fine tuning needed to render the electroweak-doublet Higgs in the $\bf{5}$-dimensional $SU(5)$ Higgs light while keeping the color-triplet components at the GUT scale, and (iii) neutrinos remain massless, as in the SM. Nevertheless, many proposals were made in the literature towards building a viable GUT by extending the GG model; for an incomplete list of realistic models that also incorporate non-zero neutrino masses, see, e.g., Refs.~\cite{Langacker:1980js,Dorsner:2005fq, Bajc:2006ia, Dorsner:2006fx, Perez:2007rm, Bajc:2007zf, Dorsner:2007fy, Schnitter:2012bz, DiLuzio:2013dda, Dorsner:2014wva, Tsuyuki:2014xja, Perez:2016qbo, Hagedorn:2016dze, Babu:2016aro, Dorsner:2017wwn, Kumericki:2017sfc, FileviezPerez:2018dyf, Saad:2019vjo, Klein:2019jgb,Dorsner:2019vgf,Dorsner:2021qwg}. One of the most minimal renormalizable extensions of the GG model was advanced recently in Ref.~\cite{Dorsner:2019vgf} and employs fields residing in the first five lowest dimensional representations of the $SU(5)$ group.   In this model, a vectorlike fermion in the $\bf{15}$-dimensional representation is assisted with a $\bf{35}$-dimensional scalar multiplet to serve three purposes: (a) the wrong fermion mass relations are corrected, (b) unification is achieved at a high enough scale to be consistent with the current experimental bounds, and (c) neutrinos receive non-zero masses via 1-loop quantum corrections. In addition to having the least number of parameters in the Yukawa sector, this model inextricably connects the origin of neutrino masses with the experimentally observed difference between the charged lepton and down-type quark masses.  Due to its minimality, this model turns out to be very predictive. The outcomes are as follows: (a) neutrinos are Majorana particles and predicted to have normal mass ordering, (b) the lightest neutrino is massless, (c) there are four  light scalar relics at or below a 100 TeV mass scale, and (d) the proton lifetime is expected to be within $\tau_p(p\to e^+\pi^0)\approx 10^{34}-10^{36}$ yrs with an upper bound of $\tau_p\lesssim 2.3\times 10^{36}$ yrs.

One can also attempt to correct the wrong fermion mass relations by adding a $\bf{45}$-dimensional Higgs field \cite{Georgi:1979df} or by considering non-renormalizable operators~\cite{Ellis:1979fg}. However, these modifications reduce the predictivity of the model. Higher-dimensional operators, however, are not enough to provide sufficient contributions to neutrino masses. In building realistic extensions along this line, the options include the following (i) a $\bf{15}$-dimensional scalar~\cite{Dorsner:2005fq} or (ii) a $\bf{24}$-dimensional fermionic~\cite{Bajc:2006ia} representation. The former (latter) option generates correct neutrino mass via a type-II~\cite{Magg:1980ut,Schechter:1980gr,Lazarides:1980nt,Mohapatra:1980yp} (type-III~\cite{Foot:1988aq} + type-I~\cite{Minkowski:1977sc,Yanagida:1979as,Glashow:1979nm,Gell-Mann:1979vob,Mohapatra:1979ia}) seesaw mechanism. The analyses of both these theories typically suggest  that proton lifetime has an upper bound of $\tau_p(p\to e^+\pi^0) \lesssim 10^{36}$ yrs~\cite{Bajc:2007zf}.

On the other hand, predictivity can be achieved by employing the \textit{single operator dominance}  \cite{Antusch:2009gu,Antusch:2013rxa},  where a single higher-dimensional operator dominates each Yukawa entry. Group-theoretical factors from GUT symmetry breaking can lead to predictions for the ratios of quark and lepton masses at the unification scale \cite{Antusch:2009gu,Antusch:2013rxa} that  can be utilized to correct the bad mass relations between the down-type quarks and the charged-leptons. For a recent analysis utilizing this concept in the context of non-SUSY $SU(5)$ GUT that predicts nucleon decay rates from specific quark-lepton Yukawa ratios at the GUT scale, see, e.g., Ref.~\cite{Antusch:2021yqe}.

\textbf{\textbullet\; \underline{SUSY $SU(5)$ GUTs:\;}} 
If supersymmetry is realized in nature and the SUSY breaking scale is not too far above the electroweak scale, then gauge coupling unification takes place close to $M_{\textrm{GUT}}=2\times 10^{16}$ GeV, and therefore the gauge-boson-induced proton lifetime is predicted to be $\tau_p\approx 10^{35}$ yrs~\cite{Hisano:1992jj}, which is above the current experimental lower limit. However,   SUSY GUTs predict new proton decay contributions arising from dangerous dimension-5~\cite{Sakai:1981pk,Weinberg:1981wj} operators of the type $QQQL$,  among which the dominant decay mode is $p\to \overline{\nu}K^+$ \cite{Dimopoulos:1981dw,Ellis:1981tv} generated by the exchange of the
colored Higgs multiplet (of mass $M_{H_c}$); these
operators are proportional to $1/M_{H_c}$. Therefore the colored Higgs fields must be superheavy. By dressing these dimension-5 operators by gauginos or Higgsinos, dimension-6 operators are generated~\cite{Nath:1985ub,Nath:1988tx,Goto:1998qg}  that cause the proton to decay  as shown in Fig.~\ref{fig:ProtonDecay} (diagram on the right).   Typically the Wino exchange dominates; however, for larger $tan\beta$, the Higgsino exchange (the one shown in Fig.~\ref{fig:ProtonDecay}) also becomes important. Also, the Higgsino exchange dominates
the Wino exchange if the Higgsino mass is much larger than the Wino mass—such a
situation is realized in, e.g., mini-split SUSY. This is because the loop diagram like the
right figure in Fig.~\ref{fig:ProtonDecay} is accompanied with a chirality flip and thus is proportional to the
mass of the exchanged particle. 
Interestingly, assuming SUSY particles of masses of order of the electroweak scale, the lifetime of the proton is typically found to be $\tau_p(p\to \overline{\nu}K^+)\lesssim 10^{30}$ yrs~\cite{Goto:1998qg, Murayama:2001ur}. On the other hand, the Super-Kamiokande experiment gives a stringent limit on this channel: $\tau_p(p\to \overline{\nu}K^+)\gtrsim 6.6\times 10^{33}$ yrs~\cite{Mine:2016mxy}. This apparent contradiction rules out minimal SUSY  $SU(5)$ GUT~\cite{Dimopoulos:1981zb,Sakai:1981gr}  with low-scale SUSY (typically for sfermion masses $\lesssim$ TeV).

\begin{figure}[t!]\centering
\includegraphics[width=0.99\textwidth]{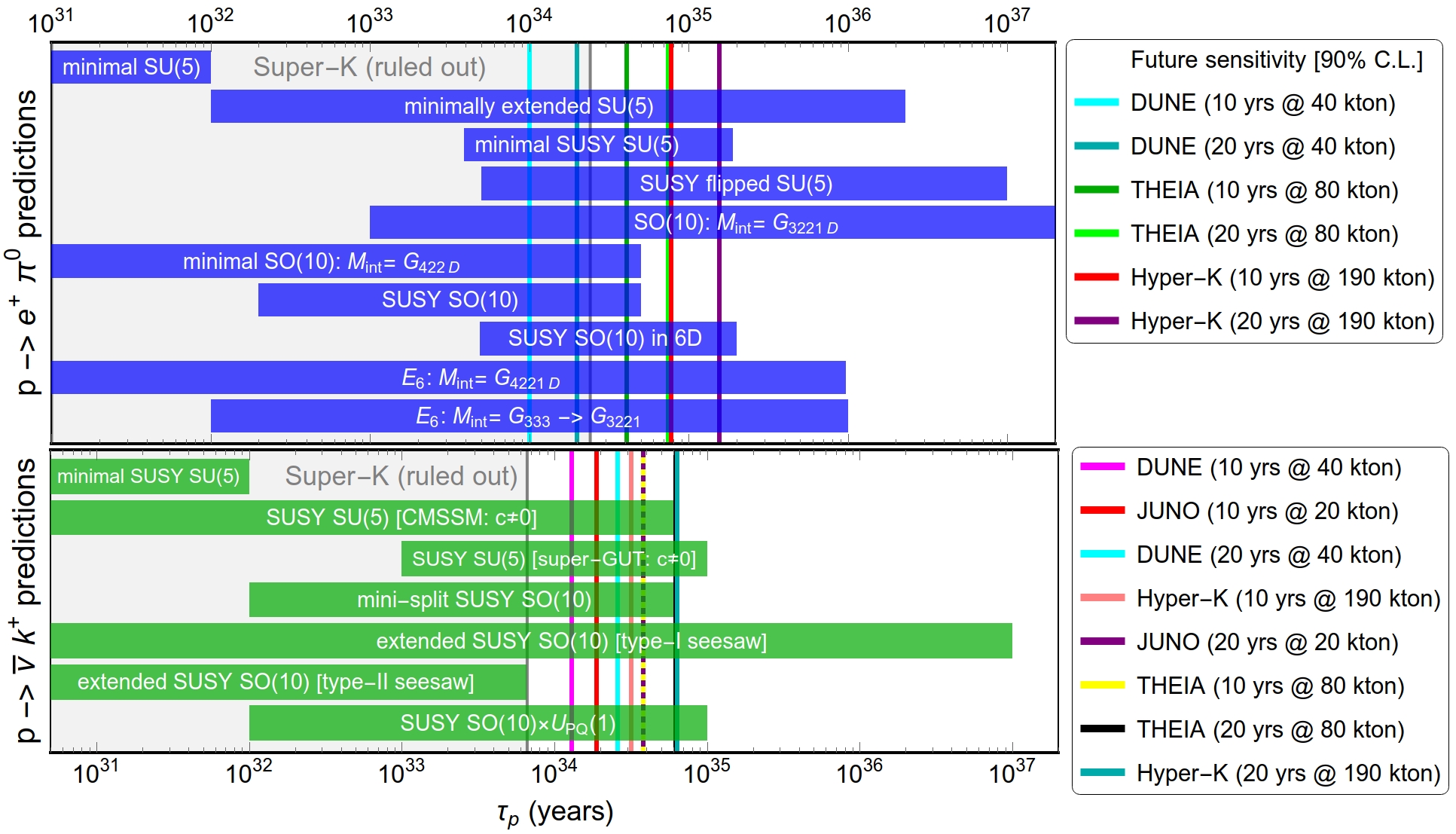}
\caption{Theoretical predictions of proton lifetime for representative GUT models are presented (for the underlying assumptions, see text). ($c$ here represents the coefficient of a Planck suppressed dimension-5 operator, for details, see Ref.~\cite{Ellis:2019fwf}.) Current Super-K data rule out the gray shaded regions. Future projections/sensitives from JUNO, DUNE, THEIA, and Hyper-K are also specified in the diagram (see Section~\ref{sec:expt} for details). }\label{fig:PREDICTIONS}
\end{figure} 

The solution to this problem resides in the Higgs boson mass which is currently accurately measured at $125.35\pm 0.15$ GeV. This is close to the upper limit of $\sim 131$ GeV predicted in supergravity unified models~\cite{Chamseddine:1982jx,Akula:2011aa} and implies that the weak SUSY scale is high lying in the several TeV region which in turn implies that
the sfermion masses could be high, even as large as 10 TeV and above. Such a scale could still be natural on the hyperbolic branch of radiative breaking of the electroweak symmetry~\cite{Chan:1997bi}. Thus the proton
lifetime exhibits a sensitive dependence on the Higgs boson mass~\cite{Liu:2013ula}. Using the precise value of the Higgs boson mass one finds for the proton decay mode $\overline{\nu}K^+$
a lifetime range of $3\times 10^{34}-2\times 10^{35}$ yrs for mSUGRA and $3\times 10^{34}-10^{36}$ yrs for SUGRA models with non-universal soft breaking (NUSUGRA) consistent with analyses that a heavy SUSY spectrum might revive the minimal $SU(5)$ GUT ~\cite{Hisano:1992jj, Goto:1998qg,  Hisano:2013exa, Nagata:2013sba,Bajc:2015ita,Babu:2020ncc,Hisano:2022qll}. Further,
 SUSY CP phases enter in proton decay and can lead to cancellations in the decay rate enhancing the lifetime~\cite{Ibrahim:2000tx}.
Similar cancellations due to specific flavor mixings are possible and can suppress proton decay rate, see, e.g., Refs.~\cite{Bajc:2002bv,Bajc:2002pg}. The cancellation mechanism
is quite general and can apply to wide class of unified models including SO(10)~\cite{Nath:2007eg}. Additionally, 
 higher-dimensional operators can affect proton decay textures which
are in general different from the textures for the fermion masses and affect proton lifetime~\cite{Nath:1996qs,Nath:1996ft} as well as can easily increase the triplet mass and thus increase the proton lifetime significantly, see, e.g., Refs.~\cite{Bachas:1995yt,Chkareuli:1998wi,Bajc:2002pg}. Recently, proton decay in minimal SUSY $SU(5)$ has been revisited in Ref.~\cite{Ellis:2019fwf} (see also Ref.~\cite{Evans:2021hyx}) where various soft SUSY-breaking conditions are imposed and  uncertainties associated with numerous phenomenological
inputs in the calculation of the proton lifetime are analyzed. For example, in constraint MSSM (cMSSM) case, assuming sparticle masses $\lesssim {O}(10)$ TeV,  the proton lifetime is found to be $\tau_p(p\to \overline{\nu}K^+) \lesssim (2-6)\times 10^{34}$  yrs which can be tested in the near future.

We emphasize that unusual decay modes, such as $p \to \mu^+ \pi^0$ and $p \to \mu^+ K^0$, can also be within the reach of the future experiments even in the minimal SUSY SU(5), if there exists flavor violation in sfermion mass matrices~\cite{Nagata:2013sba}. These decay modes can also be enhanced in flipped SU(5) GUTs with $R$ symmetry, as recently discussed in Refs.~\cite{Hamaguchi:2020tet, Mehmood:2020irm}.

\textbf{\textbullet\; \underline{$SO(10)$ GUTs:\;}} 
GUTs based on $SO(10)$ gauge symmetry~\cite{Georgi:1974my, Fritzsch:1974nn} are especially attractive since all SM fermions of each family are unified into a single irreducible $\bf{16}$-dimensional representation. Additionally,  $\bf{16}$ contains a right-handed neutrino which naturally leads to non-zero neutrino mass via a type-I seesaw mechanism. Furthermore, $SO(10)$ is free of gauge anomalies, whereas, in contrast, in $SU(5)$, the gauge anomaly due to the $\bf{5}_R$ cancels the anomaly due to the $\bf{10}_L$ of fermions (separately for each generation). The unification of all fermions of each generation into a single multiplet suggests that $SO(10)$ may serve as a fertile ground for addressing the flavor puzzle.   In fact, it turns out that $SO(10)$ GUTs are very predictive and have only a limited number of parameters to fit charged fermion as well as neutrino masses and mixings, which   have been extensively analyzed in the literature~\cite{Babu:1992ia, Bajc:2001fe,Bajc:2002iw,Fukuyama:2002ch,Goh:2003sy,
Goh:2003hf,Bertolini:2004eq, Bertolini:2005qb, Babu:2005ia,Bertolini:2006pe, Bajc:2008dc,
Joshipura:2011nn,Altarelli:2013aqa,Dueck:2013gca, Fukuyama:2015kra, Babu:2016cri, Babu:2016bmy, Saad:2017wgd, Babu:2018tfi, Babu:2018qca, Ohlsson:2019sja,Babu:2020tnf}. As shown in Ref.~\cite{Babu:2016bmy}, the most economical Yukawa sector with only $SO(10)$ gauge symmetry consists of a \textit{real} $\bf 10_H$, a \textit{real} $\bf 120_H$, and a \textit{complex} $\bf \overline{126}_H$ Higgs fields. Another widely studied class of models~\cite{Babu:1992ia}, with minimal Yukawa sector, utilizes a \textit{complex} $\bf 10_H$ and a \textit{complex} $\bf \overline{126}_H$ fields, which relies on the additional Peccei–Quinn symmetry exterior to the $SO(10)$ gauge symmetry. Moreover, from decays of the heavy right-handed neutrinos, the matter-antimatter asymmetry of the Universe~\cite{Kuzmin:1985mm,Shaposhnikov:1986jp,Shaposhnikov:1987tw} can also be generated~\cite{Fukugita:1986hr} in $SO(10)$ GUTs.

\begin{figure}[t!]\centering
\includegraphics[width=1\textwidth]{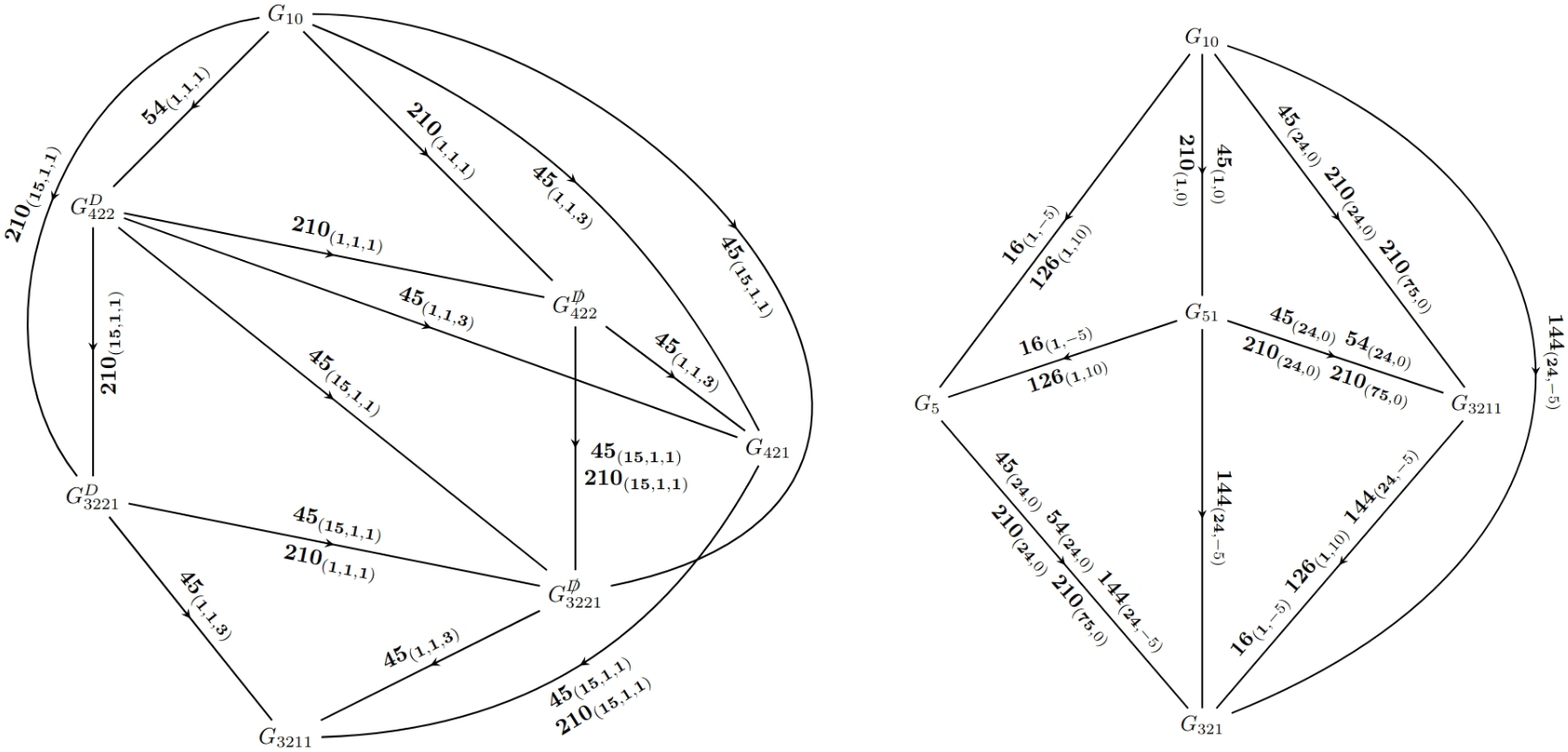}
\caption{Various symmetry breaking chains~\cite{Chang:1984qr} in  $SO(10)$ GUTs (reproduced from Ref.~\cite{Ferrari:2018rey}).  The left (right) diagram shows breaking chains along the Pati-Salam (Georgi–Glashow) route.  }\label{fig:SO10}
\end{figure}

Since the $SO(10)$ group has rank 5, whereas the SM has rank 4,   there are various possible pathways for the GUT symmetry to break down to the SM as demonstrated in Fig.~\ref{fig:SO10}. Depending on which Higgs multiplet is employed to break the GUT symmetry, at the classical level, there are typically four possibilities with minimal Higgs content for the intermediate gauge symmetry~\cite{Langacker:1980js,Lee:1994vp}:
(a) $G_{422}\times D$, (b) $G_{422}$, (c) $G_{3221}\times D$, and (d) $G_{3221}$. Here $G_{422}$ is the PS group $SU(4)_C\times SU(2)_L\times SU(2)_R$ and  $G_{3221}$ is the group of the left-right symmetric model $SU(3)_c\times SU(2)_L\times SU(2)_R\times U(1)_{B-L}$.  PS symmetry with (without) $D$-parity~\cite{Chang:1983fu} is obtained if GUT symmetry is broken by a $\bf{54}$~\cite{Lazarides:1980cc,Uzhinsky:2011qu} ($\bf{210}$~\cite{Chang:1985zq,Basecq:1988it,He:1989rb}) Higgs representation, whereas  left-right symmetry with (without) $D$-parity is achieved if a  $\bf{210}$~\cite{Chang:1985zq,Basecq:1988it,He:1989rb} ($\bf{45}+\bf{54}$~\cite{Kaymakcalan:1985us,Thornburg:1986kd,Kuchimanchi:1992aa}) Higgs is used. Depending on the intermediate gauge symmetry, these models with predominant decay mode $p\to e^+ \pi^0$ predict a lifetime in a wide range that varies in between  $10^{28}$ and $10^{40}$ yrs \cite{Lee:1994vp}. For a recent study along similar lines using a non-minimal Higgs sector in generic $SO(10)$ models see Ref.~\cite{Meloni:2019jcf, Chakrabortty:2019fov}. In these works, amounts of threshold corrections required to be consistent with present bounds on proton decay for various intermediate breaking chains are quantified.  In Ref.~\cite{Babu:2015bna},  a minimal renormalizable model with a symmetry breaking sector consisting of Higgs fields in the $\bf{10}+ \bf{54}+ \bf{\overline{126}}$ is analyzed and shown to have an upper limit on the lifetime $\tau_p(p\to e^+\pi^0)\lesssim 5\times  10^{35}$ yrs. 
A variant of this framework with the adjoint {\bf 45} scalar instead of {\bf 54} has been studied thoroughly in~\cite{Bertolini:2013vta,Kolesova:2014mfa} due to its very special quantum structure~\cite{Bertolini:2009es,Graf:2016znk,Jarkovska:2021jvw} and particular robustness with respect to the leading Planck-scale effects.   

Recently, a minimal non-renormalizable version of $SO(10)$ GUT utilizing {\bf 10}+{\bf 16}+{\bf 45} Higgs multiplets in the symmetry breaking sector was proposed in \cite{Preda:2022izo}. To be consistent with the current proton decay bounds, three scalar multiplets -- a scalar quark doublet, a weak triplet, and a color octet -- must remain lighter than $\mathcal O(10)$ TeV. With this spectrum, maximum unification scale and proton lifetime are found to be $M_{GUT}\lesssim 10^{16}$ GeV and $\tau_p \lesssim 10^{35}$ yrs, respectively. 

For another model that uses a combination of type I and type II seesaw for neutrino masses and leads to a prediction of proton lifetime in the accessible range is in Ref.~\cite{Parida:2016hln}. The model solves also the baryogenesis and DM problems of the SM.

\begin{table}[t!]
\centering
\footnotesize
\resizebox{0.99\textwidth}{!}{
\begin{tabular}{|l|l|l|l|}\hline
Model  & Decay modes & $\tau_N$ ($N=p, n$) [years] & Ref.\\ \hline \hline

\rowcolor{blue!10}
Non-SUSY minimal $SU(5)$  & $p\to e^+ \pi^0$ & $10^{30}-10^{32}$ & Georgi, Glashow~\cite{Georgi:1974sy}\\ \hline

\rowcolor{blue!10}
Non-SUSY minimally extended   & $p\to e^+ \pi^0$ & $\lesssim 2.3\times 10^{36}$ & Dor\v{s}ner, Saad~\cite{Dorsner:2019vgf}\\\rowcolor{blue!10}
$SU(5)$ (neutrino mass: 1-loop)&&&  \\ \hline

\rowcolor{blue!10}
Non-SUSY minimally extended   & $p\to e^+\pi^0$ & $10^{32}-10^{36}$ & Perez, Murgui~\cite{Perez:2016qbo}\\\cline{2-3} \rowcolor{blue!10}
$SU(5)$ (neutrino mass: 1-loop)&$p\to \overline{\nu}K^+$&$10^{34}-10^{37}$& \\ \hline

\rowcolor{blue!10}
Non-SUSY Minimal $SU(5)$ [NR] & $p\to \nu+\left(K^+, \pi^+, \rho^+\right)$ & $10^{31}-10^{38}$ & Dor\v{s}ner, Perez~\cite{Dorsner:2005fq}\\\rowcolor{blue!10}
(neutrino mass: type-II seesaw)&$n\to \nu+\left(\pi^0, \rho^0, \eta^0, \omega^0, K^0\right)$&& \\ \hline

\rowcolor{blue!10}
Non-SUSY Minimal $SU(5)$ [NR] & $p\to e^+ \pi^0$ & $\lesssim  10^{36}$ & Bajc, Senjanovi\'c~\cite{Bajc:2006ia}\\\rowcolor{blue!10}
(neutrino mass: type-III+I seesaw)&&& \\ \hline

\rowcolor{blue!10}
Non-SUSY Extended $SU(5)$  & $p\to e^+ \pi^0$ & $10^{34}-10^{40}$ & Saad~\cite{Saad:2019vjo}\\\rowcolor{blue!10}
(neutrino mass: 2-loop)&&& \\ \hline 

\rowcolor{blue!10}
Minimal flipped non-SUSY $SU(5)$ & $p\to e/\mu^+\pi^0$ & $10^{38}-10^{42}$& Arbel\'{a}ez, Kole\v{s}ov\'a, Malinsk\'y \cite{ArbelaezRodriguez:2013kxw}\\ \hline\hline

\rowcolor{green!10}
Non-SUSY Minimal $SO(10)$  & $p\to e^+ \pi^0$ & $\lesssim 5\times 10^{35}$ & Babu, Khan~\cite{Babu:2015bna}\\ \hline
\rowcolor{green!10}
Minimal $SO(10)$ with 45 Higgs & $p\to e^+ \pi^0$ & $\lesssim 10^{36}$ & Bertolini, Di Luzio, Malinsk\'{y}~\cite{Bertolini:2012im}\\ \hline

\rowcolor{green!10}
Minimal non-Renormalizable $SO(10)$  & $p\to e^+ \pi^0$ & $\lesssim 10^{35}$ & Preda, Senjanovi\'c, Zantedeschi~\cite{Preda:2022izo}\\ \hline

\rowcolor{green!10}
Non-SUSY Generic $SO(10)$   & $p\to e^+ \pi^0$ &  & Chakrabortty, King, Maji~\cite{Chakrabortty:2019fov}

\\\rowcolor{green!10}
$M_{\textrm{int}}:\;G_{422}$&&$10^{34}-10^{46}$&
\\\rowcolor{green!10}
$M_{\textrm{int}}:\;G_{422D}$&&$10^{31}-10^{34}$&
\\\rowcolor{green!10}
$M_{\textrm{int}}:\;G_{3221}$&&$10^{36}-10^{46}$&
\\\rowcolor{green!10}
$M_{\textrm{int}}:\;G_{3221D}$&&$10^{33}-10^{43}$&
\\ \hline \hline

\rowcolor{orange!10}
Non-SUSY Generic $E_6$   & $p\to e^+ \pi^0$ &  & Chakrabortty, King, Maji~\cite{Chakrabortty:2019fov}

\\\rowcolor{orange!10}
$M_{\textrm{int}}:\;G_{4221}$&&$10^{27}-10^{36}$&
\\\rowcolor{orange!10}
$M_{\textrm{int}}:\;G_{4221D}$&&$10^{27}-10^{36}$&
\\\rowcolor{orange!10}
$M_{\textrm{int}}:\;G_{333}\to G_{3221}$&&$10^{32}-10^{36}$&
\\\rowcolor{orange!10}
$M_{\textrm{int}}:\;G_{4221D}\to G_{421}$&&$10^{26}-10^{48}$&
\\\rowcolor{orange!10}
$M_{\textrm{int}}:\;G_{4221}\to G_{421}$&&$10^{25}-10^{48}$&
\\ \hline \hline

\rowcolor{cyan!10}
Minimal SUSY $SU(5)$  &  $p\to \bar{\nu}K^+$ && Dimopoulos, Georgi~\cite{Dimopoulos:1981zb}, Sakai~\cite{Sakai:1981pk} \\\rowcolor{cyan!10} 
 &$n\to \bar{\nu}K^0$ & $10^{28}-10^{32}$& Hisano, Murayama, Yanagida~\cite{Hisano:1992jj}\\ \hline

\rowcolor{cyan!10}
Minimal SUSY $SU(5)$   & $p\to \bar{\nu}K^+$ &  $\lesssim (2-6)\times 10^{34}$& Ellis et. al.~\cite{Ellis:2019fwf}\\\cline{2-3} \rowcolor{cyan!10}
(cMSSM)&$p\to e^+\pi^0$&$10^{35}-10^{40}$&\\ \hline

\rowcolor{cyan!10}
Minimal SUSY $SU(5)$   & $p\to \bar{\nu}K^+$ &  $\lesssim 4\times 10^{33}$& Babu, Bajc, Tavartkiladze~\cite{Babu:2012pb}\\\cline{2-3} \rowcolor{cyan!10}
($\bf 5+\overline 5$ matter fields)&$p\to \mu^+\pi^0/K^0, n\to \overline{\nu} \pi^0/K^0$&$10^{33}-10^{34}$&\\ \hline

\rowcolor{cyan!10}
SUGRA $SU(5)$  & $p\to \bar{\nu}K^+$ &  $10^{32}-10^{34}$& Nath, Arnowitt~\cite{Nath:1985ub, Nath:1998kg}\\ \hline
\rowcolor{cyan!10}
mSUGRA $SU(5)$ (Higgs mass constraint)  &  $p\to \bar{\nu}K^+$ & $3\times 10^{34}-2\times 10^{35}$  &Liu, Nath~\cite{Liu:2013ula}\\ \cline{2-3} \rowcolor{cyan!10}
NUSUGRA $SU(5)$ (Higgs mass constraint) &$p\to \bar{\nu}K^+$ &  $3\times 10^{34}-10^{36}$&\\ \hline

\rowcolor{cyan!10}
SUSY $SU(5)$ or $SO(10)$&$p\to e^+ \pi^0$&$\sim 10^{34.9\pm 1}$&Pati~\cite{Pati:2003qi}\\\rowcolor{cyan!10}
MSSM ($d=6$)&&&\\ \hline

\rowcolor{cyan!10}
Flipped SUSY $SU(5)$ (cMSSM) & $p\to e/\mu^+\pi^0$ & $10^{35}-10^{37}$& Ellis et. al. \cite{Ellis:2002vk,Ellis:2021vpp,Ellis:2020qad}\\ \hline

\rowcolor{cyan!10}
Split SUSY $SU(5)$   & $p\to e^+\pi^0$ & $10^{35}-10^{37}$& Arkani-Hamed, \emph{et. al.}~\cite{Arkani-Hamed:2004zhs}\\ \hline

\rowcolor{cyan!10}
SUSY $SU(5)$ in 5D  & $p\to \mu^+K^0$ & $10^{34}-10^{35}$& Hebecker, March-Russell\cite{Hebecker:2002rc}\\\rowcolor{cyan!10}
&$p\to e^+ \pi^0$ &&\\ \hline

\rowcolor{cyan!10}
SUSY $SU(5)$ in 5D variant II & $p\to \bar{\nu}K^+$ & $10^{36}-10^{39}$& Alciati \emph{et.al.}\cite{Alciati:2005ur}\\
\hline \hline

\rowcolor{red!10}
Mini-split SUSY $SO(10)$  & $p\to \bar{\nu}K^+$ & $\lesssim 6\times 10^{34}$& Babu, Bajc, Saad~\cite{Babu:2018tfi}\\ \hline

\rowcolor{red!10}
SUSY $SO(10)\times U(1)_{\textrm{PQ}}$  & $p\to \bar{\nu}K^+$ & $10^{33}-10^{35}$& Babu, Bajc, Saad~\cite{Babu:2018qca}\\ \hline

\rowcolor{red!10}
Extended SUSY $SO(10)$  & $p\to \bar{\nu}K^+$ & & \\\rowcolor{red!10}
Type-I seesaw&&$10^{30}-10^{37}$&Mohapatra, Severson~\cite{Mohapatra:2018biy}
\\\rowcolor{red!10}
Type-II seesaw&&$\lesssim 6.6\times 10^{33}$&Mohapatra, Severson~\cite{Mohapatra:2018biy}
\\
\rowcolor{red!10}
Inverse seesaw&&$\lesssim 10^{34}$&Dev, Mohapatra~\cite{BhupalDev:2010he}
\\ \hline

\rowcolor{red!10}
SUSY $SO(10)$ & $p\to \bar{\nu}K^+$ && Shafi, Tavartkiladze~\cite{Shafi:1999vm} \\\rowcolor{red!10}
with anomalous& $n\to \bar{\nu}K^0$ & $10^{32}-10^{35}$&\\\rowcolor{red!10}
flavor $U(1)$ & $p\to \mu^+K^0$ &&\\ \hline

\rowcolor{red!10}
SUSY $SO(10)$ &  $p\to \bar{\nu}K^+$ &  $10^{33}-10^{34}$&Lucas, Raby~\cite{Lucas:1996bc}, Pati~\cite{Pati:2003qi}\\ \cline{2-3} \rowcolor{red!10}
MSSM &$n\to \bar{\nu}K^0$ &  $10^{32}-10^{33}$&\\ \hline

\rowcolor{red!10}
SUSY $SO(10)$&$p\to \bar{\nu}K^+$&$10^{33}-10^{34}$&Pati~\cite{Pati:2003qi}\\\rowcolor{red!10}
ESSM &&$\lesssim 10^{35}$&\\ \hline

\rowcolor{red!10}
SUSY $SO(10)/G(224)$ &  $p\to \bar{\nu}K^+$& $\lesssim 2\cdot 10^{34}$& Babu, Pati, Wilczek~\cite{Babu:1997js, Babu:1998wi, Pati:2000wu},\\\cline{2-3} \rowcolor{red!10}
MSSM or ESSM  & \multicolumn{2}{l|}{$p\to \mu^+K^0$}&Pati~\cite{Pati:2003qi}\\ \rowcolor{red!10}
(new $d=5$)&\multicolumn{2}{r|}{$B\sim(1-50)\%$}&\\ \hline

\rowcolor{red!10}
SUSY $SO(10)\times S_4$  & $p\to \bar{\nu}K^+$ & $\lesssim 7\times 10^{33}$&  Dev, Mohapatra, Dutta, Severson~\cite{BhupalDev:2012nm}\\ \hline

\rowcolor{red!10}
SUSY $SO(10)$ in 6D & $p\to e^+ \pi^0$ & $10^{34}-10^{35}$ & Buchmuller, Covi, Wiesenfeldt~\cite{Buchmuller:2004eg}\\ \hline\hline

\rowcolor{yellow!10}
GUT-like models from & $p\to e^+ \pi^0$ & $\sim10^{36}$& Klebanov, Witten~\cite{Klebanov:2003my}\\ \rowcolor{yellow!10}
Type IIA string with D6-branes &  & &\\ \hline

\end{tabular}
}
\caption{Synopsis of the expected nucleon lifetime in various representative GUT  models (see also Ref.~\cite{Bueno:2007um}). NR here stands for non-renormalizable. For details, see text.}
\label{tab:MODELS}
\end{table}

\textbf{\textbullet\; \underline{SUSY $SO(10)$ GUTs:\;}} 
In SUSY $SO(10)$ GUTs, if the intermediate symmetry is broken by a $\bf{\overline{126}}$ Higgs that breaks ${B-L}$ by two units, in addition to giving large masses to the right-handed neutrinos (that in turn generates tiny neutrino masses), R-parity of the MSSM automatically emerges from within the $SO(10)$ symmetry. Therefore, SUSY $SO(10)$ GUTs are highly attractive as they naturally provide a DM candidate and prohibit $d=4$ baryon number violating operators without any additional symmetry assumptions.  The minimal renormalizable model~\cite{Aulakh:1982sw,Clark:1982ai,Babu:1992ia} with a symmetry breaking sector consisting of $\bf{10}+ \bf{126}+ \bf{\overline{126}}+ \bf{210}$ was found to be very successful in fitting fermion data~\cite{Bajc:2002iw,Fukuyama:2002ch,Goh:2003sy,Goh:2003hf, Bertolini:2004eq,Babu:2005ia,Bertolini:2006pe,Joshipura:2011nn,Altarelli:2013aqa,Dueck:2013gca,Bajc:2008dc}. Soon after, it was realized~\cite{Aulakh:2005bd,Bajc:2005qe,Aulakh:2005mw,Bertolini:2006pe} that the intermediate symmetry breaking scale that is required to be of order $10^{12}-10^{13}$ GeV from fits to light neutrino masses, subsequently leads to certain colored particles from various Higgs fields with masses of order  $10^{10}$ GeV that spoil the successful perturbative gauge coupling unification. Besides that, current proton decay limits completely rule out this version of the model with TeV scale SUSY spectrum.

These caveats can be resolved by extending the minimal version of the SUSY $SO(10)$ GUT.  Without introducing any new parameter in the Yukawa sector, it was shown in Ref.~\cite{Babu:2018tfi} that a minimal realistic extension is to add a $\bf{54}$ multiplet to the symmetry breaking sector.  This setup allows the intermediate breaking scale of order  $10^{12}-10^{13}$ GeV to fit neutrino data, yet preserves perturbative gauge coupling unification. The viability of this model requires  a mini-split SUSY spectrum with sfermion masses of order 100 TeV using the current experimental lower bound $\tau_p(p\to \overline{\nu}K^+)> 6.6\times 10^{33}$ yrs. Even though squarks and sleptons are far beyond the reach of LHC,  the model can be tested at colliders with the discovery of LSP (Wino) along with its charged partners. Improvement of the proton lifetime limits in the channel $p\to \overline{\nu}K^+$ by one order of magnitude will highly disfavor this version of the model (expected upper limit on the proton lifetime with ${O}(100)$ TeV sfermions  is $\tau_p \lesssim 6\times 10^{34}$ yrs).  On the other hand, implementation of  Peccei-Quinn (PQ) symmetry~\cite{Peccei:1977hh} to solve the strong CP problem (for a recent review see Ref.~\cite{DiLuzio:2020wdo})  in renormalizable SUSY $SO(10)$ with minimal Yukawa sector allows TeV scale sfermion masses as shown in Ref.~\cite{Babu:2018qca}. This is possible since Higgsino mediated proton decay rate is strongly suppressed by an additional factor of $(M_{\rm PQ}/M_{\rm GUT})^2$~\cite{Hisano:1992ne,Babu:2018qca} and the expected proton lifetime in this framework is $\tau_p \approx 10^{33}-10^{36}$ yrs.   This scenario is exciting since the proton decay suppression mechanism is inherently linked to the solution to the strong CP problem. 

The axion solution to the strong CP problem is particularly interesting since axion can be a cold DM candidate. Models in which the axion field is embedded into a scalar representation responsible for breaking the GUT symmetry show an interesting correlation between the proton decay rate and the axion mass. Models of this type were first proposed in the context of $SU(5)$ in Ref.~\cite{Wise:1981ry} and $SO(10)$ in Ref.~\cite{Lazarides:1981kz}. For recent works along this line, see, e.g., Refs.~\cite{Ernst:2018bib,DiLuzio:2018gqe,Ernst:2018rod,FileviezPerez:2019ssf}.

Another option to accommodate low scale SUSY in the $SO(10)$ context is to permit more parameters in the Yukawa sector by introducing a $\bf{120}$ multiplet on top of $\bf{10}+ \bf{\overline{126}}$~\cite{Dutta:2004zh, Mohapatra:2018biy}. With new parameters in the Yukawa sector, it is possible to realize some cancellations to save the theory from too rapid proton decay even with a TeV scale  SUSY spectrum. As shown in Ref.~\cite{Mohapatra:2018biy}, type-II seesaw case is highly disfavored by the current proton decay lower limit, whereas for type-I scenario, the proton lifetime for the channel $p\to \overline{\nu}K^+$ can reach up to $10^{37}$ yrs.

SUSY $SO(10)$ GUTs that adopt small Higgs representations, to be specific, $\bf{16}$ instead of $\bf{\overline{126}}$, do not guarantee automatic R-parity, but they still provide quite simply matter-parity which keeps the lightest SUSY particle stable to serve as DM.  Nevertheless, models of this class (consisting of $\bf{16}+\bf{\overline{16}}$, $\bf{10}$, and $\bf{45}$ Higgs fields) are particularly interesting, not only because doublet-triplet splitting occurs naturally but also  an interesting correlation of the decay rates between the two major proton decay modes $p\to \overline{\nu}K^+$ and $p\to e^0\pi^+$ emerges~\cite{Babu:2010ej}. This interdependence shows that  the empirical lower limit on the lifetime for $p\to \overline{\nu}K^+$ provides a constrained upper limit on the lifetime for $p\to e^+ \pi^0$ mode and vice versa. Based on this correlation, estimated upper limits for proton lifetimes have been obtained in the context of an updated version of Ref.~\cite{Babu:2010ej} (to be published; private
communications from the authors of Ref.~\cite{Babu:2010ej}; see also Sec. 7.2 of Ref.~\cite{Pati:2017ysg}). This updated version takes two factors into account: (a) the current LHC constraints
on SUSY spectrum by allowing for an inverted hierarchy of the sfermion masses with a light stop ($\sim$ 800 GeV to 1 TeV) and lighter higgsino/bino ($\sim$ 700-800GeV), together with heavy masses ($\sim$ 18-20 TeV) for the sfermions of the first two generations on the one hand, and (b) the appropriate normalization factors in the definitions of certain effective couplings involving identical fields, which were missed in Ref.~\cite{Babu:2010ej}, on the other hand. With this updating, the correlation mentioned above yields conservative estimates for the upper limits of $\tau_p(p\to e^+\pi^0)$ less than about $(2 - 10) \times 10^{34}$ yes, and $\tau_p(p\to \overline \nu K^+)$ less than about $(1 - 8)\times 10^{34}$ yrs. These upper limits are within the reach of the Hyper-K and Dune detectors including their planned upgrades respectively.

Moreover, GUT models with family symmetries (for an incomplete list of references, see, e.g., Refs.~\cite{Meloni:2017cig,Altarelli:2010gt,Dutta:2009bj,King:2009tj,Hagedorn:2010th,Chen:2007afa,Hagedorn:2008bc,Ishimori:2008fi,Joshipura:2009tg,Antusch:2010es,deMedeirosVarzielas:2011wx,Antusch:2013wn,Feruglio:2015jfa,Xing:2020ijf}) are particularly interesting since they tend to explain the flavor puzzle. An important aspect of this class of models is that Yukawa couplings emerge dynamically from
minimization of the flavon potential, thereby reducing the number of parameters considerably.  For example, assuming $S_4$ flavor symmetry on top of $SO(10)$ gauge symmetry, in Ref.~\cite{BhupalDev:2012nm}, in addition to explaining the hierarchies in the charged fermion masses and mixings, neutrino observables are also predicted (such as $\theta_{13}\sim 9^\circ$). Furthermore, the proton decay mode $p\to \overline{\nu}K^+$  in this model is found to have a lifetime of $\sim 10^{34}$ yrs, which is within reach of the upcoming experiments.  

It should be noted that unified models such as $SO(10)$ with appropriate symmetry breaking that produce the SM gauge group are also constrained by electroweak physics. It is then of interest to investigate the consistency of the unified models (such as Yukawa coupling unification) with the recent result from the Fermilab E989 experiment \cite{Muong-2:2021ojo} on the muon anomalous magnetic moment. The Fermilab experiment has measured $a_\mu=(g_\mu-2)/2$ with a significantly greater accuracy than the previous Brookhaven \cite{Muong-2:2006rrc} experiment and the combined Fermilab and Brookhaven experimental data gives a $4.2\sigma$ deviation \cite{Aoyama:2020ynm} from the SM. An investigation of the Yukawa coupling unification for the third generation in a class of SUSY $SO(10)$ unified models~\cite{Babu:2011tw,Aboubrahim:2021phn} shows that the $SO(10)$ model is fully consistent with the Fermilab result.

Finally we mention two new classes of $SO(10)$ models not discussed thus far. As noted  above, $SO(10)$ models require several Higgs representations to break
 the GUT symmetry to the SM gauge group. Thus a $\bf{16}$ or a $\bf{126}$ of Higgs field is needed
 to change rank, and one of $\bf{45}$, $\bf{54}$ or a $\bf{210}$ is needed to break the symmetry down further, and
 to achieve electroweak symmetry breaking and to generate quark and lepton masses an additional {\bf 10} dimensional representation is also needed. Recently a class of $SO(10)$ models have been proposed
 consisting of  $\bf{144+\overline{144}}$ of Higgs fields~\cite{Babu:2005gx,Babu:2006rp}
 which can break the symmetry down to $SU(3)_c\times U(1)_{em}$. Proton decay analysis in this model exhibits a cancellation mechanism consistent with the current experimental constraints
 and the possibility of observation in future experiment~\cite{Nath:2007eg}. 
  Another class of $SO(10)$ models relates to the doublet-triplet splitting problem which as noted earlier 
  is quite generic in grand unified models. One way to resolve it is the so called missing VEV mechanism~\cite{Dimopoulos:1981xm,Babu:1993we} (see also Refs.~\cite{Lucas:1996bc,Barr:1997hq,Babu:2010ej}) where the vacuum expectation value of a $\bf{45}$ Higgs field which breaks the $SO(10)$ symmetry lies in the $(B-L)$-preserving direction, and generates masses for the Higgs triplets but not for the Higgs doublets from a {\bf 10}–plet. This mechanism works in $SO(10)$ and has no analog in SU(5). A second approach is the missing partner mechanism~\cite{Masiero:1982fe,Babu:2006nf,Babu:2011tw}.
  For $SO(10)$ it leads to a variety of models discussed in~\cite{Babu:2011tw}.
  $B-L=-2$ violating dimension-7 and dimension-9 
 operators have been computed in this class of models~\cite{Nath:2015kaa}. Thus previous analyses using a bottom up effective low energy operator approach can now  be extended to a top down one~\cite{Nath:2015kaa}  for GUT scale baryogenesis and for $B-L=-2$ proton decay such $p\to \nu \pi^+$, $n\to e^-\pi^+$, $e^-K^+$ and $n-\bar n$ oscillations.

In short, well-motivated non-SUSY and SUSY GUTs generically predict rates of BNV processes that can be probed by next-generation experiments if not already ruled out by the current experimental data. A sketch of theoretical predictions for selected models and experimental reach of upcoming detectors are depicted in  Fig.~\ref{fig:PREDICTIONS} for non-SUSY and SUSY GUTs.  Furthermore, nucleon decay predictions for  a wide range of models are summarized in Table~\ref{tab:MODELS}. For details on theoretical assumptions associated with each model's predictions, the readers are referred to the original literature.

As a cautionary remark, it is worth noting that none of the predictions in Fig.~\ref{fig:PREDICTIONS} or in Table~\ref{tab:MODELS} is actually sharp; one typically encounters ranges stretching over several orders of magnitude. This has to do with a number of theoretical uncertainties affecting the precision of the calculations at various levels of significance. 
These can be loosely divided into three main classes corresponding to different ways the quantitative estimates based on diagrams in Fig.~\ref{fig:ProtonDecay} are influenced:
(i)~uncertainties in the determination of the masses of the relevant leptoquark fields (i.e., the GUT scale),
(ii)~uncertainties in the couplings (gauge, Yukawa) governing the GUT-scale dynamics and
(iii)~uncertainties in the relevant hadronic or nuclear matrix elements.
As for the first class, the most prominent of these effects are the uncertainties related to the generally unknown shape of the GUT-scale spectrum of the models at stakes, to the proximity of the GUT and the Planck scales enhancing the uncontrolled corrections from  higher-dimensional operators (for instance those due to gravity effects)~\cite{Hill:1983xh,Shafi:1983gz,Calmet:2008df} and to the limited precision attainable in the perturbative  accounts (see e.g. Ref.~\cite{Jarkovska:2021jvw}), all inflicting uncertainties in the GUT-scale matching conditions.   
The second class corresponds to the limits in our understanding of the flavor structure of the relevant $B$- and $L$-violating charged currents (cf.~\cite{Dorsner:2004jj,Kolesova:2016ibq}) which, indeed, gets only partially reflected in the flavor observables accessible at low energies (the quark and lepton masses and mixings).  It is also worth mentioning that unknown CP phases in the GUT Yukawa couplings~\cite{Ellis:1979hy} can change the proton decay rate significantly (see, e.g., Ref.~\cite{Ellis:2016tjc} and Ref.~\cite{Ellis:2019fwf}).   
The third class concerns the general difficulty with ab initio QCD calculations in the strongly coupled low-energy regime.  
While the modern lattice techniques have recently brought enormous progress in point (iii), cf. Sect.~\ref{sec:Lattice}, (i) and (ii) still represent a formidable challenge to any precision calculations with rather limited prospects for any significant near-future improvement.

Thus far, we have discussed gauge-mediated proton decay, which dominates if the mass scale of vector-bosons and scalar-bosons are of a similar order. This happens for the latter contribution because the first-generation Yukawa couplings suppress the scalar-mediated proton decay.   Naively, color-triplet scalar contributions become significant only if $M_S\lesssim \mathcal{O}(10^{-4}) M_V$. It should be emphasized that the scalar-mediated contributions depend highly on the Yukawa sector of the theory and are model-dependent to a large extent. For minimal $SO(10)$ GUTs, using typical Yukawa coupling structures, color-triplet masses need to be heavier than about $10^{10}-10^{11}$ GeV \cite{Patel:2022wya}. In the context of a minimal model based on $\bf 10_H$ and $126_H$ with Peccei-Quinn symmetry, when scalar-mediated contributions dominate, the proton decay spectrum is found to
be quite different from the one typically anticipated \cite{Patel:2022wya}. For example, (i) proton dominantly decays into $\overline{\nu}K^+$ or $\mu^+K^0$ for lighter $(\overline 3,1,1/3)$ or $(3,1,-1/3)$, respectively, and (ii)  $BR(p\to \mu^+\pi^0)\gg BR(p\to e^+\pi^0)$ is expected.
Both these features are distinct from gauge-mediated proton decays; hence, if scalar-mediated contributions dominate, proton decay can provide very useful insight into the Yukawa structure of the theory.

\textbf{\textbullet\; \underline{GUTs in extra spatial dimensions:\;}}
Extra spatial dimensions provide a useful avenue for GUT model building in which some of the usual problems of four-dimensional GUTs can be addressed. For example, some versions of orbifold GUTs \cite{Kawamura:2000ev,Altarelli:2001qj,Hebecker:2002rc,Hall:2001pg} use compactification to break the unified gauge symmetry avoiding the doublet-triplet splitting problem. Moreover, localizing fermion fields appropriately in the bulk of extra dimension, a natural understanding of hierarchical fermion mass spectrum can be achieved in certain versions of GUTs \cite{Feruglio:2014jla,Feruglio:2015iua,Kitano:2003cn}. In 6D GUTs, the origin of multiple families of matter fields can be realized by quantization of flux in torus constructed from the compact extra two dimensions \cite{Buchmuller:2015jna,Buchmuller:2017vho,Buchmuller:2017vut}. In these classes of GUTs, the Yukawa couplings become calculable parameters and the quarks and lepton spectrum can be obtained from a very small number of parameters. In the extra-dimensional GUT models, proton decay can have distinct features. For example, in a class of 5D models, the proton decay mediated by vector bosons has an additional suppression due to the wave-function profiles of first-generation fermions and the underlying vector bosons in the bulk \cite{Altarelli:2001qj,Dermisek:2001hp,Alciati:2005ur,Alciati:2006sw}. In SUSY orbifold GUTs, the dimension-5 operators are often suppressed due to $U(1)_R$ symmetry \cite{Alciati:2005ur,Alciati:2006sw}. In 6D models with flux compactification, the proton decay also receives a non-trivial contribution from the higher Kaluza-Klein modes of the vector bosons \cite{Buchmuller:2019ipg}. This, along with the special flavour structure of these theories, implies proton dominantly decaying into $\mu^+\, \pi^0$ which is otherwise suppressed in the 4D GUT models.

The absence of indication of low energy supersymmetry in the experimental searches so far has inspired investigations of the GUTs in the regime of split \cite{Bajc:2008dc} or high scale SUSY \cite{Joshipura:2012sr}. Gauge coupling unification in this class of models is achieved by keeping only a part of SUSY spectrum at the intermediate energy scale while the remaining MSSM fields stay close to the GUT scale. Despite having most of the spectrum at intermediate or at very high energies, this class of theories are still reasonably constrained from the perspectives of Higgs mass, stability of the electroweak vacuum, flavour transitions, dark matter and proton decay \cite{Bagnaschi:2015pwa,Mummidi:2018nph,Mummidi:2018myd,SuryanarayanaMummidi:2020ydm}. It has been shown that if TeV scale Higgssino forms almost all of the dark matter of the universe then it leads to particular ranges for the unified couplings and the scale of unification which in turn put an upper bound on the proton decay requiring the proton lifetime less than $7 \times 10^{35}$ years \cite{Mummidi:2018myd,SuryanarayanaMummidi:2020ydm}.

Finally, we point out that a public software package \texttt{SusyTCProton} for nucleon decay calculations in non-SUSY and SUSY GUTs is available, which includes, e.g., the full loop-dressing at the SUSY scale; for details, see Ref.~\cite{Antusch:2020ztu}.

\subsection{Pati-Salam Partial Unification}
The first step towards unification was made  in Ref.~\cite{Pati:1973rp} (see also Refs.~\cite{Pati:1973uk,Pati:1974yy,Pati:2017ysg}) that are based on  partial unification with non-Abelian gauge group $G_{422}=SU(4)_C\times SU(2)_L\times SU(2)_R$. By proposing the $G_{422}$ symmetry, this work~\cite{Pati:1973rp} introduced many new concepts into the literature, which include: (i) Quark-lepton unification; (ii) The first realistic model of matter and its three interactions which quantized electric charge, and thus (as was realized later) the first quantum-theoretically consistent model of magnetic monopoles; (iii) The first left-right symmetric gauge structure providing a compelling reason for the existence of the right-handed neutrino and the right-handed gauge-boson $W_R$; and (iv) The first model that questioned baryon number conservation and proton stability in the context of higher unification; while this was in the context of integer-charge quarks, nevertheless it served to remove partly the major conceptual barrier against BNV violation and proton instability that existed in the community in the early 1970s.

In this PS model,  unlike $SO(10)$, there is no gauge-mediated proton decay. In fact in the original PS model, proton decay appears only if the quarks are chosen to have integer charges. This is why PS symmetry can be realized at rather low energy scales. In minimal models, the PS breaking scale can be as low as $v_R\gtrsim 10^3$ TeV~\cite{Valencia:1994cj,Smirnov:2007hv}, which is coming from experimental limits on the branching ratios for $K_L^0 \to \mu^\pm e^\mp$ mediated by the new gauge bosons $X_\mu$ carrying $4/3$ charge under ${B-L}$. However, in the extended models, where additional fermionic degrees of freedom are introduced, the PS symmetry can even break down close to the TeV scale that has the potential to be directly probed at colliders; for a recent study, see, e.g., Ref.~\cite{Dolan:2020doe}.  In light of recent anomalies~\cite{LHCb:2015gmp,LHCb:2017avl,LHCb:2014vgu,LHCb:2019hip} in the beauty-meson decays, low scale PS models have gained a lot of attention. Within this setup, the vector leptoquark $X_\mu$ is an attractive candidate to explain some of the flavor anomalies ~\cite{Assad:2017iib,DiLuzio:2017vat,Calibbi:2017qbu,Bordone:2017bld,Blanke:2018sro,Heeck:2018ntp,Balaji:2018zna,Fornal:2018dqn,Balaji:2019kwe}.

\begin{figure}[th!]\centering
\includegraphics[width=0.65\textwidth]{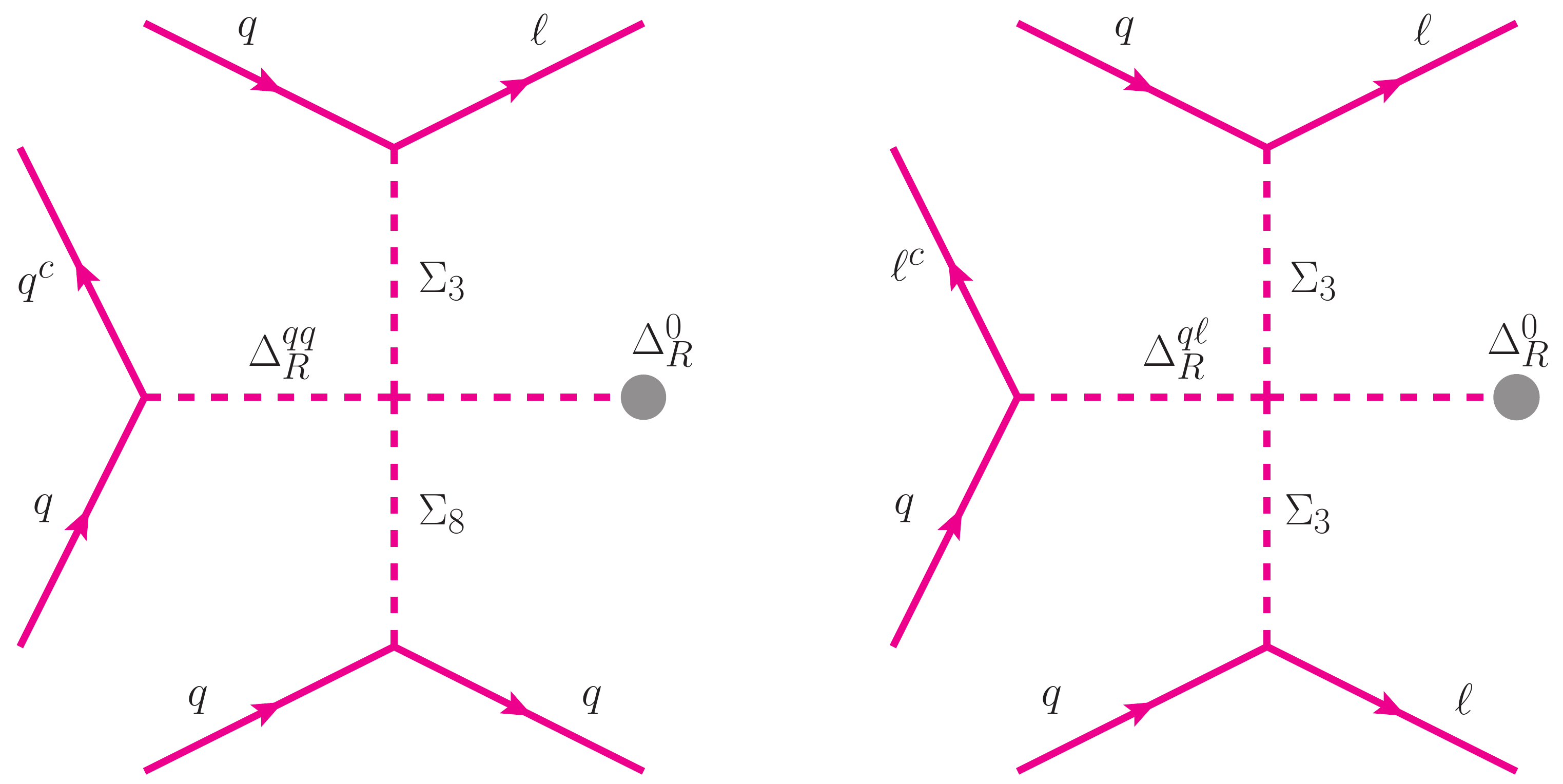}
\caption{Nucleon decay modes in the $SU(4)_C\times SU(2)_L\times SU(2)_R$  model with inclusion of color sextet Higgs fields $\Delta_R$ a la Ref.~\cite{Mohapatra:1980qe}. Here the scalar  sub-multiplets $\Delta_R^{qq}$, $\Sigma_8$, and $\{\Sigma_3, \Delta_R^{q\ell}\}$ are color sextet, di-quark, and leptoquarks, respectively;  moreover, $\Delta_R^0$ is SM singlet that acquires a VEV and breaks ${B-L}$ by two units. The Feynman diagram on the left (right) leads to nucleon decay of the type $3q\to \ell q \overline q$ ($3q\to  \ell \ell  \overline \ell$).  }\label{fig:Nucleon-PS}
\end{figure} 

Even though baryon number violation does not proceed through gauge bosons, such processes naturally arise via the scalar sector, even with the minimal Higgs content~\cite{Davidson:1978pm,Mohapatra:1980qe,Pati:1983zp,Pati:1983jk}.  In the minimal model, to generate a realistic charged fermion mass spectrum, the presence of both $\Phi({\bf 1},{\bf 2},{\bf 2})$ and $\Sigma({\bf 15},{\bf 2},{\bf 2})$ Higgs fields are required. Furthermore, $\Delta_R({\bf 10},{\bf 1},{\bf 3})$ multiplet breaks the PS symmetry to the SM group and generates non-zero neutrino mass via type-I seesaw. In this PS version, quartic terms of the form $\Sigma^2 \Delta_R^2$ in the scalar potential lead to BNV decays which are distinct from the proton decay modes discussed above (such as $p\to e^+ \pi^0$) that realize the $\Delta (B-L)=0$ selection rule.  On the contrary, these BNV nucleon decay modes  correspond to dimension-9 operators and obey a selection rule  $\Delta (B-L)=-2$ ~\cite{Pati:1983zp,Pati:1983jk,Rudaz:1992fi,ODonnell:1993kdg,Brahmachari:1994dj,Weinberg:1979sa,Weinberg:1980bf,Wilczek:1979hc,Weldon:1980gi,Saad:2017pqj}. There are two types of such decays:  (i) a single lepton $p,n\to \ell+$ mesons (such as $p\to e^-\pi^+\pi^+,~e^-\pi^+K^+,~\mu^-\pi^+K^+,~\nu \pi^+$ etc.) and (ii) three leptons $p,n\to \ell \ell \overline{\ell}+$ mesons (such as $n\to e^-e^+\nu,~e^-\mu^+\nu,~e^-\nu\overline{\nu}\pi^+$ as well as 
$p\to \nu e^+\nu,~e^-e^+\nu\pi^+,~e^- \nu\overline{\nu} \pi^+\pi^+$ etc.) in the final state as shown in Fig.~\ref{fig:Nucleon-PS}, where $\ell$ and $\overline \ell$ are lepton and antilepton, respectively.

The difference between the left and the right Feynman diagrams in Fig.~\ref{fig:Nucleon-PS} is: the right-diagram is obtained from the left-diagram by replacing the di-quark $\Sigma_8$  by a leptoquark  $\Sigma_3$.   A priori, the amplitudes for these alternative decay modes could
compete with each other. It is interesting to note that, in a certain region of the parameter space of the theory, the three lepton final states can dominate over single lepton final states; moreover,   all these BNV processes can be within the observable range. Stringent limits on three-body decays $p\rightarrow e^+(\mu^+)\overline{\nu}\nu$ have been set by the Super-Kamiokande experiment, with $\tau > 2 \times 10^{32}$~yrs~\cite{Super-Kamiokande:2014pqx}. Unlike  two-body nucleon decay channels, phase space alone provides only a crude approximation for the resulting particle spectra from these processes. For the trilepton modes, dynamics and phase space can be approximately accounted for using a general effective Fermi theory formalism of electroweak muon decay $\mu \rightarrow e^+\overline{\nu}\nu$~\cite{Chen:2014ifa}. Searches for nucleon  decay in other channels have also been conducted  and typically have present lower bounds in the range of $10^{31}-10^{33}$ yrs~\cite{ParticleDataGroup:2020ssz}.

\subsection{Neutron-antineutron Oscillation}
Oscillations of electrically neutral particles are well-known phenomena; for example, neutrino oscillations and neutral meson ($K^0-\overline K^0$, $B^0-\overline B^0$, $D^0-\overline D^0$) oscillations are all experimentally well established.  Therefore, one would naturally expect to observe neutron-antineutron ($n-\overline{n}$) oscillations.   However,  the conservation of baryon number forbids the transition from a neutron (with one unit of baryon number)  to an antineutron (with negative  one unit of baryon number). As already pointed out, in the SM there is a global symmetry which forbids  baryon number violation. However, once we go beyond the SM, there is no obvious reason to expect baryon number conservation to hold. Indeed, baryon number violation is one of the necessary conditions noted by Sakharov for the generation of a baryon asymmetry in the universe \cite{Sakharov:1967dj}.  It was observed by Kuzmin that $n$-$\bar n$ oscillations might provide a source of baryon number violation connected with this generation of baryon asymmetry in the universe \cite{Kuzmin:1970nx}.  Proton decay is mediated by four-fermion operators, which have coefficients of the dimensional form $1/M^2$, whereas $n$-$\bar n$ oscillations are mediated by six-quark operators, which have coefficients of the dimensional form $1/M^5$.  Consequently, if there were only one high scale $M$ responsible for baryon number violation, one might expect that $n$-$\bar n$ oscillations would be suppressed relative to proton (and bound neutron) decay.  However, there are theories beyond the SM where BNV nucleon decays are absent or highly suppressed and $n$-$\bar n$ oscillations are the main manifestation of baryon number violation \cite{Mohapatra:1980qe,Nussinov:2001rb}.  Observation of $n-\overline{n}$ transition would show that baryon number is violated by two units $|\Delta B|=2$, which can be naturally expected within GUTs and other extensions of the SM. However, the selection rule $|\Delta B|=2$ is again very different from $p\to e^+ \pi^0$ that follows  $\Delta B=-1$ (and $\Delta L=-1$, hence $\Delta ({B-L})=0$). Nucleon decay with a selection rule $\Delta B=-1$ are induced by dimension-6 (or dimension-5) operators and, therefore, would correspond to the existence of new physics at an energy scale of about $10^{15}$ GeV, whereas $n-\overline{n}$ transitions with $|\Delta B|=2$ are induced by dimension-9 operators, and therefore, would imply new physics around  the 100 TeV scale. Some early studies of $n$-$\bar n$ oscillations include \cite{Mohapatra:1980qe,Chang:1980ey,Kuo:1980ew,Cowsik:1980np,Rao:1982gt,Rao:1983sd}. Some recent reviews include \cite{Phillips:2014fgb,Addazi:2020nlz}.

\begin{figure}[th!]
\centering
\includegraphics[scale=0.35]{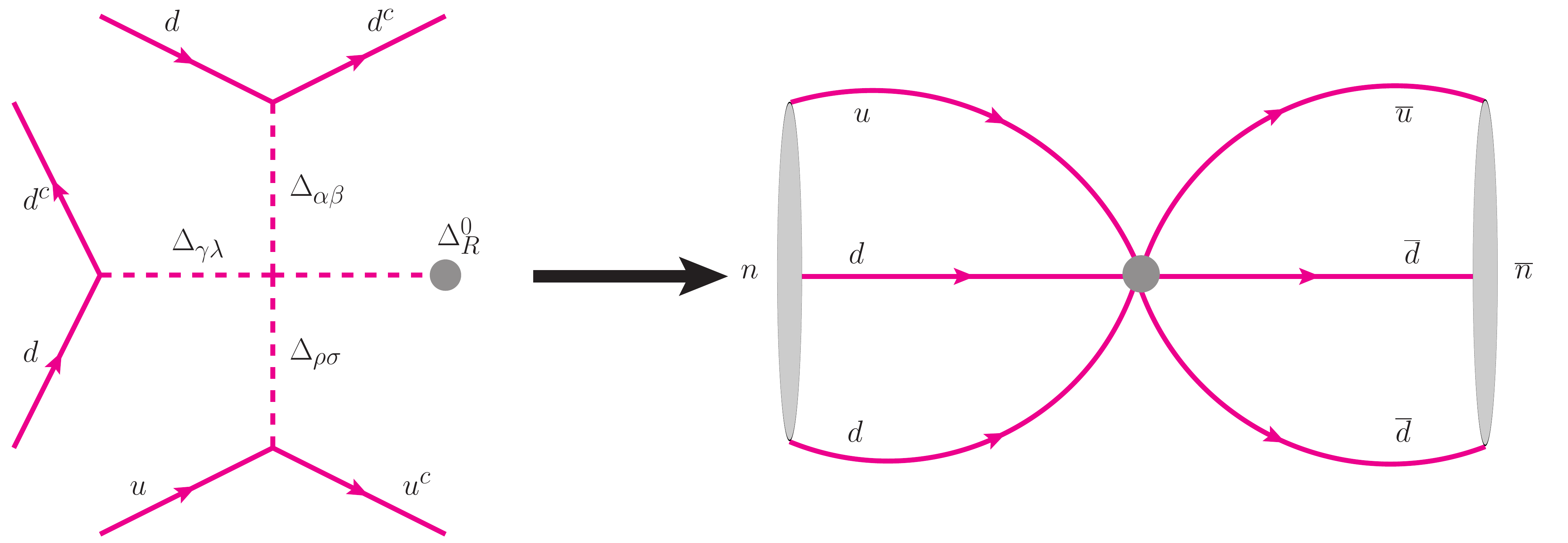}
\caption{Feynman diagram for $n-\overline{n}$ oscillation in the model of Ref.~\cite{Mohapatra:1980qe}. Here, sub-multiplets $\Delta_{R,ab}$ are color sextets and $\Delta^0_R$ acquires a VEV that breaks $SU(4)_C\times SU(2)_L\times SU(2)_R$ symmetry to the SM.}\label{nnbar}
\end{figure}

Although, in general, a theory may violate $L$ without violating $B$,  in most of the well-motivated extensions of the SM, these two are connected via fundamental symmetries. As an  example, within left-right symmetric models, $U(1)_{B-L}$ is promoted to a gauge symmetry~\cite{Marshak:1979fm}. As long as this  symmetry is exact, $\Delta B=\Delta L$ is realized. This is highly motivating since the existence of $|\Delta L|=2$ generates the Majorana neutrino mass term, which is then connected to the existence of $|\Delta B|=2$ operators. Therefore, $n-\overline{n}$ oscillation may be directly linked to neutrinoless double beta decay processes; see Section~\ref{sec:281}.

It was shown in Ref.~\cite{Mohapatra:1980qe} that if the Higgs multiplet that  breaks the $SU(4)_C\times SU(2)_L\times SU(2)_R$ group to the SM group is $\Delta_R(10,1,3)$ (instead of the Higgs set chosen in the original PS model), it generates Majorana neutrino mass via $|\Delta L|=2$ processes while at the same time yielding $n-\overline{n}$ transition via $|\Delta B|=2$ processes at an observable rate, keeping the proton stable. A term of the form $\Delta_R^4$ in the scalar potential gives rise to a cubic term among three color sextet scalars once the $B-L$ violating vacuum expectation value of the neutral component of $\Delta_R$ is inserted, which causes $n-\overline{n}$ oscillations ~\cite{Mohapatra:1980qe,Phillips:2014fgb} as shown in Fig.~\ref{nnbar}. The $n-\overline{n}$ oscillations arising in this class of models with TeV scale color sextet scalars have an {\it upper limit} of $\tau_{n\overline n}\lesssim 10^{10-11}$ sec~\cite{Babu:2008rq, Babu:2013yca}, within the reach of the next-generation experiments; see Fig.~\ref{fig:expectedlimits}. This upper bound is a consequence of the model requirements to satisfy the neutrino oscillation data via type-II seesaw, observed baryon asymmetry via the post-sphaleron baryogenesis mechanism~\cite{Babu:2006xc}, as well as the low-energy flavor changing neutral current (FCNC) constraints. Besides, TeV scale color sextets  have rich phenomenology and can be uniquely probed at the hadron colliders~\cite{Mohapatra:2007af,Chen:2008hh,Berger:2010fy,Baldes:2011mh,Chivukula:2015zma, CMS:2019gwf, Pascual-Dias:2020hxo}. There exist other simplified models of $n-\overline{n}$~\cite{Dev:2015uca, Allahverdi:2017edd, Grojean:2018fus}, which, however, do not have an upper limit on $\tau_{n\overline n}$, although they can still be tested using the collider probes of colored particles.

An interesting class of BSM theories hypothesizes that our usual four spacetime dimensions are embedded in a space with higher (spatial) dimensions, such that SM fermions have strongly localized wave functions in the extra dimensions \cite{Arkani-Hamed:1999ylh}. It has also been shown that in these extra-dimensional models, neutron-anti-neutron oscillations can occur at an observable rate while baryon-number-violating nucleon decays are suppressed far below experimental limits keeping the proton stable~\cite{Nussinov:2001rb,Girmohanta:2019fsx,Girmohanta:2020qfd}. The BNV nucleon decays are suppressed by separating quark wave function centers sufficiently far from lepton wave function centers in the extra dimensions, but this does not suppress $n$-$\bar n$ oscillations since the operators that mediate these oscillations only involve quarks.  These models can also produce a seesaw mechanism with naturally light neutrino masses and viable DM candidates \cite{Girmohanta:2020llh,Girmohanta:2021gpf}. Another study of baryon-number violation without proton decay is \cite{Arnold:2012sd}.

An experimental search for $n$-$\bar n$ oscillations was carried out using neutrons from a reactor at the Institut Laue-Langevin (ILL) and obtained a lower limit $\tau_{n\bar n} > 0.86 \times 10^8$ sec (90 \% CL) \cite{Baldo-Ceolin:1994hzw}. Neutron-antineutron oscillations also lead to instability of matter with associated $\Delta B = -2$ dinucleon decays yielding mainly multi-pion final states.  These have been searched for in deep underground nucleon decay detectors, most recently in Super-Kamiokande \cite{Super-Kamiokande:2011idx,Super-Kamiokande:2020bov}, obtaining a lower limit on a lifetime characterizing such matter instability of $\tau > 3.6 \times 10^{32}$ yr, which, when converted to a corresponding free neutron oscillation time, is $\tau_{n \bar n} > 4.7 \times 10^8$ sec. This difference is due to the suppression of $n-\overline{n}$ oscillations in matter as a result of nuclear potential differences. The oscillation time is matter ($\tau_m$) is related to free neutron oscillation time ($\tau_{n-\overline{n}})$ as $\tau_m = R\, \tau_{n-\overline{n}}^2$, with the conversion factor $R$ for Oxygen evaluated to be about $0.5 \times 10^{23}\, {\rm sec}^{-1}$ \cite{Friedman:2008es}.
Upcoming experiments at the ESS  (European Spallation Source)  and DUNE (Deep Underground Neutrino
Experiment) plan to improve these bounds to $\tau_{n\overline n}\gtrsim 10^{9-10}$ sec~\cite{Milstead:2015toa,Frost:2016qzt,Hewes:2017xtr}. These oscillation times can be translated into new physics scales of roughly $\left(\tau_{n\overline n} \Lambda^6_{\textrm{QCD}} \right)^{1/5}\sim {O}(100-1000)$ TeV, which are well above energies  directly accessible at colliders. For recent reviews on this subject, see Ref.~\cite{Phillips:2014fgb,Addazi:2020nlz}. Existing upper bounds on the rates for the dinucleon decays $nn \to \pi^0\pi^0$, $nn \to \pi^+\pi^-$, and $np \to \pi^+\pi^0$ imply upper bounds on the rates for dinucleon decays to dileptons $nn \to e^+e^-$, $nn \to \mu^+\mu^-$, $nn \to \nu\bar\nu$, and $np \to \ell^+ \nu_\ell$, where $\ell=e,\mu$. These have been calculated in \cite{Girmohanta:2019cjm,Nussinov:2001rb}.

\subsection{Other $B-L$ violating Processes}
\label{sec:BLtheor}
Nonzero neutrino masses, if these are Majorana fermions, are evidence for new physics that violates lepton number by two units. GUTs, as argued earlier, will ultimately lead to a connection between lepton-number and baryon-number violating new physics. While the details depend on both the mechanism behind nonzero neutrino masses and the underlying GUT, it is often the case that the lepton number violating physics is a subset of different phenomena correlated with $\Delta (B-L)=2$ phenomena. Some connections have been explored at the effective operator level, in a number of papers; e.g., Ref.~\cite{Rao:1983sd,deGouvea:2014lva}. 

Other than neutrinoless double-beta-decay ($\Delta L =2$) and $n\bar{n}$-oscillations ($\Delta B=2$), there are several other physics processes that are mediated by $\Delta(B-L)=2$ new physics. For example, there are $|\Delta L|=2$ meson decays, such as $K^+ \to \pi^-\mu^+\mu^+$ and 
$K^+ \to \pi^-\mu^+e^+$. Initial upper bounds on these decays were set in \cite{Littenberg:1991ek} and searches for them have been carried out at BNL \cite{Appel:2000tc,Littenberg:2000fg} and recently by the NA62 experiment at CERN \cite{NA62:2019eax,NA62:2021zxl}, yielding the upper limits $BR(K^+ \to \pi^-\mu^+\mu+) < 4.2 \times 10^{-11}$ and $BR(K^+ \to \pi^-\mu^+e^+) < 4.2 \times 10^{-11}$ (90 \% CL).  $|\Delta L|=2$ baryon decays include, e.g., $\Xi^- \to p\mu^-\mu^-$, on which an initial upper bound was set in \cite{Littenberg:1991rd}. A Fermilab experiment searched for this decay and set the upper limit $BR(\Xi^- \to p \mu^-\mu^-) < 4.0 \times 10^{-8}$ \cite{HyperCP:2005sby} (90 \% CL). Other decays violating 
$B$ and $L$ include $B\to L M$ where $B$ is a baryon, $L$ is a lepton (neutrino or negatively-charged lepton) and $M$ is a mesonic state. These include $n\to \mu^-\pi^+$, $p\to \nu\pi^+$, $n\to \nu\rho^0$, and $p\to e^-\pi^+K^+$.

The strongest bounds on these $\Delta(B-L)=2$ processes come from searches for nucleon decay. These are nicely summarized in Ref.~\cite{ParticleDataGroup:2020ssz}, and some are very strong, of order $10^{32}$ years. It is also the case that all of the strongest bounds on $\Delta(B-L)=2$ neutron and proton decay come from twentieth century experiments, including Frejus~\cite{Frejus:1991ben} and IMB~\cite{Seidel:1988ut,McGrew:1999nd}, keeping in mind that it is not possible to distinguish a nucleon decay into a neutrino from that into an antineutrino. The DUNE experiment, thanks to its tracking, calorimetric, and particle-ID capabilities, is well positioned to be sensitive to $\Delta(B-L)=2$  with lifetimes that approach $10^{34}$~years~\cite{DUNE:2020ypp}. 

\subsection{Effective Field Theory of $B$ Violation}

\begin{figure}[!htb]
	\centering
	\renewcommand\thesubfigure{\roman{subfigure}}
	\begin{subfigure}[t]{3cm}
		\includegraphics[scale=0.5]{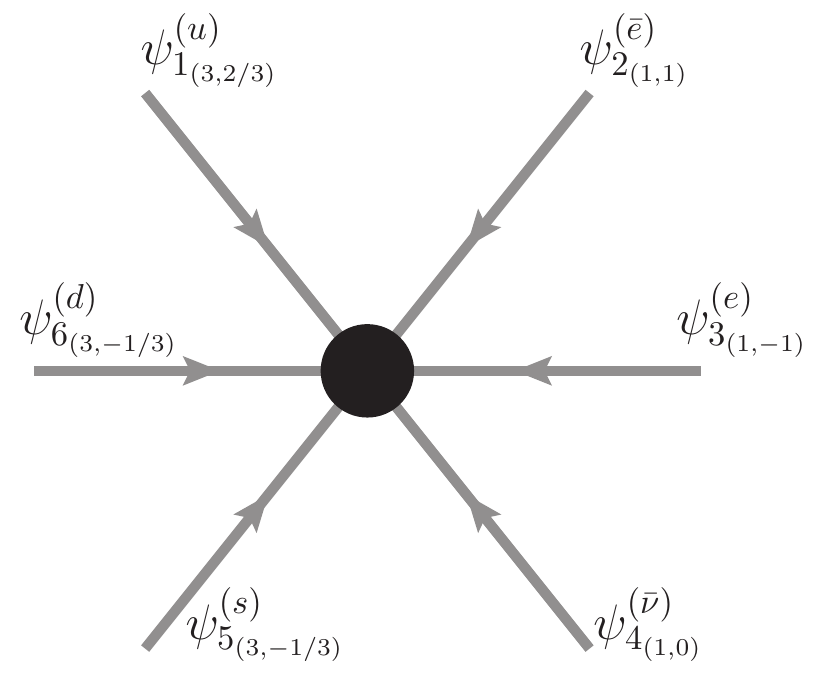}
	\end{subfigure}\hspace{2cm}
	\begin{subfigure}[t]{3cm}
		\includegraphics[scale=0.5]{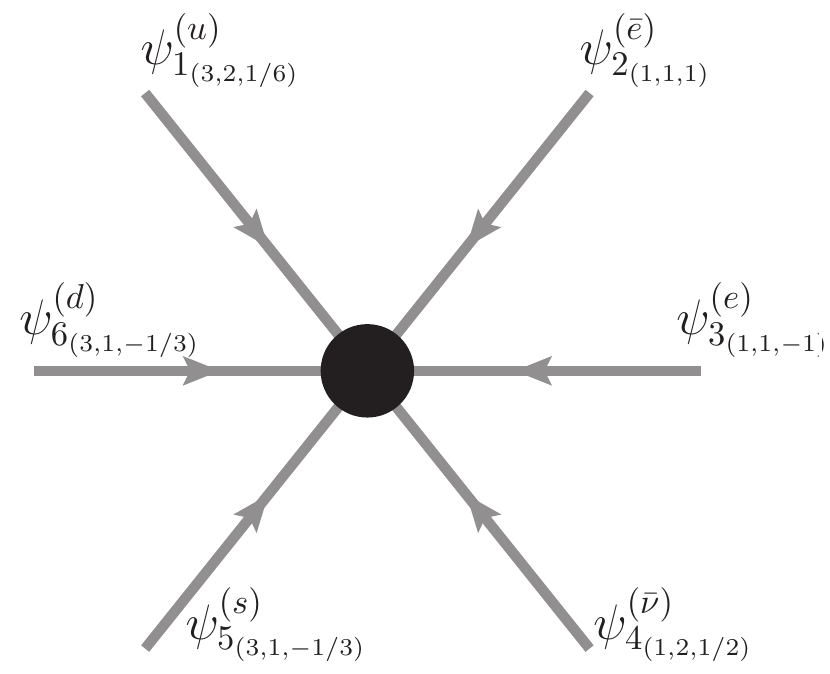}
		\caption{$ p \,\rightarrow\, K^+\, e^+\, e^-\,\nu$}
	\end{subfigure}\hspace{2cm}
	\begin{subfigure}[t]{4cm}
		\includegraphics[scale=0.5]{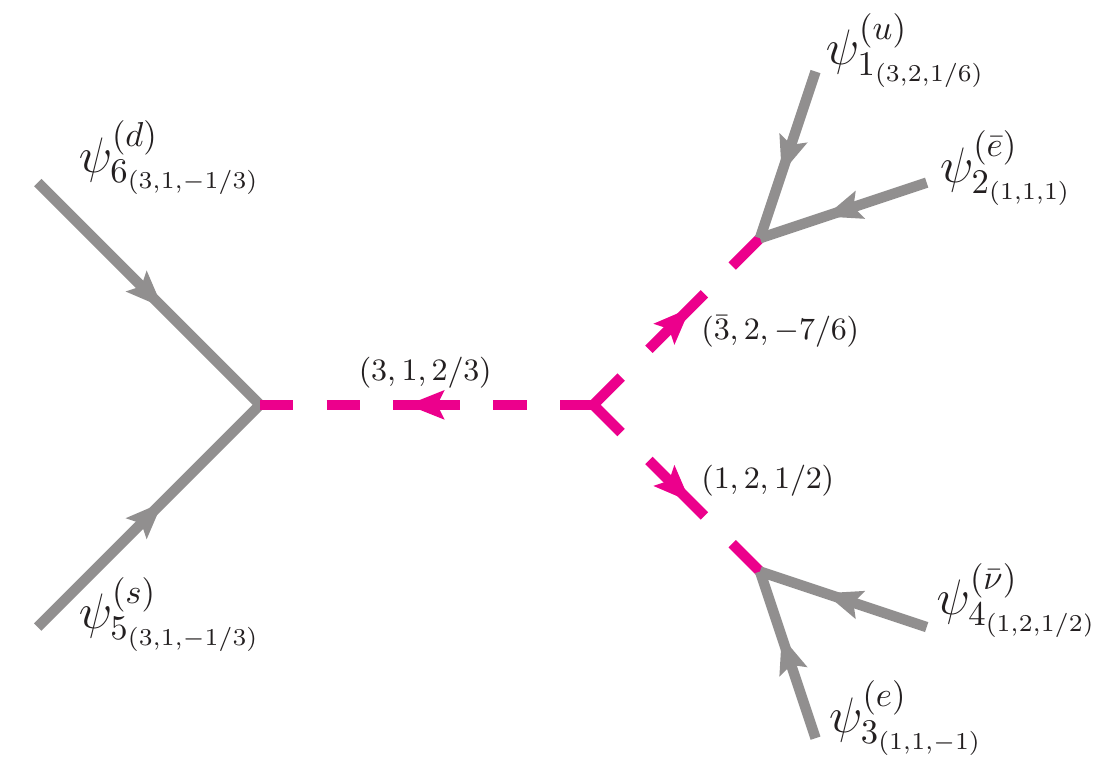}
	\end{subfigure}\newline
	\begin{subfigure}[t]{3cm}
		\includegraphics[scale=0.5]{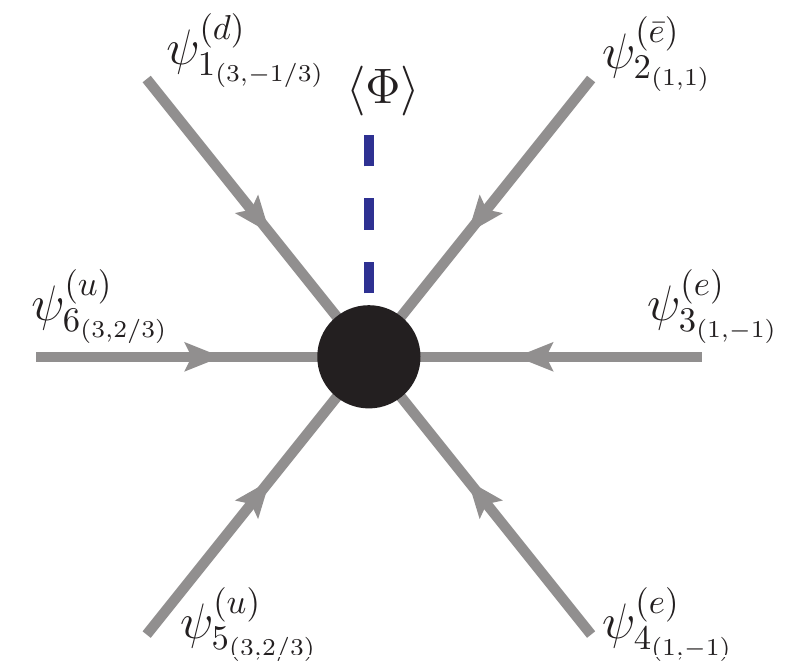}
	\end{subfigure}\hspace{2cm}
	\begin{subfigure}[t]{3cm}
		\includegraphics[scale=0.5]{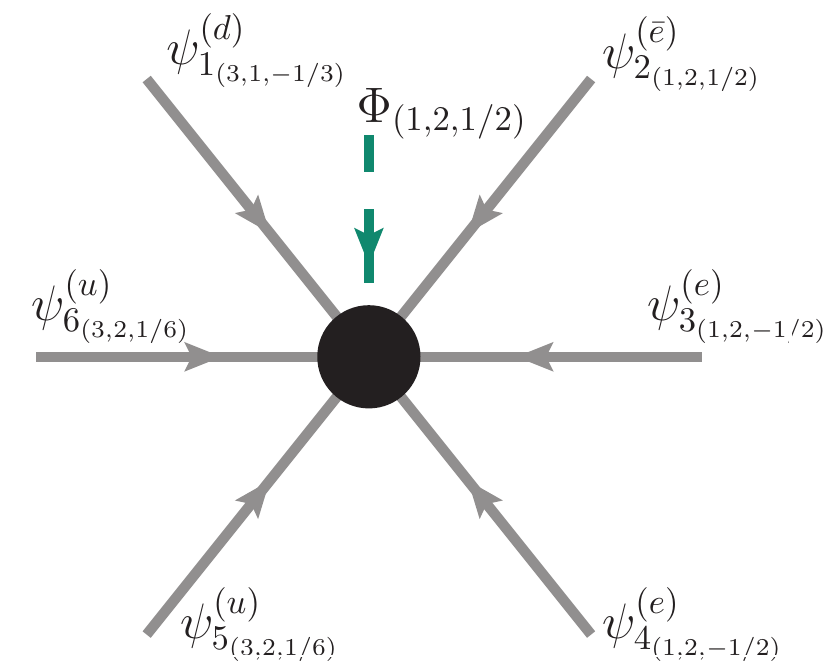}
		\caption{$p\,\rightarrow\, e^+\, e^+\, e^-$}
	\end{subfigure}\hspace{2cm}
	\begin{subfigure}[t]{4cm}
		\includegraphics[scale=0.5]{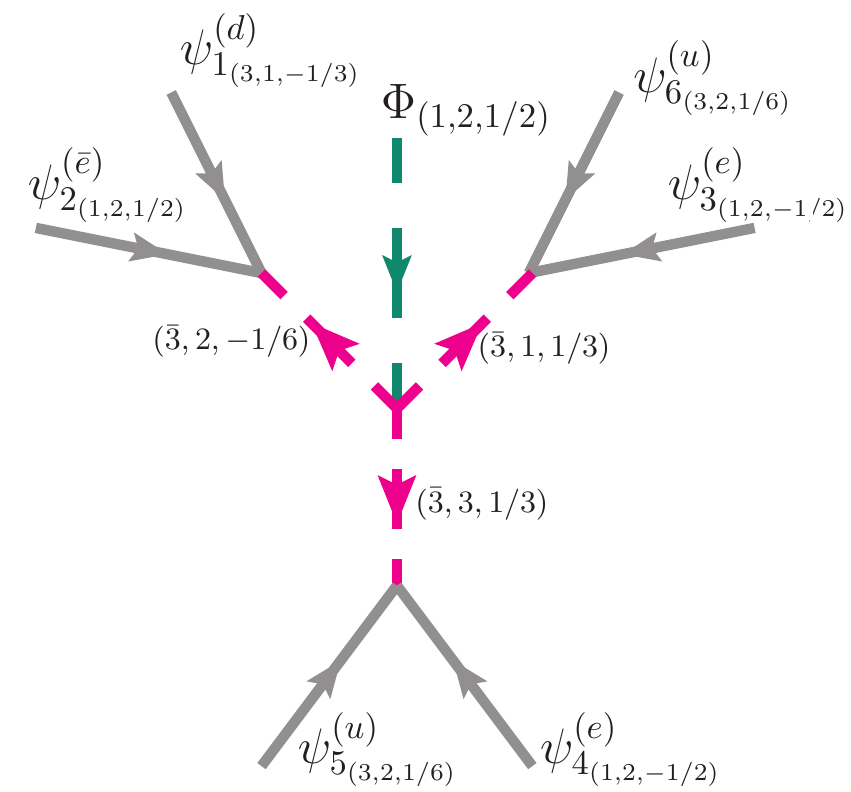}
	\end{subfigure}\newline\newline
	\captionsetup[subfigure]{labelformat=empty}
	\begin{subfigure}[t]{3.5cm}
		\caption{  \hspace{-4cm}{ Low-energy EFT  }   }
	\end{subfigure}
	\begin{subfigure}[t]{3.5cm}
		\caption{  \hspace{-1cm}{ SMEFT }   }
	\end{subfigure}
	\begin{subfigure}[t]{3.5cm}
		\caption{   \hspace{3cm}{ BSM }   }
	\end{subfigure}
	\caption{The low-energy EFT contact interactions of mass dimension 9 enveloping two distinct modes of proton decay, the corresponding SMEFT counterparts, and their realizations in terms of BSM propagators. For the LEFT diagrams, $\psi^{(f)}_{(R_C, Q)}$ denotes the light fermions $f \in \{ \nu,\, e,\, u,\,d,\,s,\,c,\,b \}$  with $R_C$ being the $SU(3)$ representation, and $Q$ is the electromagnetic charge. For the SMEFT diagrams, $\psi^{(f)}_{(R_C, R_L, Y)}$ denotes the SM fermions $f \in \{q,\,l,\,e,\,d,\,u \}$ with $R_L$ being the $SU(2)$ representation, and $Y$ is the hypercharge. $\Phi$ is the SM Higgs and $\langle\Phi\rangle$ is its vacuum expectation value.}
	\label{fig:pdecay}
\end{figure}

The most suitable framework for studying low-energy, below electroweak-scale phenomena is the Low Energy Effective Field Theory (LEFT) \cite{Jenkins:2017jig}, which differs from the SM with respect to the internal symmetry as well as the degrees of freedom. In going from the SM to the LEFT, the heavy degrees of freedom namely, the Higgs boson $h$, the electroweak gauge bosons $W_{\mu}^{\pm}$ and $Z_{\mu}$ and the top-quark $t$ are integrated out, whereas the internal symmetry is spontaneously broken from $SU(3)_C\times SU(2)_L \times U(1)_Y$ to $SU(3)_C \times U(1)_{\text{em}}$.      	

LEFT operators can describe a wide variety of fermion-fermion interactions, also encompassing the BNV ones. The best examples are operators describing the proton decay into mesons and leptons or entirely into three charged leptons \cite{Hambye:2017qix,Fonseca:2018aav}. In the SM baryon and lepton number conservation emerge as accidental symmetries. This suggests that the violation of these symmetries within the LEFT is a sign of beyond the SM (BSM) physics. The backdrop for conducting analyses on and to test the veracity of BSM models is the Standard Model Effective Field Theory (SMEFT) and it also acts as the necessary bridge between the low and the high energy sectors. 

A complete matching between a BSM model and the LEFT would involve an intricate and meticulous multi-step procedure of integrating out heavy fields and matching parameters. But a simpler way to catch a glimpse of possible BSM origins of the rare processes encapsulated by LEFT operators is to find their embedding within SMEFT operators and then based on symmetry arguments unfurl them into Feynman diagrams consisting of heavy propagators \cite{Chakrabortty:2020mbc}. 

We have highlighted two examples in Fig.~\ref{fig:pdecay}, where the first column presents LEFT contact operators of mass dimension 9 describing proton decay processes $(i)\, p \,\rightarrow\, K^+\, e^+\, e^-\,\nu$ and $(ii)\,p\,\rightarrow\, e^+\, e^+\, e^-$ The second column shows the relevant SMEFT operators of the same or higher mass dimension that describe the same contact interaction as the LEFT ones \cite{Chakrabortty:2020mbc}. Finally, the last column reveals the BSM heavy field propagators that can mediate such processes. 

\begin{figure}[th!]\centering
\includegraphics[width=0.7\textwidth]{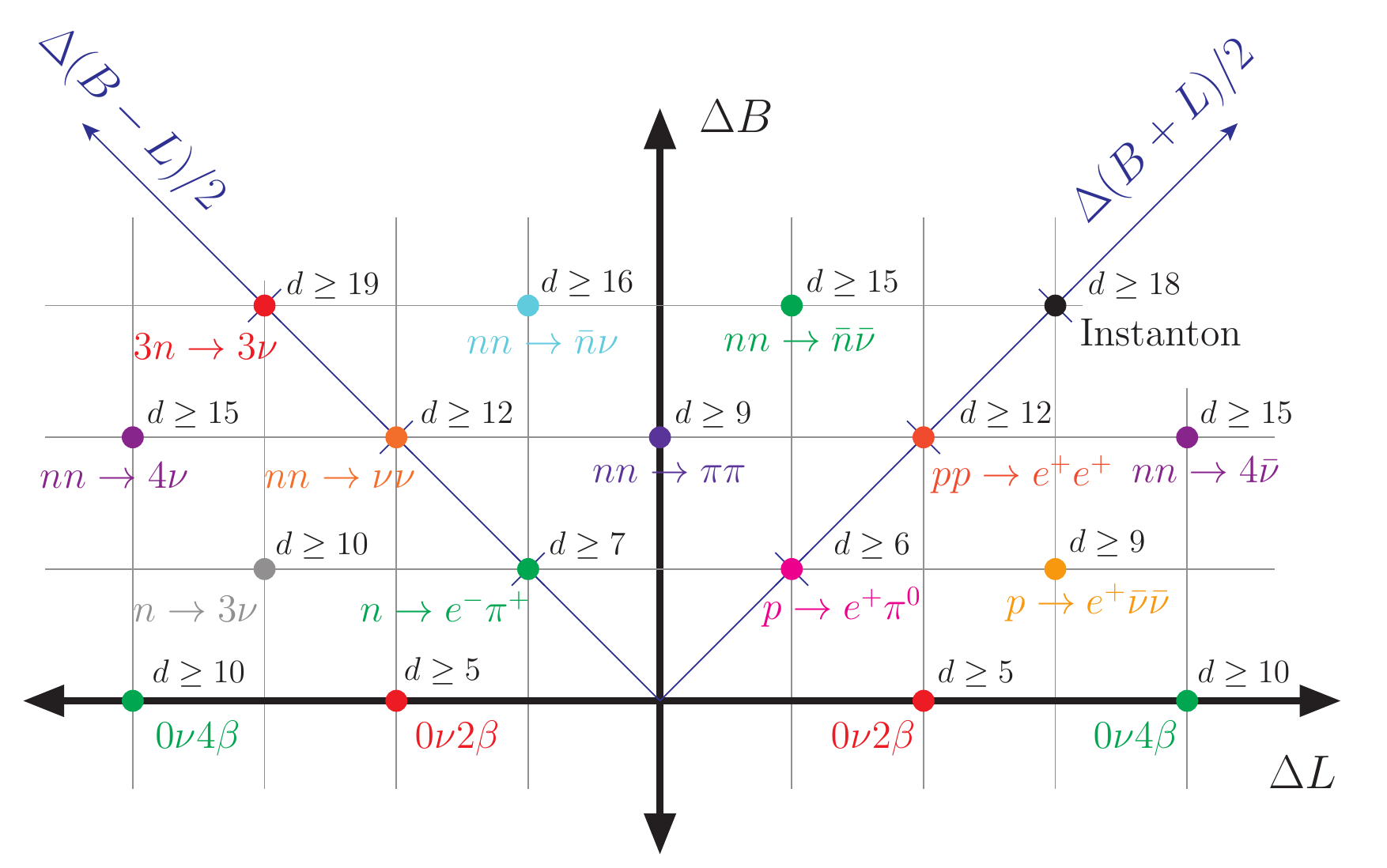}
\caption{Process examples with baryon and lepton number violation by $\Delta B$ and $\Delta L$ units, respectively.  ``Instanton'' refers to processes that break the same quantum numbers as non-perturbative electroweak instantons. $0\nu2(4)\beta$ refers to
neutrinoless double (quadruple) beta decay.  The minimal mass dimension $d$ of the underlying effective operator is shown. Operators also carry flavor. Reproduced from Ref.~\cite{Heeck:2019kgr}.}\label{fig:BLproc}
\end{figure}

The $\Delta B \neq 0$ operators of the SMEFT provide a model-independent framework for exploring  $B$-violation, both for $B$-violating nucleon decay~\cite{Weinberg:1979sa,Wilczek:1979hc,Weinberg:1980bf,Abbott:1980zj} and for $n$-$\bar n$ oscillations \cite{Mohapatra:1980qe,Kuo:1980ew,Rao:1982gt,Rao:1983sd,Caswell:1982qs}. The flavor structure of the  operators at different dimensions allows one to establish nucleon decay-mediating processes that can dominate. This has been explored in further detail in more recent works such as Refs.~\cite{Kobach:2016ami,Fornal:2018dqn,Heeck:2019kgr,Girmohanta:2019fsx,Babu:2012iv,Babu:2012vb}. In SMEFT, $\Delta B = 1$ appears at dimension-6. Any flavor $\Delta B = 1$ term leads to nucleon decay, including particles heavier than the proton such as charm or the tau lepton that can induce proton decay through off-shell contributions~(see e.g.~Ref.~\cite{Hou:2005iu}). These contributions are strongly constrained by two-body nucleon decays such as $p \rightarrow e^+ \pi^0$. At dimension $>6$, non-trivial lepton number $\Delta L \neq 0$ allows to enforce dominance of some operators (e.g. Ref.~\cite{Weinberg:1980bf}). $\Delta B$ and $\Delta L$ are connected through the dimensionful operators (e.g. \cite{Rao:1983sd,Kobach:2016ami}). Higher dimensional operators can also often lead to multi-body channels, such as $n \rightarrow K^+\mu^+e^-e^-$ at dimension-9, or multi-nucleon decays with $\Delta B > 1$~\cite{Heeck:2019kgr}. One can use limits on rates for $p \to \ell^+M$ and $n \to \bar\nu M$, where $M$ denotes a pseudoscalar or vector meson to obtain indirect limits on rates for $p \to \ell^+ \ell'^+ \ell'^-$, $n \to \bar\nu\ell^+\ell'^-$, $p \to \ell^+\nu\bar\nu$, and $n \to \bar\nu\bar\nu\nu$ \cite{Girmohanta:2019xya}. Fig.~\ref{fig:BLproc} displays characteristic examples of processes with distinct $\Delta B$ and $\Delta L$ structures.

\subsection{Discrete Symmetries and Supersymmetry} 

 $B$-violation processes appear in many extensions of the SM, a notable example being supersymmetric (SUSY) theories. Already in the MSSM realization nucleon decay-mediating dimension-4 operators $QLd^c$ and $u^cd^cd^c$ appear, where $Q$, $L$ are left-chiral quark and lepton doublets and $u^c$, $d^c$ are the $u$-type, $d$-type superfields, respectively. To forbid rapid proton decay through these interactions, models often impose a ${Z}_2$ symmetry called R-parity (matter parity)  ~(e.g. Ref.~\cite{Dimopoulos:1981yj}). However, at dimension-5, one encounters nucleon decay-mediating $QQQL$, which can be forbidden by ``proton hexality'' $Z_6$ symmetry that contains $R$-parity as a subgroup~\cite{Dreiner:2005rd}. Since all global symmetries are expected to be violated at some level~\cite{Banks:2010zn}, it is appealing to consider discrete gauge symmetries. Such symmetries can appear as remnants of spontaneously broken local ${U}(1)$ symmetries~\cite{Krauss:1988zc}. Thus, care must be taken to ensure anomaly cancellation. Favorable discrete symmetries that allow for rich phenomenology, resolve theoretical puzzles (e.g. $\mu$-problem) and forbid dangerously rapid nucleon decay have been identified, such as those of Refs.~\cite{Babu:2002tx,Dreiner:2005rd,Lee:2010gv,Chen:2014gua,Hicyilmaz:2021khy}. Discrete (gauge) symmetries can also be considered in the context of GUT models.

Unlike the SM, the baryon and lepton numbers are no longer accidental symmetries of the classical Lagrangian in the MSSM. In the most general supersymmetric theory, gauge-invariant terms that violate baryon and lepton numbers are allowed, which are
\begin{align}
&W_{\Delta L=1}= \lambda_{ijk}L_iL_j\overline e_k +
\lambda_{ijk}^{\prime} L_iQ_j\overline d_k
+ \mu^{\prime}_iL_iH_u, \label{LMSSM}
\\
&W_{\Delta B=1}= \lambda^{\prime\prime}_{ijk}\overline d_i \overline d_j \overline u_k,\label{BMSSM}
\end{align}
where standard notation of chiral superfields of the MSSM are used and $i,j,k$ are family indices. The chiral supermultiplets carry baryon (lepton) numbers $ B=+1/3$ for $Q_i$ and $B=-1/3$ for $\overline u, \overline d$ ($L=+1$ for $L_i$ and $L=-1$ for $\overline e_i$). Subsequently, the terms in Eq.~\eqref{LMSSM}  violate total lepton number by one unit, and terms in Eq.~\eqref{BMSSM} violate total baryon number by one unit. If both $\lambda^\prime$ and $\lambda^{\prime\prime}$ are present and are unsuppressed, then the lifetime of the proton would be extremely short. In the low-energy MSSM model, one forbids these terms by imposing a new discrete symmetry, known as ``$R$-parity''~\cite{Farrar:1978xj}.  One can allow some of the R-parity-violating (RPV) terms while still avoiding excessively rapid proton decay, and these could lead to $|\Delta L|=2$ decays such as $K^+ \to \pi^- \mu^+\mu^+$ and $K^+ \to \pi^-\mu^+e^+$; thus upper bounds on the branching ratios for these decays can be used to obtain upper bounds on the coefficients of certain RPV terms \cite{Littenberg:2000fg}.   

Here we consider the possibility of only baryon number violation. Therefore, the only non-zero term in the above superpotential we allow is $\overline d_i \overline d_j \overline u_k$. The associated coupling  $\lambda^{\prime\prime}_{ijk}$ is antisymmetric in the first two flavor indices, leading to in total nine couplings $\lambda^{\prime\prime}_{dsu}, \lambda^{\prime\prime}_{dbu},$ and so on. Due to this antisymmetric nature, coupling to three quarks of the first generation is absent; this is why $n-\overline n$ oscillation is highly suppressed. However, there exist stringent bounds on these couplings coming from dinucleon decay~\cite{Barbieri:1985ty,Dimopoulos:1987rk,Brahmachari:1994wd,Goity:1994dq,Barbier:2004ez,Chemtob:2004xr,Smirnov:1996bg}.  For example, let us consider the coupling $\lambda^{\prime\prime}_{dsu}$, which violates baryon number by one unit and strangeness by one unit; however, it conserves ${B-S}$. Hence  it is easy to understand that the best bound comes from dinucleon decay into two mesons of identical strangeness  via dimension-9 operators. Using the current limits of dinucleon decay to two kaons, $pp\to K^+ K^+$ for instance, Super-K provides a lower limit of $> 1.7\times 10^{32}$ yrs~\cite{Super-Kamiokande:2014hie} (for pion final states, limits of similar order exist from Super-K~\cite{Super-Kamiokande:2015jbb}), and one obtains the following constraint on the $R$-parity-violating coupling, 
\begin{align}
\lambda^{\prime\prime}_{dsu}< 10^{-16} \left(\frac{m_S}{\Lambda}\right)^{5/2},    
\end{align}
where $m_S$ represents a common mass scale of SUSY particles and $\Lambda$ is the hadronic scale. Dinucleon decays also appear in non-SUSY models, e.g., dinucleon decay to  two kaons and dinucleon decay to leptons occur in a class of models studied in Ref.~\cite{Arnold:2012sd}.

\subsection{Implications of BNV for other BSM Physics}

\subsubsection{Majorana Nature of Neutrinos} \label{sec:281}
All fermions in the SM are charged, except neutrinos. Therefore, charged fermions are Dirac particles, whereas the nature of the neutrinos, whether Dirac or Majorana, is an unresolved question in particle physics. As is well known, observation of neutrinoless double beta decay  would establish the Majorana nature of neutrinos; for a recent review see Ref.~\cite{Dolinski:2019nrj}.  However, if neutrinos have normal mass ordering with non-degenerate eigenvalues,  the prospects of observing such decays are bleak in the near-future experiments. Interestingly, observation of BNV processes can confirm the Majorana nature of neutrinos. This provides additional motivation for experiments searching for nucleon decays and neutron-antineutron oscillation.   

\begin{figure}[th!]
\centering
\includegraphics[width=0.5\textwidth]{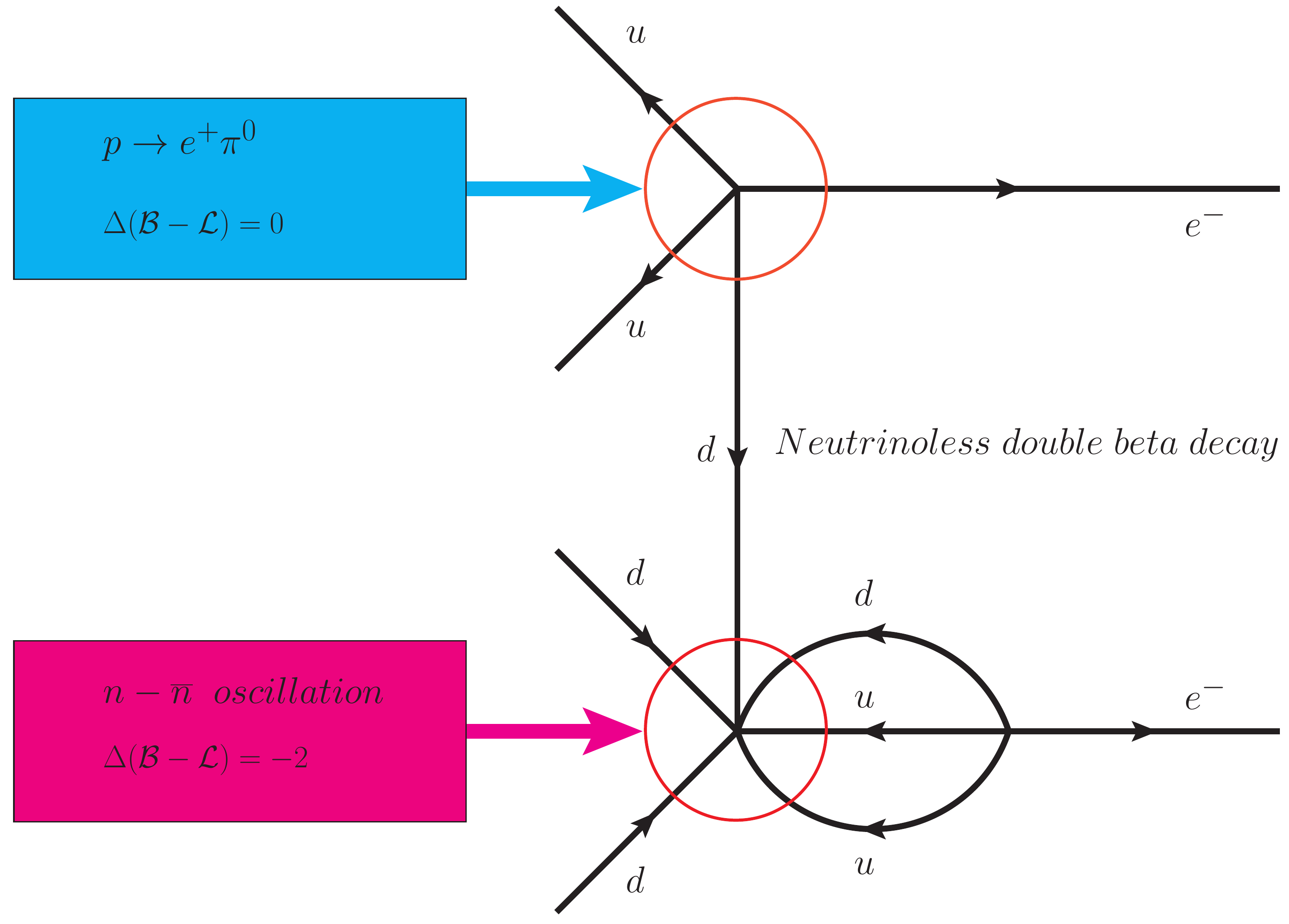} 
\caption{ Neutrinoless double beta decay diagram originating from proton decay and $n-\overline{n}$ oscillation obeying selection rules $\Delta (B-L)=0$ and $\Delta (B-L)=-2$, respectively. Observation of the latter two processes would guarantee the Majorana nature of neutrinos. }
\label{fig:betadecay}
\end{figure}

As suggested in Ref.~\cite{Babu:2014tra}, baryon number violation in two processes with at least one obeying the selection rule $\Delta (B-L)=\pm 2$ can infer  Majorana nature of neutrinos. As an example, consider two BNV processes: (i) $p\to e^+ \pi^0$ ($|\Delta B|=1$) and (ii) $n-\overline{n}$ oscillation ($|\Delta B|=2$). The former and the latter processes correspond to $\Delta (B-L)=0$ and $\Delta (B-L)=- 2$, respectively. Suppose both these BNV processes are observed in the experiments. In that case, neutrinoless double beta decay is guaranteed to exist just by combining these two vertices as depicted in Fig.~\ref{fig:betadecay}. This would subsequently confirm the Majorana character of neutrinos. One can get to the same conclusion if instead (i) $p\to e^+ \pi^0$ ($\Delta (B-L)=0$) and (ii) $n\to e^-\pi^+$ ($\Delta (B-L)=-2$)  processes are considered~\cite{Babu:2014tra}.  Since nucleon decay in well-motivated GUT models is within  reach of ongoing and upcoming experiments, one can hope to resolve this outstanding puzzle (i.e., Majorana or Dirac) by observing BNV processes, even if $0\nu\beta\beta$ is not directly observed by then.

\begin{figure}[th!]
\centering
\includegraphics[width=0.35\textwidth]{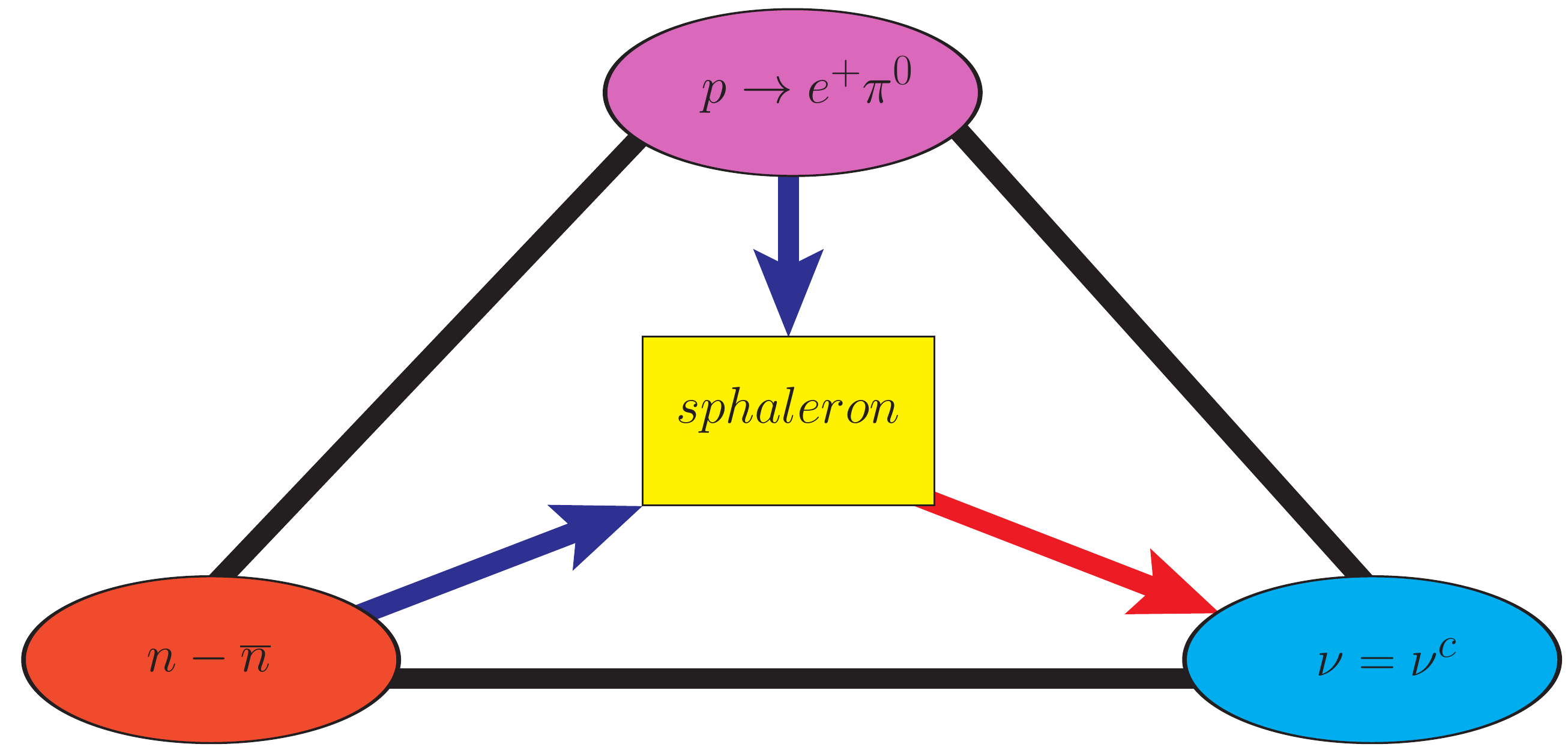} 
\caption{ ${B-L}$ triangle~\cite{Babu:2014tra} demonstrating how Majorana mass for neutrinos can be confirmed if BNV processes are observed. }
\label{fig:triangle}
\end{figure}

By following the arguments given above, the presence of Majorana mass for neutrinos can be directly inferred via the weak instanton effects. Non-perturbative instanton/sphaleron configurations of weak interactions that lead to solutions to an effective operator involving twelve SM doublet fermions can be written as: $[uddudd] \cdot [uude] \cdot [\nu\nu]$.  Of this operator, the first, second, and third pieces correspond to $n-\overline{n}$ oscillation, proton decay $p\to  e^+\pi^0$, and Majorana neutrino mass, respectively.  Therefore, if $p\to  e^+\pi^0$ and $n-\overline{n}$ oscillation are observed, the sphaleron configuration would imply that neutrinos have Majorana masses; this non-trivial connection known as the ``${B-L}$ triangle'' \cite{Babu:2014tra} is demonstrated in Fig.~\ref{fig:triangle}. 

\subsubsection{Leptogenesis}
In the SM, baryon number and lepton number are accidental symmetries of the classical Lagrangian. However, $B$ and $L$ are exactly conserved only in perturbation theory and are not respected by non-perturbative effects. Sphaleron processes which are effective non-perturbative interactions constructed out of twelve left-handed SM fermions are the key to leptogenesis~\cite{Fukugita:1986hr}. These interactions change baryon number $B$ and lepton number $L$ by a multiple of three ($\Delta B=\Delta L= 3N_{CS}$, where the integer $N_{CS}$ is the Chern-Simons number characterizing the sphaleron gauge field configuration), but on the contrary, preserve ${B-L}$. Sphaleron processes are in thermal equilibrium in a wide range of temperatures: $T_{\rm EW}\sim 10^{2}~ \textrm{GeV} \lesssim T \lesssim T_{\rm sph}\sim 10^{12} ~\textrm{GeV}$~\cite{Kuzmin:1985mm}; owing to  these interactions, in Sakharov's conditions~\cite{Sakharov:1967dj} for baryogenesis, lepton number violation can replace baryon number violation.

The existence of sphaleron processes typically divides models of baryogenesis into three categories depending on the energy scale when asymmetry is generated: (i) Leptogenesis: high scale $L$ asymmetry generation after inflation that is converted to $B$ asymmetry through sphaleron processes~\cite{Fukugita:1986hr}, (ii) Electroweak baryogenesis: phase transition dynamics at the electroweak scale coupled with new physics  CP-violating sources near the EW scale (for a recent review see Ref.~\cite{Morrissey:2012db}), (iii) Post-sphaleron baryogenesis: new physics $B$-violating interactions occurring below the electroweak phase transition scale~\cite{Babu:2006xc,Babu:2008rq,Gu:2011ff, Babu:2013yca}.  Here we briefly discuss leptogenesis scenarios (for reviews on leptogenesis, see, e.g., Refs.~\cite{Buchmuller:2004nz,Nir:2007zq,Davidson:2008bu,Pilaftsis:2009pk,DiBari:2012fz,Fong:2012buy,Chun:2017spz,Dev:2017trv, Bodeker:2020ghk}) that are inherently connected to neutrino mass generation via the seesaw mechanism~\cite{Minkowski:1977sc, Yanagida:1979as,Glashow:1979nm,Gell-Mann:1979vob,Mohapatra:1979ia}.

The seesaw mechanism is the most natural candidate to explain tiny neutrino masses within the GUT framework or BSM frameworks with a local $B-L$ symmetry~\cite{Marshak:1979fm}. In this scenario, tiny neutrino masses arise due to the heaviness of right-handed partners. The heavy right handed neutrinos have Majorana masses breaking $B-L$ by two units which then gives a Majorana masses of light neutrinos. Neutrino mass in type-I seesaw is given by,
\begin{align}
m^\nu= -m_D M^{-1}_R m_D^T,     
\end{align}
where, $m_D=v/\sqrt{2}\; Y_\nu$ with $Y_\nu$ being the Dirac neutrino Yukawa coupling, and $M_R$ is the right-handed neutrino mass. For order one Yukawa coupling for the neutrinos, one then obtains:
\begin{align}
M_3\sim 10^{15}~ \textrm{GeV}\sim M_{\rm GUT},\;\; \Rightarrow m^\nu_3\sim 0.01 ~ \textrm{eV},  
\end{align}
consistent with neutrino oscillation data. In the vanilla leptogenesis scenario when $N_1$, the lightest of the right-handed neutrinos,  decays into lepton-Higgs pairs, a  lepton asymmetry in these CP-violating out-of-equilibrium decays are generated, which is then partially converted to  baryon
asymmetry by the sphaleron processes and the right amount~\cite{Planck:2018nkj} of matter-antimatter asymmetry of the Universe can be reproduced. Leptogenesis and its connection to absolute neutrino mass and neutrino mixing parameters 
in the context of $SO(10)$ GUTs are studied in Refs.~\cite{Altarelli:2013aqa,Fong:2014gea,DiBari:2020plh,Mummidi:2021anm}.

The $CP$-asymmetry from the right-handed neutrino decays can be estimated to be $\epsilon \sim {m_\nu M_R}/{(16\pi v^2)}$. A sufficiently large $CP$-asymmetry $\epsilon \gtrsim 10^{-7}$ to generate adequate baryon asymmetry imposes the so-called Davidson-Ibarra bound $M_R \gtrsim 10^9$ $\text{GeV}$ \cite{Davidson:2002qv}. However, if a pair of right-handed neutrinos are quasi-degenerate in mass, the $CP$ asymmetry can be resonantly enhanced \cite{Covi:1996fm,Pilaftsis:1997jf, Pilaftsis:2003gt} (see also Refs. \cite{Akhmedov:1998qx,Pilaftsis:1997jf,Drewes:2017zyw,Dev:2017wwc}). Subsequently, the lower bound on $M_R$ comes only from the requirement that sufficient baryon asymmetry is induced at $T > T_{sph}$. In this resonant leptogenesis scenario, right-handed neutrinos can have masses as low as sub-GeV to TeV range and low-scale seesaw models of this type have the virtue of being directly probed in experiments~\cite{Chun:2017spz}.   For implications of low-scale resonant leptogenesis from GUT, see e.g., Ref.~\cite{Fong:2021tqj}.

\subsubsection{Baryogenesis}
\begin{figure}[th!]
\centering
\includegraphics[width=0.45\textwidth]{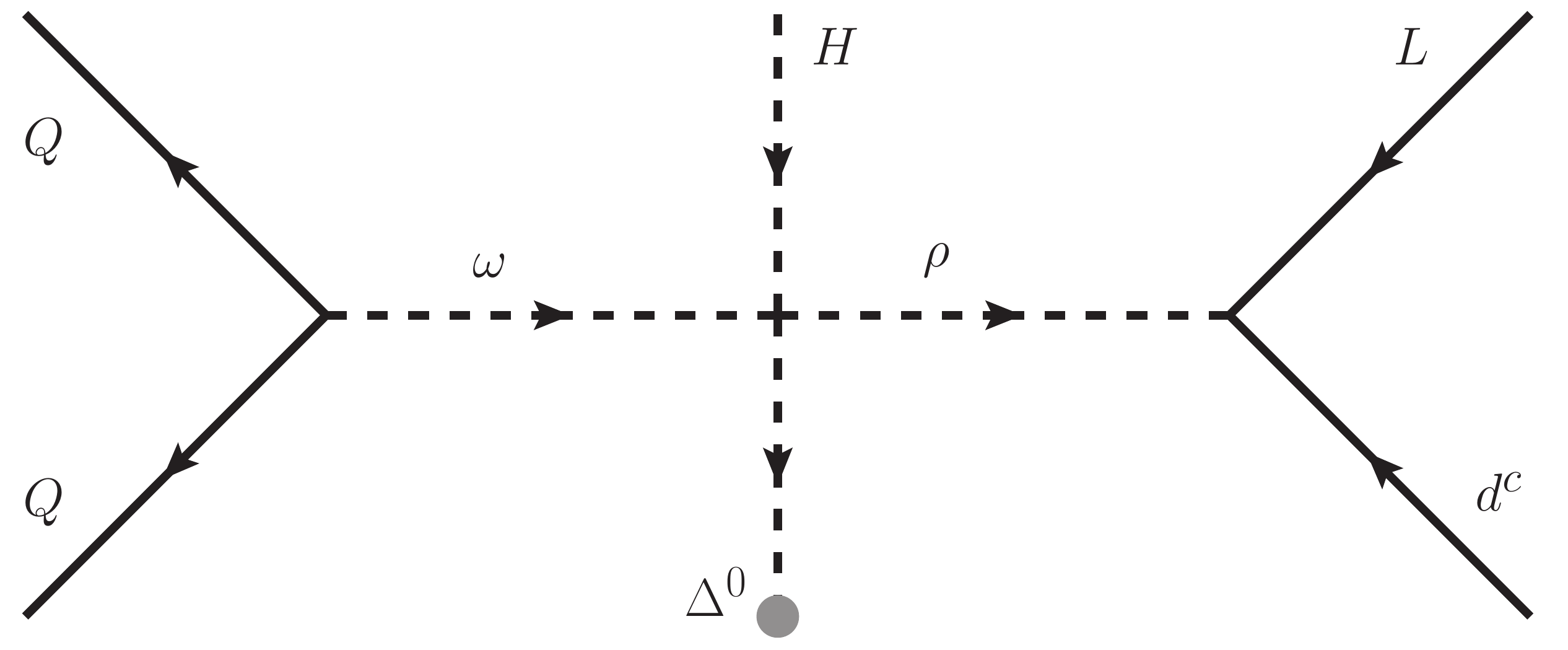} 
\caption{ 
Effective ${B-L}$ violating dimension-7 nucleon decay operators induced by symmetric Yukawa couplings of $\bf 10$ and $\bf \overline{126}$ Higgs fields of $SO(10)$.}
\label{fig:Baryogenesis}
\end{figure}

The two leading proton decay modes, $p\to e^+ \pi^0$ and $p\to \overline{\nu}K^+$ as extensively discussed above, both conserve ${B-L}$  symmetry (i.e., $\Delta ({B-L})=0$). For this reason, the minimal $SU(5)$ model does not have any way to explain the origin of matter even though it has both baryon number violation as well as CP violation. Sphaleron processes would wash out any baryon asymmetry generated in such models. However, going beyond these $d=6$  operators, the next-to-leading BNV operators correspond to $d=7$,
an explicit computation of which for a class of $SO(10)$ models is given in Ref.~\cite{Nath:2015kaa}, which break this symmetry by $\Delta ({B-L})=-2$ and generate nucleon decay modes such as $p\to \nu\pi^+$, $n\to e^-\pi^+$, $n\to e^-K^+$. As mentioned above, spontaneous breaking of ${B-L}$ in $SO(10)$ GUTs is directly related to the neutrino mass generation. 
Intriguingly, in a class of $SO(10)$ GUTs (applicable to both non-SUSY and SUSY), these $d=7$ nucleon decay modes for which the lifetime can be in the experimentally accessible range are inherently linked~\cite{Babu:2012vb, Babu:2012iv}  to the matter-antimatter asymmetry in the Universe.  Due to their ${B-L}$ breaking nature, electroweak sphaleron processes would not wash out such a GUT-scale-induced baryon asymmetry; hence asymmetry generated would survive down to low temperatures. The vacuum expectation value (VEV) of $\bf \overline{126}$ Higgs breaks the ${B-L}$ generator of the $SO(10)$ and provides a large Majorana mass for right-handed neutrinos as well as generates trilinear scalar couplings that induce $d=7$ nucleon decay operators; an example diagram is presented in Fig.~\ref{fig:Baryogenesis}. In this figure, a trilinear coupling $\omega H \rho^\ast$  is induced via VEV insertion of a $\Delta(1,1,0)$ scalar that breaks the ${B-L}$ generator. Here, the quantum numbers of these fields are: $\omega(3,1,-1/3), H(1,2,1/2), \rho(3,2,1/6)$ and this term originates from a quartic coupling in the scalar potential of the form ${\bf 126}^4$, which is gauge invariant. From the left (right) vertex in the Feynman diagram Fig.~\ref{fig:Baryogenesis}, it can be understood that $\omega$ ($\rho$) has ${B-L}=-2/3$ (${B-L}=+4/3$); this clearly dictates that the associated trilinear coupling  leads to the decay $\omega\to H^\ast \rho$ which violates ${B-L}$ by $-2$. Thus, when tree as well as loop diagrams for the decay $\omega\to H^\ast \rho$ are combined, a net ${B-L}$ asymmetry is generated. This GUT-scale-induced baryon asymmetry in $\omega\to H^\ast \rho$ decay that has deep correlations with neutrino mass generation and nucleon decay  is shown~\cite{Babu:2012vb, Babu:2012iv} to correctly reproduce the observed baryon-to-entropy ratio in the Universe. 

\subsubsection{Dark Matter}
The natural appearance of DM in SUSY GUTs has already been mentioned. Here we briefly discuss some DM candidates in non-SUSY GUTs. There
are a large number of DM candidates in the literature, and a comprehensive review of these models is beyond the scope of this white paper. 

It is remarkable that GUTs based on $SO(10)$ gauge symmetry automatically contain matter parity $P_M=(-1)^{3(B-L)}$, as a discrete subgroup that can naturally stabilize the DM without the need for imposing \textit{ad hoc} symmetries by hand.  To be specific,  stability of the DM is guaranteed by the discrete subgroup ${Z}^{3(B-L)}_2$ of $U(1)_{B-L}\subset SO(10)$ ~\cite{Martin:1992mq, Kadastik:2009dj,Kadastik:2009cu,Frigerio:2009wf} that typically arises when the intermediate group ${G}_{\textrm{int}}$ breaks to the SM group ${G}_{\textrm{SM}}$~\cite{Mambrini:2015vna}: 
\begin{align}
SO(10)\to    {G}_{\textrm{int}} \to  {G}_{\textrm{SM}}\times {Z}_2.
\end{align}
The smallest representation in SO(10) that can realize this symmetry breaking pattern (\textit{i.e.}, with a remnant $Z_2$ symmetry) is $\mathbf{126}$, which can also be used to generate Majorana masses for right-handed neutrinos as discussed above.  

Under the matter parity, only the $\bf 16$- and $\bf 144$-dimensional representations are odd, whereas the rest of them are even (here, we restrict ourselves to representations not bigger than $\bf 210$). Given the fact that the SM Higgs doublet residing in a $\bf 10$-dimensional representation is even while  a $\bf 16$-dimensional representation containing 
fermions of each generation is odd, the lightest component of an additional even fermion or an odd scalar multiplet will be stable due to the residual ${Z}_2$ symmetry ~\cite{Mambrini:2013iaa,Mambrini:2015vna,Nagata:2015dma,Boucenna:2015sdg,Arbelaez:2015ila,Bandyopadhyay:2017uwc,Ferrari:2018rey}. However, this stability will be lost if a Higgs in the $\bf 16$-dimensional or $\bf 144$-dimensional representation acquires a VEV.   This leads to the list of possible DM candidates in $SO(10)$ GUTs given in Table~\ref{tab:DM}.  DM candidates have also been presented in the 
context of extra-dimensional models (e.g., \cite{Girmohanta:2020llh,Girmohanta:2021gpf}.
\begin{table}[th!]
\begin{center}
\small{
\begin{tabular}{|l|c|c|}
\hline
& {Fermions} &{Scalars} \\ \hline
\begin{tabular}{c} DM multiplet \\$SU(2)_L\times U(1)_Y$    \end{tabular} &   
\begin{tabular}{c} Even  $SO(10)$ \\ multiplet \end{tabular}   & 
\begin{tabular}{c} Odd  $SO(10)$ \\ multiplet \end{tabular}  \\ \hline
\hline
\quad$1_0$  & $\bf{45}$, $\bf{54}$, $\bf{126}$, $\bf{210}$ 
 & $\bf{16}$, $\bf{144}$ 
\\
\hline
\quad$2_{\pm1/2}$  &  $\bf{10}$, $\bf{120}$, $\bf{126}$, $\bf{210}$, $\bf{210}'$ 
&  $\bf{16}$, $\bf{144}$ 
 \\
\hline
\quad$3_0$ & $\bf{45}$, $\bf{54}$, $\bf{210}$ 
& $\bf{144}$  \\
\hline
\quad$3_{\pm1}$  & $\bf{54}$, $\bf{126}$ 
 & $\bf{144}$ \\

\hline 
\end{tabular}}
\end{center}
\caption{Possible DM candidates in $SO(10)$ GUTs.   $SU(2)_L\times U(1)_Y$ multiplet that contains a neutral component is presented in the first column. The second (third) column lists representations that are even (odd) under $P_M$ and contain a DM candidate are listed in the second (third) column. The triplet candidates with hypercharge $\pm 1$ are shown for completeness; however, they are not viable as a DM candidates. }\label{tab:DM}
\end{table}

\subsubsection{Flavor Violation}
It is widely known that  low-energy SUSY models with arbitrary mixings in the soft breaking parameters would lead to unacceptably large  flavor-violating effects (e.g., \cite{Hall:1985dx,Gabbiani:1996hi}).  It is essential to assume flavor universality in the mechanism that breaks SUSY to be consistent with experimental observations.  Starting with flavor-universal SUSY breaking boundary conditions at a high scale, running effects of the renormalization group equations (RGEs) can still induce sizable flavor mixings in soft breaking parameters that  lead to extensive flavor violations at low energies. Many constraining flavor-violating processes arise in the leptonic sector and are inherently connected to neutrino parameters. In $SO(10)$ GUTs, Dirac neutrino
Yukawa couplings $Y_\nu$ induce observable lepton flavor violations (LFVs) that put strong constraints on SUSY breaking parameters~\cite{Calibbi:2006nq, Ciuchini:2007ha}.   Present experimental bounds  on some of the most important LFV processes are~\cite{ParticleDataGroup:2020ssz}:
\begin{align}
&{\rm BR}(\mu\to e\gamma) < 4.2\times 10^{-13},\;\;\;
{\rm BR}(\mu\to eee) < 1.0\times 10^{-12},\\ 
&{\rm BR}(\tau\to e\gamma) < 3.3\times 10^{-8},\;\;\;
{\rm BR}(\tau\to eee) < 2.7\times 10^{-8},\\
&{\rm BR}(\tau\to \mu\gamma) < 4.4\times 10^{-8},\;\;\;
{\rm BR}(\tau\to \mu\mu\mu) < 2.1\times 10^{-8}.
\end{align}

\begin{figure}[th!]
\centering
\includegraphics[scale=0.45]{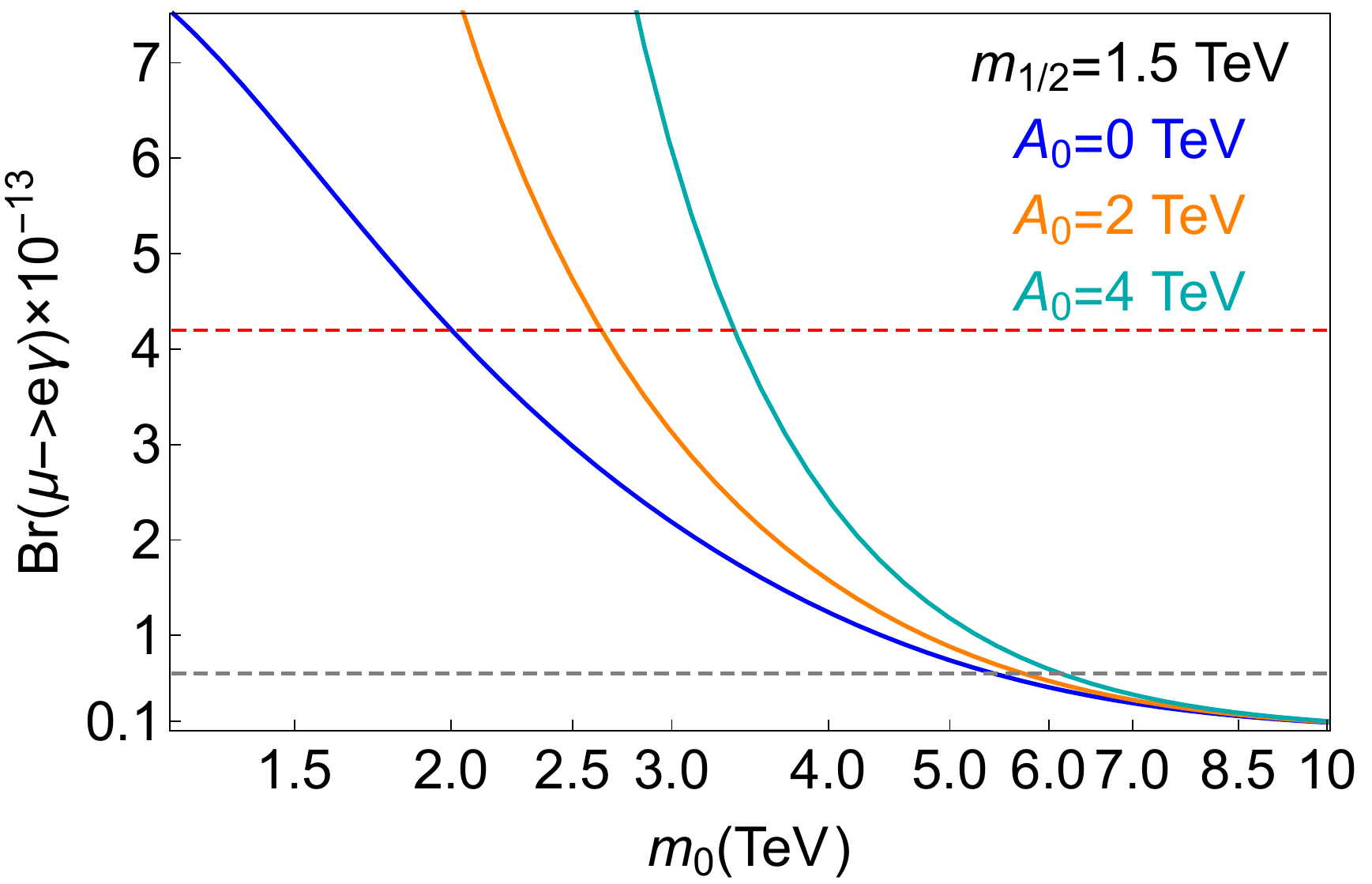}
\includegraphics[scale=0.45]{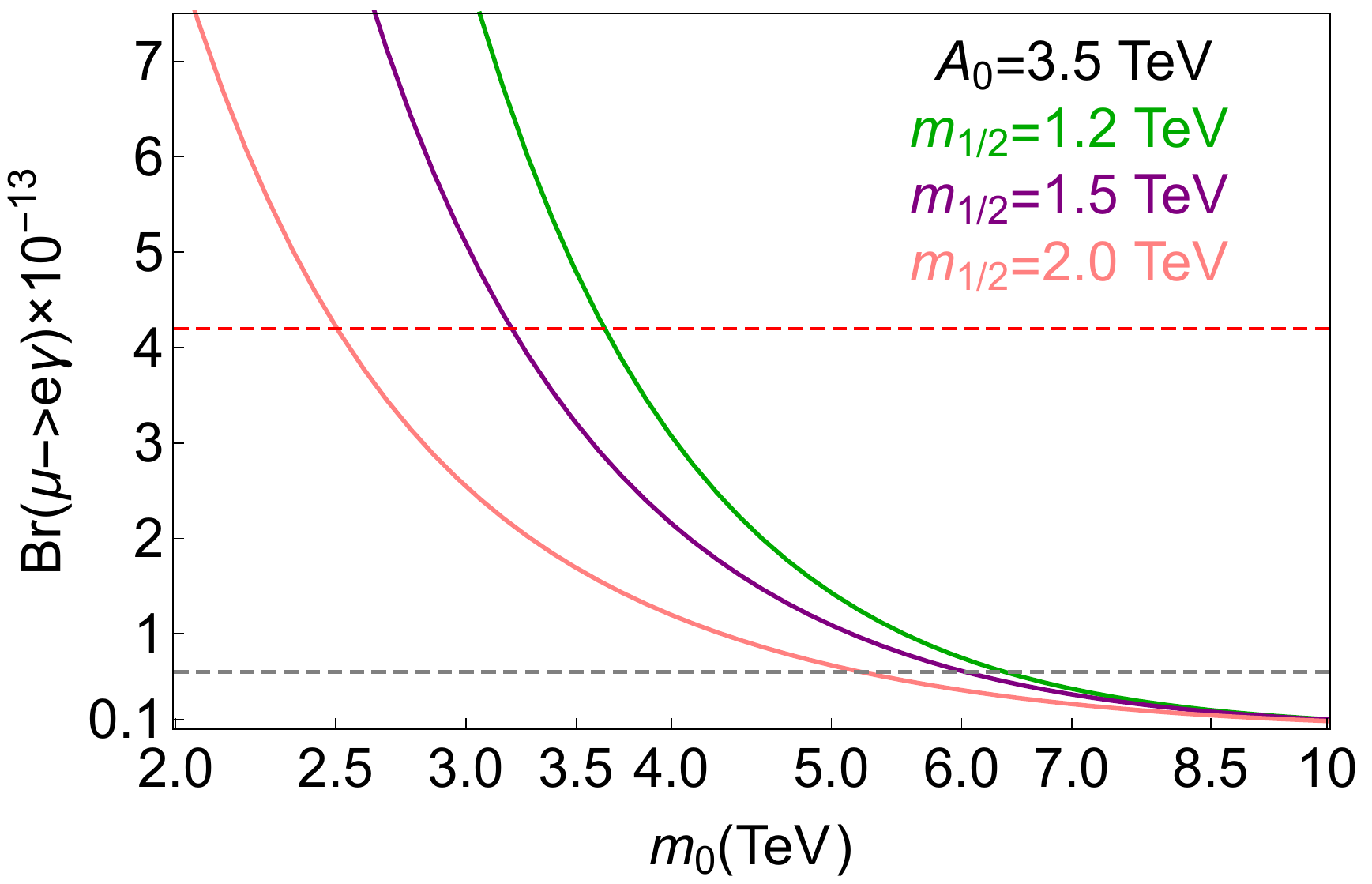}
\caption{Branching ratio of $\mu \to e\gamma$ is presented for  SUSY $SO(10)\times U(1)_{\textrm{PQ}}$ model of Ref.~\cite{Babu:2018qca}.  The red horizontal dashed line corresponds to the current  upper bound ${\rm BR}(\mu \to e\gamma) < 4.2\times 10^{-13}$  by the MEG experiment  and the  gray horizontal dashed line represents the projected sensitivity ${\rm BR}(\mu \to e\gamma) < 6\times 10^{-14}$ by the    MEG II experiment ~\cite{MEG:2016leq}. }\label{fig:LFV}
\end{figure}
In the RG evolution of the left-handed  slepton soft-masses $(m^2_{\widetilde L})_{ij}$, flavor violating off-diagonal entries are induced that are proportional to  $Y_\nu Y_\nu^\dagger$:
\begin{align}
(\Delta^{\ell}_{LL})_{ij}=-\frac{3m_0^2+A^2_0}{8\pi^2}\sum^3_{k=3} (Y_{\nu})_{ik} (Y^{\ast}_{\nu})_{kj} \text{ln}\left(\frac{M_{\rm GUT}}{M_{R_k}}\right),
\end{align}
where $M_{R_k}$ are the heavy right-handed neutrino masses. Assuming scalar mass universality and gaugino mass unification as is usually done in cMSSM, there exist only a few parameters in the SUSY breaking sector:  $\{m_0, m_{1/2}, A_0, \tan\beta, \text{sgn}(\mu) \}$. Here  $m_0$ and $m_{1/2}$ are the universal soft SUSY-breaking (SSB) scalar and gaugino masses, respectively; $A_0$ is the universal tri-linear coupling and $\mu$ the Higgs mass term, which is determined from  electroweak symmetry breaking condition.  The fit to fermion spectrum in a specific GUT fully determines $\tan\beta$ as well as $Y_{\nu}$ and the right-handed neutrino masses $M_{R_k}$. Consequently, one can  compute the rates for LFV  processes as  functions of the remaining   SUSY breaking parameters:   $\{m_0, m_{1/2}, A_0\}$ (in this way, the gaugino, slepton, and squark masses are expressed in terms of $\{m_0, m_{1/2}, A_0\}$). Such correlations within a minimal $SO(10)$ GUTs are presented in Fig.~\ref{fig:LFV}; for details see  Ref.~\cite{Babu:2018qca}. Quark flavor violating processes such as $K^0-\overline K^0$ mixing, $b\to s\gamma$, etc.  also put strong constraints on SUSY breaking parameters. For a systematic study of constraints on Yukawa couplings due to GUT relations and finding correlations between constraints on flavor violation between the lepton and quark sectors, see Ref.~\cite{Ciuchini:2007ha}.

\subsubsection{Gravitational Wave Production}
After the great success of the direct observation of gravitational waves (GWs) by the LIGO collaboration~\cite{LIGOScientific:2016dsl}, GW detection has been considered to be a powerful probe of new physics that complements experiments in particle physics. 
BNV effects are usually associated with new symmetries (e.g., those preserving $B-L$) at high scales. Spontaneous breaking of these symmetries may generate gravitational radiation in the early Universe, which appears as cosmic GW background today. The GW exploration will be very helpful to understand these processes and any new physics behind them. Here we discuss potential sources of GWs generated from BNV-related physics: 
GWs via cosmic strings generated from intermediate symmetry breaking in GUTs and those via first-order phase transition. 

The production of cosmic strings is another important prediction for most GUTs in addition to proton decay. GUTs beyond $SU(5)$ provide a series of intermediate symmetries before breaking down to the SM. The spontaneous breaking of the GUT symmetry to the SM gauge symmetry occurs in several steps, and topological defects are produced accompanied by the symmetry breaking. Those defects include monopoles, domain walls and cosmic strings~\cite{Kibble:1976sj}. The former two are problematic as they would  come to dominate the energy density of the Universe. This problem is solved by including a period of inflation after their production. 
The production of cosmic strings is usually associated with the breaking of a $U(1)$ symmetry, which appears as a subgroup of $SO(10)$ or larger groups~\cite{Jeannerot:2003qv}. These strings, if produced after inflation, evolve to a network. The network follows a scaling solution during the Hubble expansion and the energy density does not overclose the Universe~\cite{Kibble:1984hp, Bennett:1987vf, Allen:1990tv}. 

Strings in the network intersect to form loops. Cusps and oscillations of loops release energy via gravitational radiation, which forms a stochastic background today~\cite{Vilenkin:1984ib,Caldwell:1991jj,Hindmarsh:1994re}. 
The GW density parameter $\Omega_{\rm GW} \equiv \rho_{\rm GW}/\rho_{c}$, where $\rho_{c}$ is the critical energy density of the Universe, appears with a characteristic spectrum that drops in the low frequency, peaks in the middle and flattens in the high frequency~\cite{Cui:2018rwi}. $\Omega_{\rm GW}$ is directly determined by the string tension $\mu$, which represents the energy per unit length of the string. In the high frequency band, $\Omega_{\rm GW}$ follows a simple correlation with the string tension, $\Omega_{\rm GW}(f) \propto (G\mu)^{1/2}$, where $G$ is the Newton constant~\cite{Auclair:2019wcv}. In the future, a series of undergoing or planned GW observatories such as LIGO~\cite{LIGOScientific:2019vic}, SKA~\cite{Janssen:2014dka}, LISA~\cite{Audley:2017drz}, Taiji~\cite{Guo:2018npi}, TianQin~\cite{Luo:2015ght}, BBO~\cite{Corbin:2005ny}, DECIGO~\cite{Seto:2001qf}, Einstein Telescope~\cite{Sathyaprakash:2012jk} (ET), Cosmic Explorer~\cite{Evans:2016mbw} (CE), MAGIS~\cite{Graham:2017pmn}, AEDGE~\cite{Bertoldi:2019tck}, and AION~\cite{Badurina:2019hst} have the potential to explore $\Omega_{\rm GW}$ in a wide range $\sim 10^{-15}{\text -}  10^{-7}$, referring to $G \mu \sim 10^{-19}{\text -} 10^{-11}$ and the new physics scale $\sim \sqrt{\mu} \sim 10^{9}{\text -} 10^{13}$~GeV following the result of Nambu-Goto string simulation~\cite{Cui:2018rwi}. 

The measurement of the cosmic GW background provides a novel way to probe grand unification~\cite{Buchmuller:2013lra,Buchmuller:2019gfy}. It is worth mentioning that the new physics scale probed by GW measurements is usually not the GUT scale $M_X$, but an intermediate scale between $M_X$ and the electroweak scale~\cite{King:2020hyd}.
This scale is not arbitrary, but correlated with $M_X$ via the unification of gauge couplings. Measuring proton decay in neutrino experiments can determine both $M_X$ and the intermediate scale as well as $G\mu$.
For example, given the lower bound $\tau(p \to \pi^0 e^+) \gtrsim 1.6 \times 10^{34}$ yrs from Super-K~\cite{Super-Kamiokande:2016exg},
restrictions on $G\mu$ in the following typical breaking chains are obtained,
\begin{eqnarray}
SO(10) \to G_{3221}^{D \!\!\!\! /} \to {\cal G}_{\rm SM}:&&
G \mu \simeq 2.0 \times 10^{-17} \,,\nonumber\\
SO(10) \to G_{422}^D \to G_{3221}^{D \!\!\!\! /} \to {\cal G}_{\rm SM}:&&
G \mu \simeq 2.0 \times 10^{-17} {\text -} 8.4 \times 10^{-15} \,, \nonumber\\
SO(10) \to G_{422}^D \to G_{3221}^D \to G_{3221}^{D \!\!\!\! /} \to {\cal G}_{\rm SM}:&&
G \mu \simeq 2.0 \times 10^{-17} {\text -} 1.3 \times 10^{-10} \,,\label{eqn:chains}
\end{eqnarray}
where $D$ in the upper case refers the left-right matter parity symmetry and the minimal particle content consistent with the SM and neutrino masses have been considered~\cite{King:2021gmj}. Once any proton decay signal is confirmed in next-generation neutrino experiments, GW measurement provides a good opportunity to study the details of GUT breaking. Note that the the spontaneous breaking of the lowest intermediate gauge symmetries for all chains in Eq.~\ref{eqn:chains} also leads to the $U_{B-L}$ breaking. It provides a window to connect GW signals with the Majorana nature of neutrinos and to test leptogenesis via seesaw mechanism~\cite{Dror:2019syi}.
  
For the symmetry breaking scale below $10^{8}$~GeV, a strong first-order  phase transition might provide a compelling source of observable GW signals. 
During the phase transition, bubbles of the broken phase nucleate and expand in the Universe.
The collision of the bubbles and the resulting motion of the ambient cosmic fluid provides a source of a stochastic background of GWs that can be observable at GW observatories~\cite{Caprini:2019egz}.

The GW spectrum is in general given by the sum of three contributions: bubble wall collisions, sound waves and hydrodynamic turbulence (see Ref.~\cite{Caprini:2015zlo,Caprini:2019egz} for recent reviews). Despite the energy budget, the spectrum has a generic feature: a peak in the middle and polynomial suppressions in both low and high frequencies. It is distinguishable with that from cosmic strings as there is no plateau in the high frequency band.  
The frequency at the peak depends on the phase transition temperature linearly. 
In addition to the temperature, two more parameters are crucial to determine the GW spectrum: $\alpha$ the ratio of the released latent heat during the phase transition to the total energy density and $\beta/H_\star$ the ratio of the inverse duration of the phase transition to the Hubble rate. $\alpha$ dominates the GW strength. Enhancing $\alpha$ by one order of magnitude can enhance the GW strength by several orders of magnitude. $\beta/H_\star$ influences both the strength and frequency of GW. Faster the phase transition proceeding, higher frequency of the GW signal is expected. Theoretical efforts have been made to enhance $\alpha \to {\cal O}(0.1)$ and reduce $\beta/H_\star \to {\cal O}(10^2{\text -} 10^3)$ such that part of the parameter space of $B$- or $L$-violating new physics can be reached by the exploration of next-generation GW interferometers (see, e.g., discussions in~\cite{Bian:2019szo,Haba:2019qol,Hasegawa:2019amx,Marzo:2018nov,Jinno:2016knw,DiBari:2021dri,DiBari:2020bvn}).

\subsection{Lattice Developments} 
\label{sec:Lattice} Rates of proton decays and neutron oscillations depend on nucleon and nuclear matrix elements of effective baryon number-violating operators. 
Prior to development of lattice QCD methods, these matrix elements were initially calculated with various nucleon models.
However, eliminating theory uncertainties completely requires ab initio QCD calculations on a lattice with physical quark parameters. 
Calculations of proton decay amplitudes have been pursued since simulating nucleons on a lattice became possible, with methodology gradually improving over the last three decades.
Neutron oscillation amplitudes were studied only relatively recently, when realistic lattice QCD calculations were already feasible.

Apart from the usual systematic effects inherent in lattice QCD calculations (discretization, finite volume, etc), calculations of BNV amplitudes are complicated by the need to preserve chiral symmetry of the quarks.
Since the effective proton decay and neutron oscillation operators contain chiral quark fields, using a simple fermion discretization such as Wilson action may lead to undesirable mixing of operators on a lattice.
This mixing would complicate renormalization and introduce additional systematic effects.
Fortunately, formidable progress has been made towards lattice calculations with chiral fermions at the physical point~\cite{Blum:2014tka,Boyle:2015exm,Blum:2019ugy}.

Amplitudes of $p\to\bar{\ell}\pi$, $p\to\bar{\ell}K$ decays may be computed on a lattice \emph{directly} as matrix elements between the proton and a meson.
Such calculations have initially been performed in quenched QCD with Wilson valence quarks~\cite{Aoki:1999tw} and DWF quarks~\cite{Aoki:2006ib}, and later in unitary QCD with $N_f=2+1$ flavors of DW fermions~\cite{Aoki:2013yxa,Aoki:2017puj}.
All these calculations have been performed with unphysical masses of light quarks corresponding to pion masses $m_\pi\gtrsim300\,\mathrm{MeV}$, and required some form of chiral extrapolation to the physical point.
Although these amplitudes did not exhibit strong pion mass dependence, applying chiral perturbation theory (ChPT) to baryons is often questioned, and proton decay amplitudes have not been computed in ChPT beyond the leading order (LO).
In the framework of the chiral-bag proton model, it has also been suggested that the proton decay matrix elements may depend dramatically on the light quark masses~\cite{Martin:2011nd}.
These deficiencies have been addressed in the recent work~\cite{Yoo:2021gql}, in which the proton decay amplitudes have been calculated with 10-20\% precision using chirally symmetric DW fermions at the physical point, albeit with relatively coarse lattice spacing $a\gtrsim0.14\,\mathrm{fm}$.
The physical-point results are largely in agreement with previous calculations, and no suppression at light-quark masses has been observed.

The proton-to-meson transition amplitudes can also be \emph{``indirectly''} estimated from proton-to-vacuum proton decay constants $\alpha,\beta$~\cite{Claudson:1981gh} using the leading-order chiral Lagrangian.
These decay constants have been computed using quenched QCD with Wilson
quarks~\cite{Hara:1986hk,Bowler:1987us,Tsutsui:2004qc} and Domain Wall quarks~\cite{Aoki:2006ib}, as well as in unitary QCD with dynamical DW fermions~\cite{Aoki:2008ku,Aoki:2017puj,Yoo:2021gql}.
The ``indirect'' estimates of proton decay amplitudes are typically higher than the ``direct'' results, which is likely attributable to pion loop effects.
A calculation with next-to-leading order ChPT is highly desirable to understand this discrepancy.
The proton-to-vacuum decay constants are also highly important to analysis of  $p\to \mu^- e^+ e^+$ and $p\to e^-\mu^+\mu^+$ proton decays.

A particular BSM theory yields predictions for the effective scale of the physics contributing to $n$-$\bar n$ oscillations and predictions for
the coefficients of the various six-quark operators ${\cal O}_i$ contributing to these oscillations.  To calculate the expected rate of 
$n$-$\bar n$ oscillations in a given model, one then needs to compute the matrix elements $\langle \bar n | {\cal O}_i | n\rangle$.  These matrix elements have dimensions of (mass)$^6$, and since at the hadronic level, the only important mass scale in the problem is $\Lambda_{QCD} \simeq 250$ MeV, it follows that the matrix elements are $\sim \Lambda_{QCD}^6$. Early calculations of the $n$-$\bar n$ matrix elements were performed using the MIT bag model \cite{Rao:1982gt,Rao:1983sd}.  Recently, these matrix elements have been calculated using lattice QCD~\cite{Rinaldi:2018osy,Rinaldi:2019thf}. 
This calculation has been performed on a lattice with chirally symmetric action at the physical point, and the operators have been renormalized using two-loop perturbative anomalous dimension~\cite{Buchoff:2015qwa}.  As examples, for one operator, denoted $Q_5$ in \cite{Rinaldi:2018osy}, the lattice QCD calculation (normalized at 2 GeV) yielded a value in agreement with the MIT (fit A) value, while for other operators, the lattice QCD calculation yielded values larger in magnitude by up to an order of magnitude.  Since the general amplitude for $n$-$\bar n$ oscillations has contributions from several different operators and since the matrix elements of these operators have different signs, there can be destructive 
interference, so that one cannot make a statement about the overall size of the $n$-$\bar n$ transition amplitude from a knowledge of individual matrix elements of operators without knowing the details of a given UV model predicting which operators occur and the values of the coefficients of these operators. Nevertheless, the lattice QCD calculations of these matrix elements are valuable inputs for the analysis of the predictions of a given BSM model for $n$-$\bar n$ oscillations. 

Finally, lattice calculations of proton decay and neutron oscillation amplitudes inside nuclei may soon become feasible~\cite{Detmold:2019ghl,Cirigliano:2019jig}.
Although lattice simulation of nuclei requires substantially more resources due to the sign problem, development of exascale computing resources over the next decade is expected to make ab initio calculations of nuclear matrix elements possible.

%% file: Sections/Experiment.tex
\section{Experimental Overview}
\label{sec:expt}




This section summarizes results and sensitivities for baryon number violation searches for currently running, planned, and proposed neutrino detectors, as well as some avenues for potential improvement in nuclear modeling and detection techniques for future detectors.  A comparison of current limits and future sensitivities for important modes are shown in Table~\ref{tab:proton}.

\subsection{Super-K}
\label{sec:sk} 
The Super-Kamiokande neutrino detection experiment represents the 2nd generation of large water Cherenkov neutrino and nucleon decay experiments, following Kamiokande and IMB which ran in the 1980s and 1990s. Super-K (also SK) is currently running and has been in operation since 1996. The experiment addresses major neutrino topics such as atmospheric neutrino oscillation, solar neutrino oscillation, gravitational collapse supernova bursts in the Milky Way galaxy, searching for the diffuse background of supernova neutrinos, and indirectly searching for dark matter via annihilation or decay to neutrinos. The SK detector is also the far detector for the T2K long-baseline neutrino oscillation experiment. The interaction target is 22.5 to 27.2 ktons of ultrapure water in a cylindrical stainless steel tank located 1 km underground in the Kamioka mine in Japan. The target mass is viewed by $11,000$ 50-cm photomultiplier tubes, and is surrounded by an optically isolated active veto of roughly 2-meters of water viewed by $1,800$ PMTs. Since 2020, the detector water system has been upgraded to allow gadolinium sulfate in solution to enhance the capture and detection of neutrons\cite{Super-Kamiokande:2021the}. This Super-Kamiokande detector naturally enables the search for a variety of baryon number violation processes in the same target mass, with only atmospheric neutrinos as a competing background. Foremost are searches for nucleon decay, often cited as one of two possible meanings embedded in the Super-Kamiokande name: {\bf N}ucleon {\bf D}ecay {\bf E}xperiment. 

\begin{figure}[h]
    \centering
    \includegraphics[width=0.90\columnwidth]{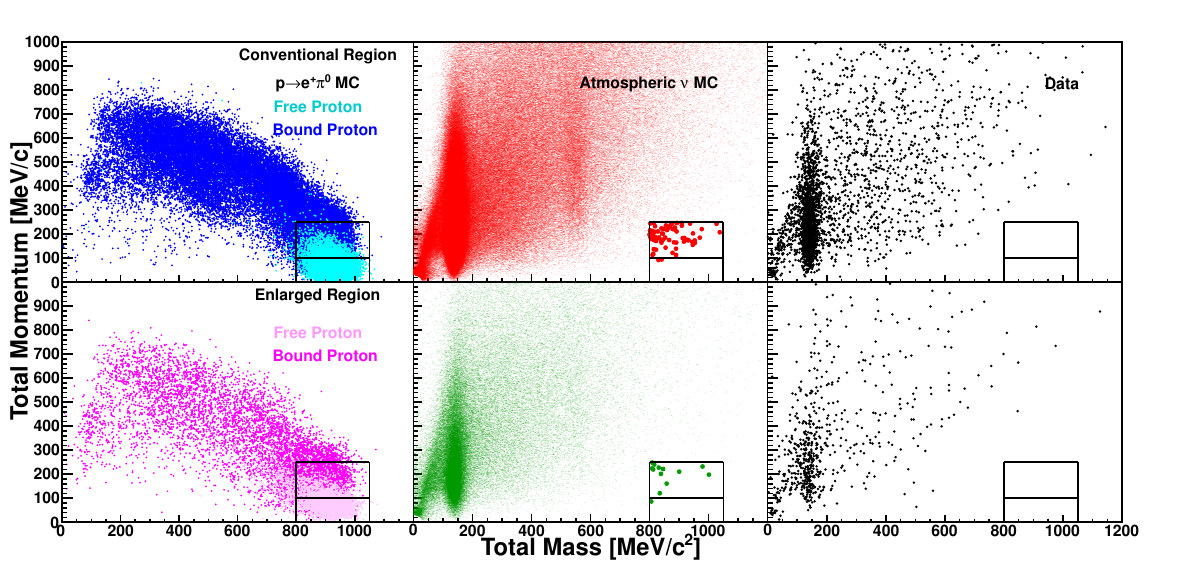}
    \caption{The search for $p \rightarrow e^+\pi^0$ in 450 kton-y of Super-Kamiokande data\cite{Super-Kamiokande:2020wjk}. The leftmost panels show the results of simulated proton decay, where the nuclear effects that degrade the signal for proton decay in $^{16}$O are distinct from the low momentum events from the decay of the free hydrogen nuclei. The center panels show the predicted background due to atmospheric neutrino interactions. The rightmost panels show the SK experimental data. The upper row represents the original fiducial volume, used in the majority of Super-K publications, requiring the interaction vertex to be 200 cm from the PMT plane. The lower row represents the additional fiducial volume that is between 100 cm and 200 cm from the PMT plane.}
    \label{fig:SKepi0}
\end{figure}

Unfortunately for fans of new physics, the SK experiment has not detected any signs of nucleon decay. The extensive exposure, 450 kt-yr in a recent publication, sets a high bar for future experiments to overcome. For example, the lifetime lower limit for $p \rightarrow e^+\pi^0$ is now at $2.4\times10^{34}$~y based on a recent analysis that uses several improvements over older publications\cite{Super-Kamiokande:2020wjk}. The improvements include: background reduction by tagging events where neutron capture on hydrogen is detected nearby the interaction, expansion of the fiducial volume from 22.5 to 27.2 ktons, and dividing the search region into a section where the nearly background-free decay of the free proton in H$_2$O is accentuated. The simplicity and clarity of this search is illustrated in the three-panels of Fig.~\ref{fig:SKepi0}, showing the low background region at high invariant mass and low net momentum (note the extremely low expected background of the free proton region), and the absence of any data events in that region.

Overall, the Super-K experiment has published leading limits on 30 baryon number violating processes including: an extensive survey of antilepton plus meson final states that conserve $(B-L)$, dinucleon decay modes that violate baryon number by two units, and three-body decay modes including those with fully leptonic final states. Special attention has been given to decay channels favored by SUSY, which are distinguished by the presence of a kaon in the final state. The most recent published limit for the key decay mode $p \rightarrow \bar{\nu} K^+$ is $5.9\times 10^{33}$ years based on an exposure of 260 kt-yr\cite{Super-Kamiokande:2014otb}, but a preliminary result of $8\times10^{33}$ years has been reported in conferences. In addition, Super-K holds the leading limit on the interesting $\Delta B=2$ process of neutron oscillation into anti-neutron within the oxygen nucleus\cite{Super-Kamiokande:2020bov}. This may be converted to an effective free $n-{\bar{n}}$ lifetime of $4.7\times 10^8$ s, which is comparable to (slightly exceeding) the leading free neutron experiment\cite{BaldoCeolin:1989qd}. Table~\ref{tbl:SKlims} lists a sample of published limits.

\begin{table}[ht]
\centering
\caption{Selected baryon number violating searches by Super-Kamiokande.}
\begin{tabular}[t]{lcccc}
\hline
Channel & Comment & Exposure & Limit & Reference\\
\hline
$p \rightarrow e^+\pi^0$ & $d=6$ operators, e.g. SU(5) & 450 kt$\cdot$y & $2.4\times 10^{34}$ y & \cite{Super-Kamiokande:2020wjk} \\
$p \rightarrow \mu^+\pi^0$ & flipped SU(5) & 450 kt$\cdot$y & $1.6\times 10^{34}$ y & \cite{Super-Kamiokande:2020wjk} \\
$p \rightarrow \nu K^+$ & $d=5$ SUSY operators & 260 kt$\cdot$y & $5.9\times 10^{33}$ y & \cite{Super-Kamiokande:2014otb} \\
$p \rightarrow \mu^+ K^0$ & SUSY SO(10) & 173 kt$\cdot$y & $1.6\times 10^{33}$ y & \cite{Super-Kamiokande:2012zik} \\
$pp \rightarrow K^+K^+$ & RPV SUSY & 92 kt$\cdot$y & $1.7\times 10^{32}$ y & \cite{Super-Kamiokande:2014hie}\\
$p \rightarrow e^+e^+e^-$ & lepton flavor symmetries & 370 kt$\cdot$y & $3.4\times 10^{34}$ y & \cite{Super-Kamiokande:2020tor} \\
$n \rightarrow \bar{n}$ & $\Delta B = 2$ & 370 kt$\cdot$y & $3.6\times 10^{32}$ y & \cite{Super-Kamiokande:2020bov} \\
$np \rightarrow \tau^+\nu$ & extended Higgs sector & 273 kt$\cdot$y & $2.9\times 10^{31}$ y & \cite{Super-Kamiokande:2015pys} \\
$n \rightarrow \nu\gamma$ & radiative & 273 kt$\cdot$y & $5.5\times 10^{32}$ y & \cite{Super-Kamiokande:2015pys}  \\
$p \rightarrow e^+ \nu\nu$ &  Pati-Salam & 273 kt$\cdot$y & $1.7\times 10^{32}$ y & \cite{Super-Kamiokande:2014pqx} \\
\hline
\end{tabular} \label{tbl:SKlims}
\end{table}%

Major improvement in sensitivity to lifetimes much beyond the limits listed here await the next generation of more massive detectors. Until then, operation of the Super-K experiment with concomitant data analysis will provide incremental progress but valuable methodological developments. Ongoing studies of nucleon decay include continued development of reconstruction techniques, refinement of the intranuclear simulations relevant to both signal and background, first studies of algorithms for neutron capture on gadolinium, and new searches for novel modes. 

\subsection{Neutron-Antineutron transformations in NOvA }

\begin{figure}
\centering
\includegraphics[width=0.7\columnwidth]{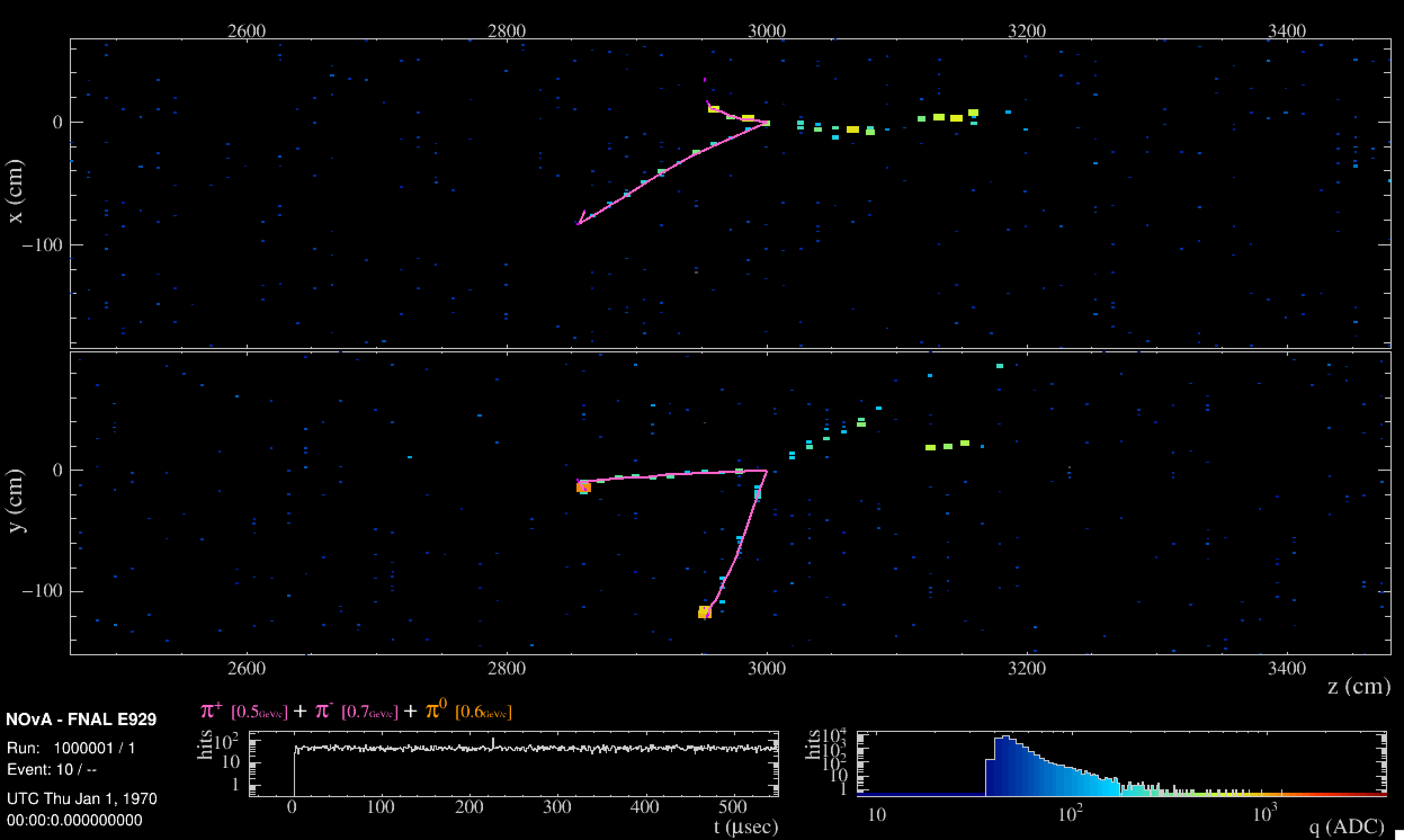}

\caption{\label{fig:novannbar} Simulation of neutron-antineutron oscillation in 
a carbon nucleus in the NOvA Far Detector, yielding $\pi^+\pi^-\pi^0$.  The 
purple lines show the true trajectories of the charged pions.}

\end{figure}

The NOvA Far Detector is sensitive to the spontaneous conversion of neutrons to
antineutrons.  Since all neutrons in NOvA are bound in nuclei, the resulting
antineutron would immediately annihilate on a neutron or proton, typically
yielding three to six pions.  The experimental signature is therefore a star
with approximately zero net momentum and a visible energy typically between
0.8\,GeV and 1.5\,GeV, depending on the mode (see Fig.~\ref{fig:novannbar}).

Neutron-antineutron oscillations are suppressed in nuclei by a factor $R$ such
that $t_\mathrm{free} = \sqrt{t_{bound} / R}$, where $t_\mathrm{free}$ is the
free neutron oscillation time.  The suppression factor varies by nuclide.
About half of the neutrons in NOvA are bound in carbon nuclei, with the
majority of the remainder in chlorine.  Although no calculations exist for
these elements, one calculation of $R$ for oxygen gives
$0.517\times10^{23}\,\mathrm{s}^{-1}$~\cite{Super-Kamiokande:2011idx,Friedman:2008es}.  Likely
the suppression factor for carbon is somewhat lower than for the larger oxygen
nucleus, so there is a possibility that the effective suppression for NOvA as a
whole is lower than for a water-based detector, allowing NOvA to set stronger
limits on $t_\mathrm{free}$, all other things being equal.

NOvA must confront a significant background due to its surface location with
only 3.6~meters water-equivalent overburden.  Since 2018, NOvA has run a
dedicated trigger searching for neutron-antineutron-like events.  By the end of
the planned NOvA run in 2026, 112~kt-years of data will have been collected.
To reduce background from cosmic rays, the trigger requires candidates to be
contained, to have a similar extent in $z$ as they do in $y$ (unlike
downward-going cosmic rays), to have only short tracks that could be produced
by pions of the expected energy, to have a total physical size and number of
hits consistent with a neutron-antineutron oscillation event, and finally to be
symmetric between the $xz$ and $yz$ views.

Backgrounds to this search include atmospheric neutrinos and cosmic rays.  On
the surface, cosmic rays are the more challenging background.  We expect that
the most difficult to exclude will be from neutrons and gammas produced in the
overburden or rock berms adjacent the detector.  Two approaches are being
pursued to study these backgrounds:  first, a data-driven approach using the
energy sidebands above and below the expected signal visible energy, and  second,
a simulation using CORSIKA~\cite{CORSIKA:2021czu}.

The analysis to search for neutron-antineutron events in triggered data is
under development.  Further information available to an offline event
classifier includes more sophisticated measures of event symmetry, calibrated
hit energies, event time duration, and various reconstructed track variables.
Using a CNN to distinguish between neutron annihilation events and cosmic rays,
as is planned for DUNE~\cite{Jwa:2020mtz}, is also a possibility.  If successful in 
suppressing cosmic ray backgrounds below the level of atmospheric neutrino
backgrounds, NOvA will achieve a sensitivity on the free neutron oscillation time
of somewhat longer than $10^8$\,s at 90\% C.L.

\subsection{MicroBooNE}

The MicroBooNE detector is an 89 ton active Liquid Argon Time Projection Chamber (LArTPC) detector located on-surface and exposed to neutrinos from the Booster Neutrino Beamline (BNB) and Neutrinos at Main Injector (NuMI) beamline at Fermilab. The excellent spatial and calorimetric resolution offered by MicroBooNE's LArTPC, also shared by the future Deep Underground Neutrino Experiment (DUNE), enables precise measurements of neutrino scattering as well as beyond-Standard Model (BSM) searches, including intranuclear neutron-antineutron transitions. In a LArTPC, this process is characterized by a unique and striking star-like signature containing multiple final-state pions from the antineutron annihilation with a nearby nucleon. The pions are visible as tracks (from charged pions) or showers (from neutral pion electromagnetic decays) pointing back to the annihilation vertex, with approximately zero net momentum and total energy of about twice the nucleon mass. 

Although MicroBooNE is too small in size and too overwhelmed with cosmogenic background to perform a competitive search, it aims to perform the first ever search for neutron-antineutron ($n-\bar{n}$) transitions in a LArTPC, which will serve as a proof-of-principle demonstration of LArTPC capabilities in searching for this process. The analysis MicroBooNE has developed ~\cite{ub:nnbarpublicnote2022} makes use of state-of-the-art reconstruction tools, including deep learning methods developed for LArTPC image data analysis, to look for neutron-antineutron transition signatures in data collected during neutrino beam-off times. The signal events are simulated using the GENIE Neutrino MC Generator (GENIE) v.3.00.04 with hA-Local Fermi Gas (hA-LFG) as a default model, while backgrounds are expected to be contributed by cosmogenic activity during multiple 2.3 millisecond intervals of exposure (corresponding to a total of 372~seconds) and are evaluated using a data-driven approach. 
The signal interaction is simulated per one exposure interval of 2.3~ms, which also contains multiple reconstructed clusters of cosmogenic activity. 

Pre-selection is applied to all clusters reconstructed during 2.3~ms intervals (``events'') based on a Boosted Decision Tree (BDT) score, to control the background rate and remove obvious backgrounds (cosmics). The BDT is trained using only the topological information derived from the extent of the reconstructed clusters in channel and time space. 
As shown in Tab.~\ref{tab:CV_combined_efficiency}, pre-selection reduces the cosmic background clusters such that the number of $n-\bar{n}$ clusters and cosmic clusters per 2.3~ms interval become of the same order. The pre-selected clusters are subsequently used to train a sparse Convolutional Neural Network (CNN) to differentiate signal from background clusters. For this step, 2D projections of clusters from MicroBooNE's three planes (U, V, Y) are formatted in such a way so as to retain only the important pixels, out of the full image, in the form of a collection of spacepoints containing information of wire position, time-tick and charge deposition per spacepoint. 
Considering statistical-only sensitivity as a figure of merit, an optimized CNN selection provides 73.6$\pm$0.034 \% signal efficiency and 8.77e-3$\pm$7.4e-4 \% cosmic background efficiency. As shown in Tab.~\ref{tab:CV_combined_efficiency}, the final selection highly suppresses the cosmic background clusters while the number of signal clusters remains high.


\begin{table}[!h]
    \vskip 1cm
    \centering
    \begin{tabular}{|c|c|c|}
    \hline\hline
    \textbf{Entities} & \textbf{$n-\bar{n}$} & \textbf{Cosmic} \\
    \hline\hline
    Events & 1,633,525 & 1,618,827 \\
    \hline
    Reconstructed clusters & 1,684,516 & 14,857,224 \\
    \hline
    Clusters (after pre-selection) & 1,455,214 & 1,283,074 \\
    \hline
    Clusters (after final-selection) & 1,207,153 & 142 \\
    \hline
    Events (after final-selection) & 1,202,281 & 142 \\
    \hline\hline
    Selection efficiency (\%) & 73.6 & 8.77e-3 \\ 
    \hline\hline
    \end{tabular}
    \caption{The number of entities for each reconstruction and selection stage. The numbers are evaluated using a simulation sample that corresponds to roughly 10 times the MicroBooNE 372 second exposure. The selection efficiency indicates the ratio between the ``Events'' and ``Events (after final-selection)''. }
    \label{tab:CV_combined_efficiency}
\end{table}

The preliminary sensitivity for MicroBooNE's $n-\bar{n}$ search is calculated assuming 372 seconds of exposure, corresponding to 3.13~neutron$\cdot$years, and considering statistical uncertainties only, and is shown in Tab.~\ref{tab:preliminary_sensitivity}. The sensitivity is calculated using the frequentist-based Rolke method~\cite{Rolke}. Although the $n-\bar{n}$ search in MicroBooNE is not competitive compared to existing limits from SNO or Super-Kamiokande, this analysis serves as a demonstration for the capability of future, larger and well-shielded LArTPC's, including the future DUNE, to perform such searches for baryon number violation with significantly higher exposure and thus higher sensitivity.  The MicroBooNE analysis is ongoing, to incorporate effects of systematic uncertainties.

\begin{table}[!h]
    \centering
    \vskip 1cm
    \begin{tabular}{|c|c|c|}
    \hline\hline
    \textbf{Source of Uncertainty} & \textbf{Sensitivity (Rolke)}\\
    \hline\hline
    Statistical-only  &  3.09 e+25 yrs \\
    \hline\hline
    \end{tabular}
    \caption{90\% C.L.~sensitivity for argon-bound $n-\bar{n}$ transition lifetime assuming 372 seconds of exposure in MicroBooNE, considering only statistical uncertainties.}
    \label{tab:preliminary_sensitivity}
\end{table}

\subsection{Hyper-K}

Continuing the neutrino and nucleon decay physics program in Kamioka, Japan, Hyper-Kamiokande (HK) is a third-generation water Cherenkov detector currently under construction roughly 8~km south of Super-Kamiokande. 
The detector will use a 187~kton water target, roughly 8~times that of its predecessor, and will observe natural neutrinos from the sun, core collapse supernovae, and atmospheric neutrinos as well as a 1.3 MW neutrino beam from an upgraded J-PARC accelerator.  
The experiment is expected to begin operations in spring of 2027 and is expected to improve on BNV searches at SK (c.f. Section~\ref{sec:sk}) by an order of magnitude or more using similar analysis techniques. 

Importantly, Hyper-Kamiokande will be instrumented with improved 50~cm photomultiplier tubes with increased quantum and collection efficiencies, resulting in twice the photon detection efficiency of the sensors used in SK. Further, HK's sensors will have roughly half the timing resolution for single photoelectron signals.
Both of these features positively impact searches for nucleon decays at HK.
Notably, atmospheric neutrino backgrounds can be reduced by 30\% relative to SK's achievement~\cite{Super-Kamiokande:2020wjk} by the augmented ability to tag the faint light from the 2.2~MeV gamma ray produced by neutron capture on hydrogen. 
Focusing specifically on the $p \rightarrow \bar{\nu} K^{+}$ mode, the improved PMT timing resolution allows Hyper-Kamiokande to better separate light from the below-Cherenkov-threshold $K^{+}$'s decay products and that from photons produced by the recoiling oxygen nucleus.
This improves the detection efficiency of the $K^{+} \rightarrow \mu^{+} \nu_{\mu}$ mode from 9.1\%~\cite{Super-Kamiokande:2014otb} to 12.7\% in HK~\cite{Hyper-Kamiokande:2018ofw}.
Altogether Hyper-Kamiokande's discovery potential after  10 years will exceed $3\sigma$ if the effective lifetime of the proton decay into this channel is less than $2\times 10^{34}$ years. 
For the other ``flagship'' mode, $p \rightarrow e^{+} \pi^{0}$, a lifetime less than $6\times 10^{34}$ years, i.e. slightly more than twice the current SK limit, will lead to a $3\sigma$ detection. 

Unified theories predict branching ratios for the various possible proton decay modes, indicating that observations of multiple channels and therefore comprehensive coverage of the possibilities is important for determining the symmetries of the underlying model.  Table~\ref{tbl:hkpdksens} lists Hyper-Kamiokande's median sensitivity to several of these modes, including some $|\Delta (B-L)| = 2|$ and $\Delta B = 2$ modes.  
It should be noted that for modes in which the initial state cannot be fully reconstructed from the decay products, such as $p \rightarrow e^{+} \nu\nu$, the search becomes background limited. 
In such cases the sensitivity increases with only the root of the exposure, resulting in only a factor of a few improvement in existing limits. 
Accordingly, the challenge for HK going forward is to leverage its improved detector to reduce backgrounds to all modes. 
Currently efforts to reduce backgrounds include adoption of improved particle reconstruction algorithms, improved detector calibration and light collection with additional photosensors, and introducing enhanced neutron tagging methods, either via algorithms or future gadolinium doping. 
\begin{table}
\begin{center}
\begin{tabular}{l|c|l|c}
\hline
Mode        & Sensitivity (90$\%$ CL) [years] &  Mode & Sensitivity (90$\%$ CL) [years]  \\
\hline
\hline
$p \rightarrow e^{+} \pi^{0}$       & 7.8 $\times 10^{34}$ &
$p \rightarrow \overline{\nu} K^{+}$ & 3.2 $\times 10^{34}$ \\
\hline
$p \rightarrow \mu^{+} \pi^{0}$ & 7.7$\times 10^{34}$ &
$p \rightarrow \mu^{+} \eta^{0}$& 4.9$\times 10^{34}$ \\
\hline
$p \rightarrow e^{+} \rho^{0}$& 0.63$\times 10^{34}$  &
$p \rightarrow \mu^{+} \rho^{0}$& 0.22$\times 10^{34}$ \\
\hline
\hline
$p \rightarrow e^{+}   \nu\nu$       & 10.2 $\times 10^{32}$  &
$p \rightarrow \mu^{+} \nu\nu$       & 10.7 $\times 10^{32}$  \\
\hline
$p \rightarrow e{+} X        $       & 31.1 $\times 10^{32}$  & 
$p \rightarrow \mu^{+} X     $       & 33.8 $\times 10^{32}$   \\
\hline
\hline
$np \rightarrow e^{+} \nu    $       &  6.2 $\times 10^{32}$ &
$np \rightarrow \mu^{+} \nu  $       &  4.2 $\times 10^{32}$ \\
\hline
$np \rightarrow \tau^{+} \nu $       &  6.0 $\times 10^{32}$             &  
$n \rightarrow e^{-}K^{+} $       & 1.0 $\times 10^{34}$       \\
\hline
\hline
\end{tabular}
\end{center}
\caption{\label{tbl:hkpdksens}Sensitivities at 90\% C.L. to various single nucleon decay modes after 10 years of Hyper-Kamiokande operations and after 20 years in the case of dinucleon decay modes. }
\end{table}

\subsection{DUNE}

The Deep Underground Neutrino Experiment (DUNE) promises one of the largest highly instrumented fiducial detector masses of any future large underground facility~\citep{DUNE:2020lwj,DUNE:2020txw}. With $40\,$kt of liquid argon (LAr) some 1500~m below Lead, South Dakota to shield against cosmic ray backgrounds, DUNE's immense wire readout particle ionization charge-collection system across four separate modules forms its three-dimensional LAr time projection chambers, allowing scientists to exploit bubble-chamber quality images~\citep{MicroBooNE:2020fxd,DUNE:2020cqd,Totani:2020poo} for world-leading precision physics studies of the SM and beyond. With potentially MeV-scale precision~\citep{Acciarri:2018myr}, the ability to distinguish $\gamma$ and $e$ species~\citep{Acciarri:2020lhp}, low cosmic $\mu$ backgrounds, and very low LAr ionization kinetic energy thresholds for even heavy charged species such as protons ($p$s)~\citep{DUNE:2020txw,DUNE:2020ypp}, the overall physics potential of DUNE goes far beyond its initial purpose as a $\nu$ detector built to better constrain and measure oscillation parameters such as $\delta_{CP}$. Indeed, the bubble-chamber-like capabilities of DUNE allow for observation of complex event topologies with potentially high multiplicities. Combined with state of the art detector reconstruction and particle identification (PID) methodologies, as well as a gargantuan number of intranuclear nucleons, there is great potential for the DUNE far detector to unlock the secrets behind rare processes.

Sensitivity to several of these processes has been studied~\cite{DUNE:2020fgq} using the full DUNE simulation and reconstruction analysis chain, including the impact of nuclear modeling and FSI on a Boosted Decision Tree (BDT)-based selection algorithm. With an expected 30\% signal efficiency, including anticipated reconstruction advances, and an expected background of one event per Mt$\cdot$yr, a 90\% confidence level (CL) lower limit on the proton lifetime in the $p \to \bar{\nu}K^{+}$ channel of $1.3\times 10^{34}$~years can be set, assuming no signal is observed for a 400 kt-year exposure. Another potential mode for a baryon number violation search is the decay of the neutron into a charged lepton plus meson, i.e., $n \to e^{-}K^{+}$. The lifetime sensitivity for a 400 kt-year exposure is estimated to be $1.1\times 10^{34}$~years.  Neutron-antineutron ($n \to \bar{n}$) oscillation is a baryon number violating process that has never been observed but is predicted by a number of BSM theories. The expected limit for the oscillation time of free neutrons for a 400 kt-year exposure is calculated to be $5.53 \times 10^8$~s.\footnote{This analysis used an intranuclear suppression factor for $n\rightarrow\bar{n}$ in ${}^{56}Fe$~\citep{Friedman:2008es} with a slightly inflated uncertainty rather than the suppression factor for ${}^{40}Ar$, which has only recently been calculated~\citep{Barrow:2019viz} and will be applied in future analyses.}

Reconstruction of these events, which have final state particle kinetic energy of order 100 MeV, is a significant challenge, made more difficult
by final-state interactions (FSI), which generally reduce the energy of observable particles.  The dominant background for these searches is from atmospheric neutrino interactions. For example, a muon from an atmospheric $\nu_{\mu}n \to \mu^{-}p$ interaction may be indistinguishable from
a muon from $K \to \mu \to e$ decay chain from $p \to \bar{\nu}K^{+}$ decay, such that identification of the event relies on the kaon-proton discrimination.  Neutron-antineutron oscillations can be detected via
the subsequent antineutron annihilation with a neutron
or a proton.  The annihilation
event will have a distinct, roughly spherical signature of
a vertex with several emitted light hadrons (a so-called
“pion star”), with total energy of twice the nucleon
mass and roughly zero net momentum. Reconstructing
these hadrons correctly and measuring their energies
is key to identifying the signal event. As with nucleon decay, nuclear
effects and FSI make the picture more complicated, as FSI can reduce the multiplicity of pions and make the pions less energetic.

\subsection{JUNO}

JUNO is a next-generation liquid scintillator detector under construction in southern China~\cite{JUNO:2015zny,JUNO:2022hxd}. It consists of a 20 kton liquid scintillator target inside an acrylic spherical vessel surrounded by $17,612$ 20-inch and $25,600$ 3-inch photomultiplier tubes (PMTs). Its choice of technology, which affords it a good timing resolution and a low energy threshold, combined with its large size, unprecedented for a detector of this type, give it unique capabilities in the search for nucleon decays. 

JUNO will be particularly sensitive to the $p \rightarrow \bar{\nu} K^+$ decay channel. The main reason is that, as highlighted in Sec.~\ref{sec:kaonmodestheia}, this decay unleashes a three-fold sequence of events, each of which is detectable in a liquid scintillator detector such as JUNO: a prompt signal from the $K^+$'s loss of kinetic energy, a short delayed signal ($\tau = 12.4$~ns) from its decay daughter (most commonly a $\mu^+$), and a long delayed signal from the daughter's decay (most commonly into a Michel positron with $\tau = 2.2~\mu$s). This threefold coincidence provides a powerful handle to suppress backgrounds, which in JUNO are dominated by atmospheric neutrino interactions. 

\begin{figure}[ht!]
    \centering
    \includegraphics[width=0.46\textwidth]{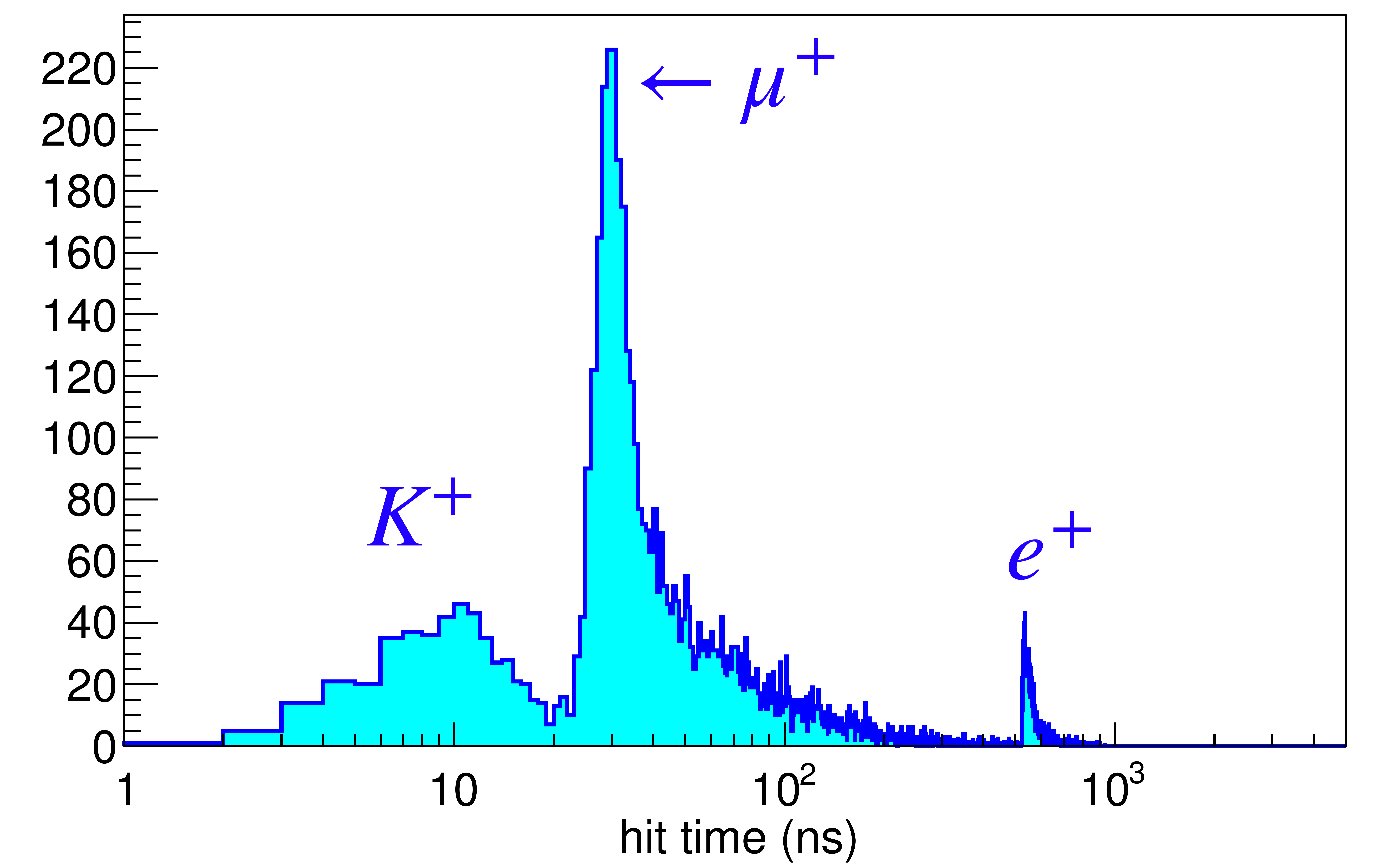}
    \includegraphics[width=0.49\textwidth]{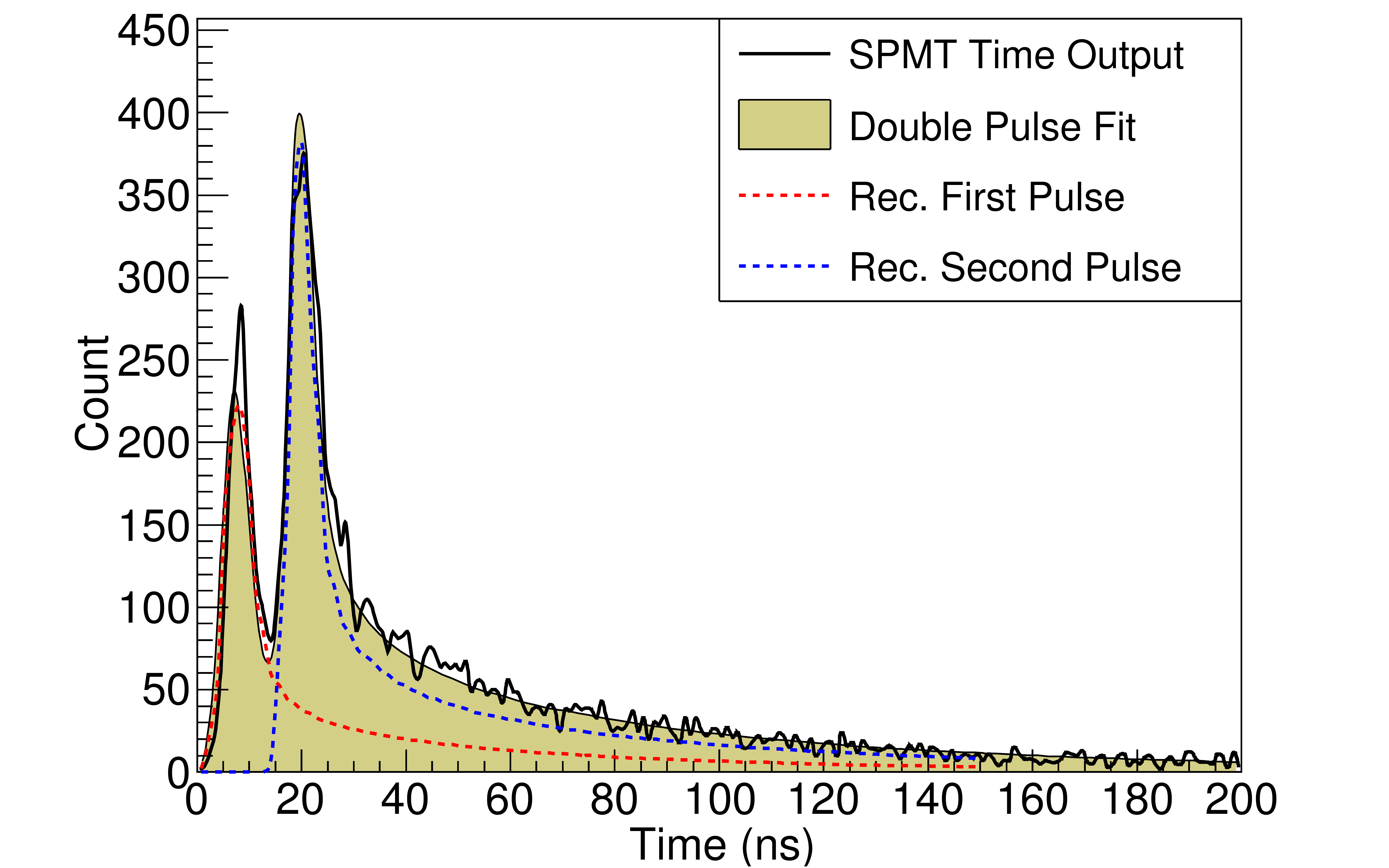}
    \caption{Left: Example of hit time distribution in JUNO for a $p \rightarrow \bar{\nu} K^+$ decay as seen with the 3-inch PMTs. Despite the proximity between the first two signals, the threefold coincidence is clearly visible. Right: Example of multi-pulse fitting in a case where the time separation between the two first pulses is 11~ns. Both images are obtained from Ref.~\cite{JUNO:2022hxd}.}
    \label{fig:junopdecay}
\end{figure}

The sensitivity to the $p \rightarrow \bar{\nu} K^+$ decay channel has been studied using JUNO's custom Monte Carlo simulation framework~\cite{Zou:2015ioy} including a realistic detector performance. A modified version of the GENIE 3.0~\cite{Andreopoulos:2015wxa} generator that accounts for final state interactions and residual nucleus de-excitations is used to generate $p \rightarrow \bar{\nu} K^+$ decays and atmospheric neutrino backgrounds. The short $K^+$ lifetime causes the energy depositions of the $K^+$ and its subsequent decay daughter to overlap in time. However, the 3-inch PMTs, whose transit time spread is around 1.5~ns, allow in many cases to disentangle these two signals after the time-of-flight correction has been applied, as illustrated on the left panel of Fig.~\ref{fig:junopdecay}. Fitting the two pulses simultaneously~\cite{KamLAND:2015pvi}, as illustrated on the right panel of Fig.~\ref{fig:junopdecay}, allows to reconstruct their time separation and energy deposition. The distribution of the best-fit time separations is broader for the signal than for the atmospheric neutrino background, allowing to discriminate between these two with high efficiency~\cite{Undagoitia:2005uu}. 

There are additional handles that further enhance the signal-to-background separation. Among them is the use of a fiducial volume cut, the consideration of a muon veto to suppress cosmogenic backgrounds, and the consideration of the visible energy of the candidate signals, among others. A preliminary selection using all these criteria yields an efficiency of 31\% to $p \rightarrow \bar{\nu} K^+$ decays, with only 0.3 background events in a period of 10 years. A Feldman-Cousins~\cite{Feldman:1997qc} estimation of the sensitivity to this decay mode yields a lower limit for the proton lifetime of $8.34\times10^{33}$ years at 90\% C.L. with 10 years of data, in absence of a signal. 

It is worth noting that JUNO is also expected to have good sensitivity to other decay modes, namely $p \rightarrow \mu^+ \mu^+ \mu^-$ and the $n \rightarrow 3\nu$ invisible mode. JUNO will have an energy resolution of $3\%$ at 1~MeV, which is unprecedented for a liquid scintillator detector. This will be an important asset in those searches where the signal is a mono-energetic energy deposition, as explained in Sec.~\ref{sec:theiainvisibles}. JUNO's sensitivity to these modes is still under evaluation, and the results are expected to be released soon. 

\subsection{THEIA}
THEIA is a detector concept that utilizes advances in photon detection technology
(fast timing, chromatic separation) with water-based liquid scintillator to create
a highly scalable detector (up to the 100 kton scale) with better energy resolution
than a pure water detector. Due to the large mass and presence of scintillation
light, THEIA would have sensitivity to several modes of nucleon decay that is either
comparable with or better than next-generation detectors. Additional details are available
in \cite{THEIA:2019}.
\subsubsection{$p \rightarrow e^+ \pi^0$ and related modes}
Decay modes which have pions in the final state are subject to inelastic intranuclear
scattering for bound protons, which causes the pion to be reabsorbed about 60\% of the
time. This dominates the total efficiency to see this decay mode in water as well
as water-based liquid scintillator. For THEIA the dominant background comes from
atmospheric neutrino interactions, and is independent of depth. Compared with
other water Cherenkov detectors, THEIA would have improved neutron
tagging (90\% efficiency), better energy resolution, and sensitivity to below-Cherenkov threshold charged particles. These features can be all used to better reject atmospheric neutrino backgrounds.
\begin{figure}[h]
    \centering
    \includegraphics[width=0.48\textwidth]{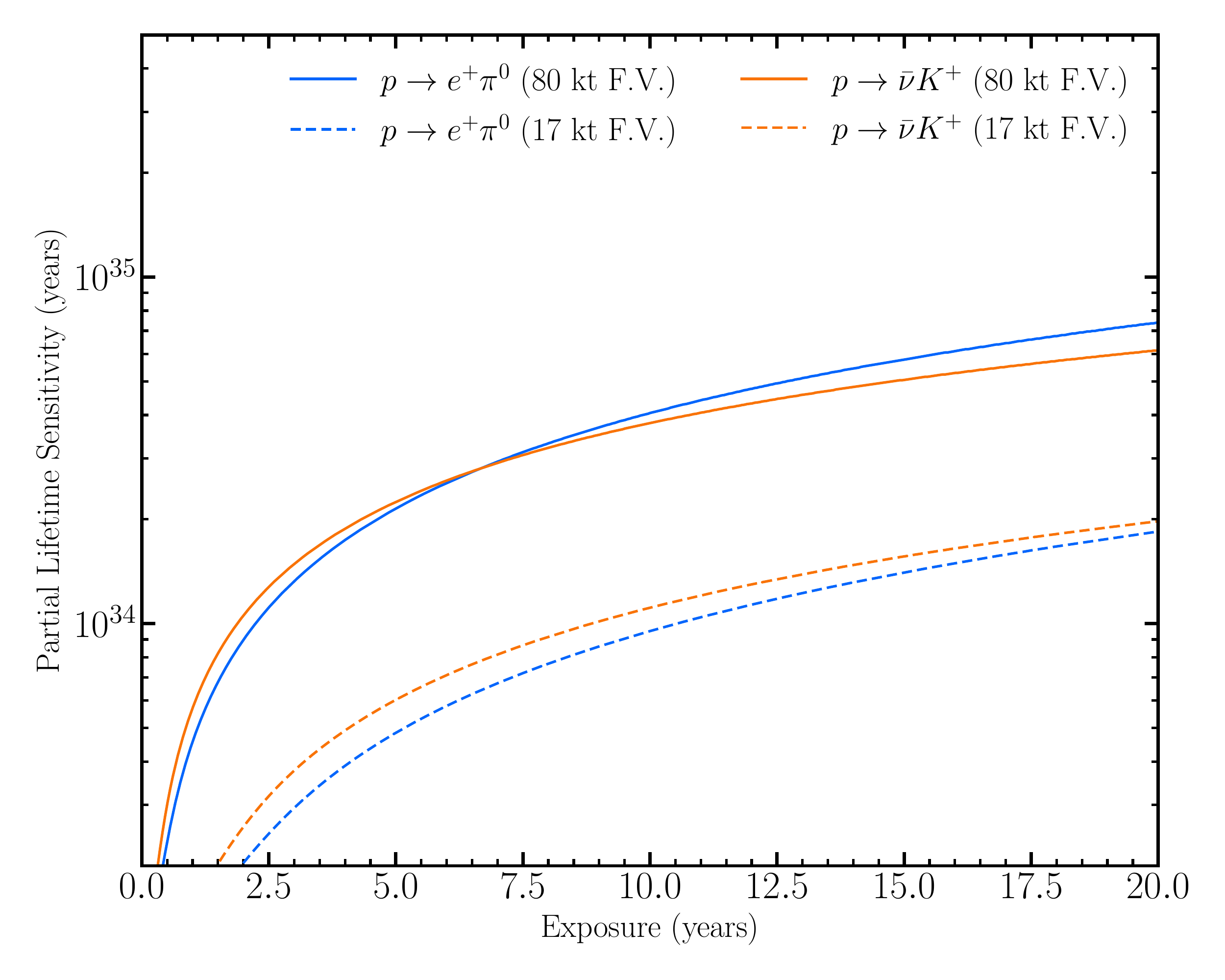}
    \includegraphics[width=0.48\textwidth]{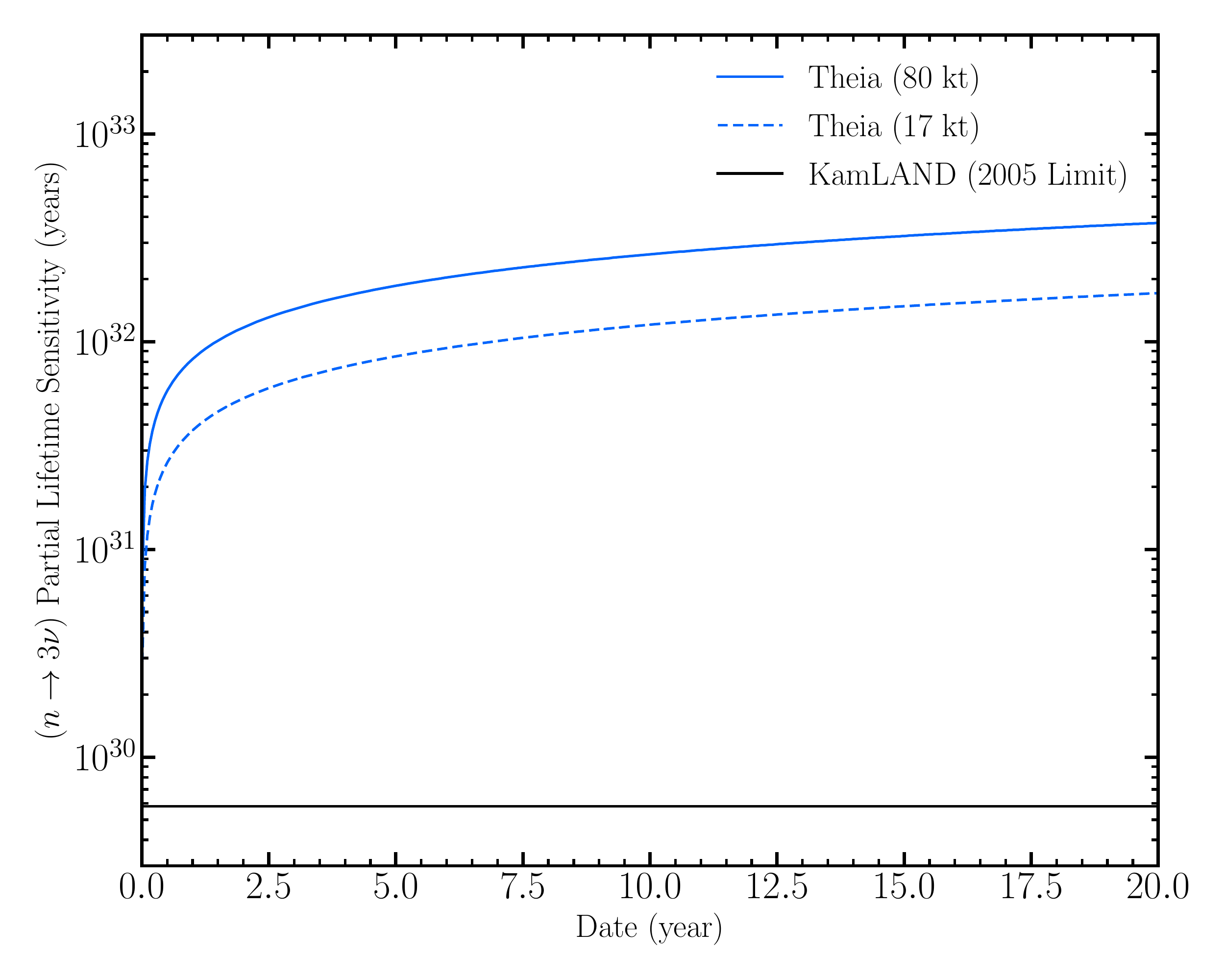}
    \caption{
    (Left) Median sensitivity curves for $p \rightarrow e^+\pi^0$ and
    $p \rightarrow \bar{\nu}K^+$, and (Right) $n \rightarrow 3\nu$ for
    the two considered THEIA fiducial volumes.
    }
    \label{fig:epi_sensitivity}
\end{figure}
\subsubsection{$p \rightarrow \bar{\nu}K$ and related modes}\label{sec:kaonmodestheia}
Kaon decay modes are less affected by intranuclear effects and produce a three-fold
coincidence signal in the detector through the subsequent decays of their daughter
particles~\cite{ParticleDataGroup:2020ssz}.
\begin{itemize}
    \item $K^+ \rightarrow \mu^+ \nu_\mu~~(63.56\%)$
    \item $K^+ \rightarrow \pi^+\pi^0~~(20.67\%)$
    \item $K^+ \rightarrow \pi^+\pi^+\pi^-~~(5.58\%)$
    \item $K^+ \rightarrow \pi^0e^+\nu_e~~(5.07\%)$
    \item $K^+ \rightarrow \pi^+\pi^0\pi^0~~(1.76\%)$
\end{itemize}
In a pure water Cherenkov detector the kaon is below the energy Cherenkov threshold
and does not produce a detectable signal; however, the multiple pions can produce
identifiable multi-ring signals, and the subsequent decays of the pions and muon
create a coincidence trigger. In THEIA, the kaon would produce scintillation light
as well, though much of the kaon signal would overlap with it's decay products due
to it's 12-ns lifetime \cite{JUNO:2015zny}. Since these first two signals would be difficult to distinguish,
it is safe to assume that THEIA would perform no better than JUNO (pure scintillator)
in terms of signal efficiency, but could be built to a much larger scale.
\begin{table}[]
    \centering
    \begin{tabular}{c|c|c}
        Signal & Efficiency & Bkg [/Mton$\cdot$yr]\\
        \hline \hline
        $p \rightarrow e^+\pi^0$ & 40\% & 0.3 \\
        $p \rightarrow \bar{\nu}K^+$ & 55\% & 2.5 \\
        $n \rightarrow 3\nu$ & 44\% & 95.6 \\
    \end{tabular}
    \caption{Detection efficiency and background rates for the three considered modes of nucleon decay in
    THEIA.} 
    \label{tab:pdecay_nuk}
\end{table}
\subsubsection{$n \rightarrow 3\nu$ and related modes}\label{sec:theiainvisibles}
Finally, THEIA would have sensitivity to a class of low-energy modes known as ``invisible'' modes, where
all of the final state particles cannot produce light emissions (such as all-neutrino final states).
When these occur in a multi-nucleon nucleus then the nucleus will often be left in an excited state
and will release deexcitation $\gamma$s and nucleons. In an $^{16}$O nucleus this manifests as a 6.18-MeV $\gamma$
with a relatively high branching ratio (44\%). The primary background for this signal comes from
solar neutrinos, internal radioactivity, and cosmogenic activation of oxygen to $^{16}$N.
THEIA would be the only large-scale water Cherenkov detector available to
look for this decay due to its large mass and great depth. Leading limits on invisible neutron and dineutron decay
are set by KamLAND \cite{Araki:2005jt}, and on invisible proton and diproton
decay by SNO+ \cite{Anderson:2018byx}.

\subsection{Detection Techniques in LArTPCs}

The search for baryon number violation is a prime goal of particle physics and is being carried out in large underground detectors.   Most of the current lifetime limits are affected by backgrounds, predominantly from the 100 events per kiloton year which arise from atmospheric neutrino interactions.    The current best limits come from water detectors, but a promising way to reduce backgrounds is with a large liquid argon time projection chamber (LArTPC), almost all  $^{40}Ar$.  Large LArTPCs provide the capability to image charged particles with mm-scale resolution and thus explore signatures from particle and nuclear physics at the same time~\citep{Djurcic:loi2021}.
The concurrent detection of light in a photo-detector system provides an opportunity to identify signatures from nucleon decay that are not present in neutrino interactions.   Analysis strategies can then be tuned to both reduce background and increase efficiency.  A photon-detection system can also measure energy calorimetrically, working as a crosscheck of the energy measured by the ionizing particles in a TPC and thus improving the energy resolution when both measurements are used together. 
There are about one hundred nucleon decay modes
accessible to large underground liquid argon TPCs, similar to the DUNE Far Detector~\citep{DUNE:2020lwj}, or to the considered module of opportunity liquid argon TPCs, under consideration~\citep{mood_workshop:2019bnl}, such as Q-Pix based LArTPC~\citep{QPix:2021snp}.  We see a valuable opportunity to combine signatures from particle and nuclear physics and take advantage of the fast and precise timing capabilities of a photon-detection system and the detailed pattern recognition capabilities of a LArTPC~\citep{uboone:2020tnp, DUNE:2020cqd} to improve the sensitivity of future nucleon decay searches.
As an example, consider the mode $p \rightarrow \bar{\nu} K^+$ ~\citep{Super-Kamiokande:2014otb}. For this mode, one looks for the $K^+$ signature via a short ionization track, followed primarily by $K^+ \rightarrow \mu^+ \nu \mu$ decay which makes an observable $\mu^+$ track. The $\mu^+$ decays to an observable positron.  There are also other less likely $K^+$ decay modes.  The kaon and its decay products can be reconstructed as images and the decay chain could be tested for kinematic consistency.  Three potential backgrounds include 1) the large number of quasi-elastic atmospheric $\nu_{\mu}$ interactions where the recoil proton is misidentified as a $K^+$, 2)  production of $K^+$  by atmospheric neutrinos, and 3) neutrino production of $K_L$ outside the detector which enter the LArTPC and charge exchange to a $K^+$.

\begin{figure}
\centering
\includegraphics[width=0.7\columnwidth]{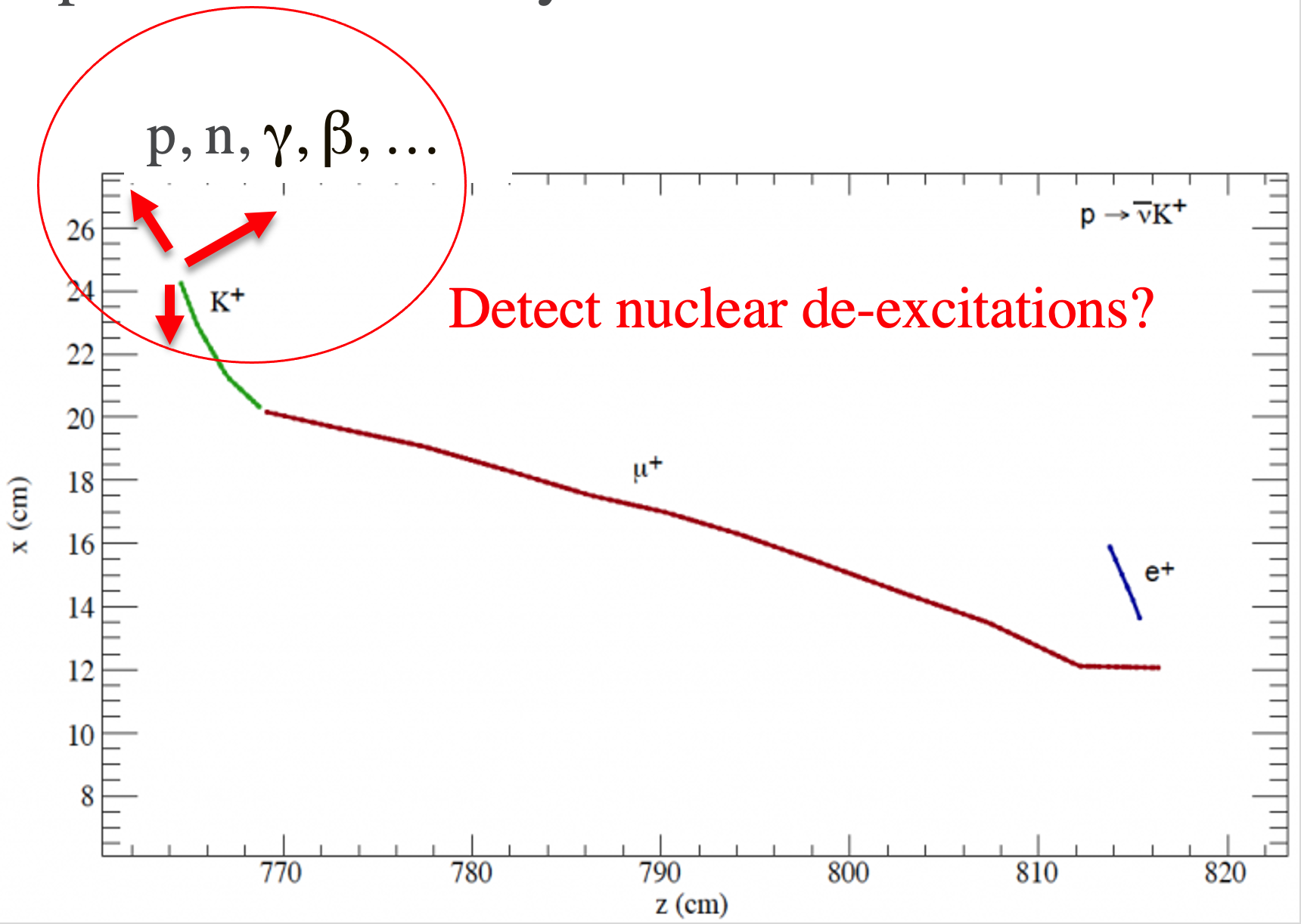}
\caption{\label{fig:nucl_cartoon} Search for the decay products from nuclear de-excitation of the residual radioactive daughter nuclei left after the disappearance of proton from $^{40}$Ar nucleus.
.}
\end{figure}

There are several additional signatures of $p \rightarrow \bar{\nu} K^+$  decay to investigate, mostly requiring LArTPC mm-scale resolution and a low-energy charge detection threshold ideally combined with a fast timing from the photon system:
\begin{itemize}
\item The $^{40}Ar$ could often become an excited state of $^{39}Cl$, with emission of de-excitation gamma rays. This nuclear process needs additional studies since any proton inside the $^{40}Ar$ can decay. The proton that decays might be deeply bound within the nucleus and the nucleus would be, at a minimum, left with an excitation energy much higher than the neutron separation energy and maybe even other separation energies. There is a possibility of even higher excitation if part of the decay energy is transferred to the nucleus. On the LArTPC response side, the detection of MeV-scale gamma-rays was demonstrated~\citep{Acciarri:2018myr} with efficiency of 50\% and energy resolution of 24\% at 0.5 MeV, and with an efficiency of almost 100\% and energy resolution of 14\% at 0.8 MeV. If the gammas from $^{40}Ar \rightarrow ~^{39}Cl$ de-excitation are measured by the photon system, in addition to the TPC charge detection, we would have a new unique time-tag of the proton decay process, see Fig~\ref{fig:nucl_cartoon}.
\item The product $^{39}Cl$ decays with a 56 minute half life to $^{39}Ar$ (feasible to tag in a deep detector).  Then, $^{39}Ar$ decays 93\% of the time to excited states and will give you characteristic gamma rays, roughly 54\% of the time a 1.27 MeV gamma, 39\% of the time a 1.52 MeV gamma and 46\% of the time a 250 keV gamma (this one typically comes with the 1.27 MeV gamma).
\item The $K^+$ lifetime is 12 ns, and a photon system with good timing could distinguish it from its decay products, as well as improve the identification of Michel electrons, from the 2 $\mu s$ decay of $\mu$’s. In recently published results of the ProtoDUNE-SP liquid argon time projection chamber performance~\citep{Acciarri:2018myr}, a time resolution of 14 ns was achieved in the case of a single photon-detector channel. It is expected that photon-detector time resolution will improve as a square root of the number of independent channels, driving the resolution well-below the 12 ns lifetime of $K^+$.
\item Bound neutron decay to excited states of $^{39}Ar$ might also make de-excitation gamma.
\end{itemize}

Our intent is, therefore, to further study these signatures and to examine how the signal sensitivity will be further enhanced, and how backgrounds could be rejected more efficiently with additional information from nuclear de-excitations and with precise timing from photon-detectors in TPCs.

\subsection{Effect of Different Nuclear Model Configurations on Sensitivity to Intranuclear Neutron-Antineutron Transformations}

Sensitivity studies for baryon number violating processes are currently relegated to simulations using generated Monte Carlo (MC) event samples. For example, using the GENIE MC event generator~\citep{Andreopoulos:2009rq} and DUNE detector simulation and reconstruction software packages such as LArSoft~\citep{Snider:2017wjd}, a first foray into sensitivity investigations for separating intranuclear $n\rightarrow\bar{n}$ from (predominately neutral-current) atmospheric neutrino backgrounds in the DUNE far detector was considered in~\citep{Hewes:2017xtr}. However, there remains much work to be done~\citep{Barrow:2021odz}.
The dependencies of convolutional neural networks (CNN's) and other automated (machine learning) methods such as multivariate boosted decision trees' (BDTs') responses to various topological inputs is not entirely clear. The nature of these algorithms' responses to underlying choices in what are generally considered to be broadly consistent nuclear model configurations (NMCs) must be further studied to assess effective uncertainties, including those originating from disparate models of nuclear Fermi motion and final state interactions (FSIs) from (non)stochastic intranuclear cascades~\citep{Golubeva:2018mrz,Barrow:2019viz}. Given that the event triggers for $n\rightarrow\bar{n}$ 
will likely utilize the signal's expected region of interest (ROI) as informed primarily by the MC simulation of the process and its separation from background via automated methods, such dependencies are highly important, particularly when physical correlations are being ignored;
this is but one important component of the larger, still developing paradigm, which should include improved reconstruction and PID.

\begin{figure}[t!]
    \centering
    \includegraphics[width=0.30\columnwidth]{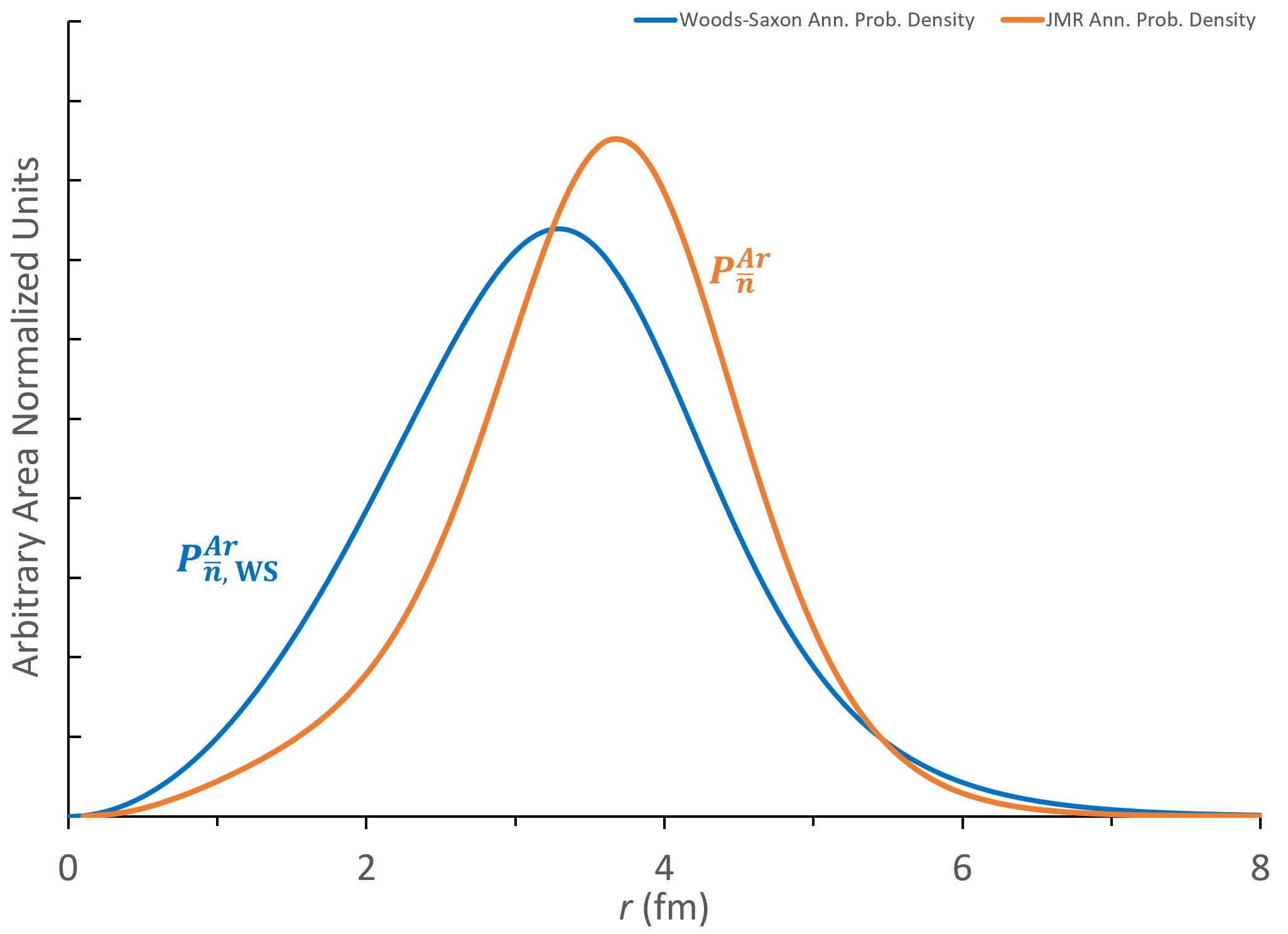}
    \includegraphics[width=0.30\columnwidth]{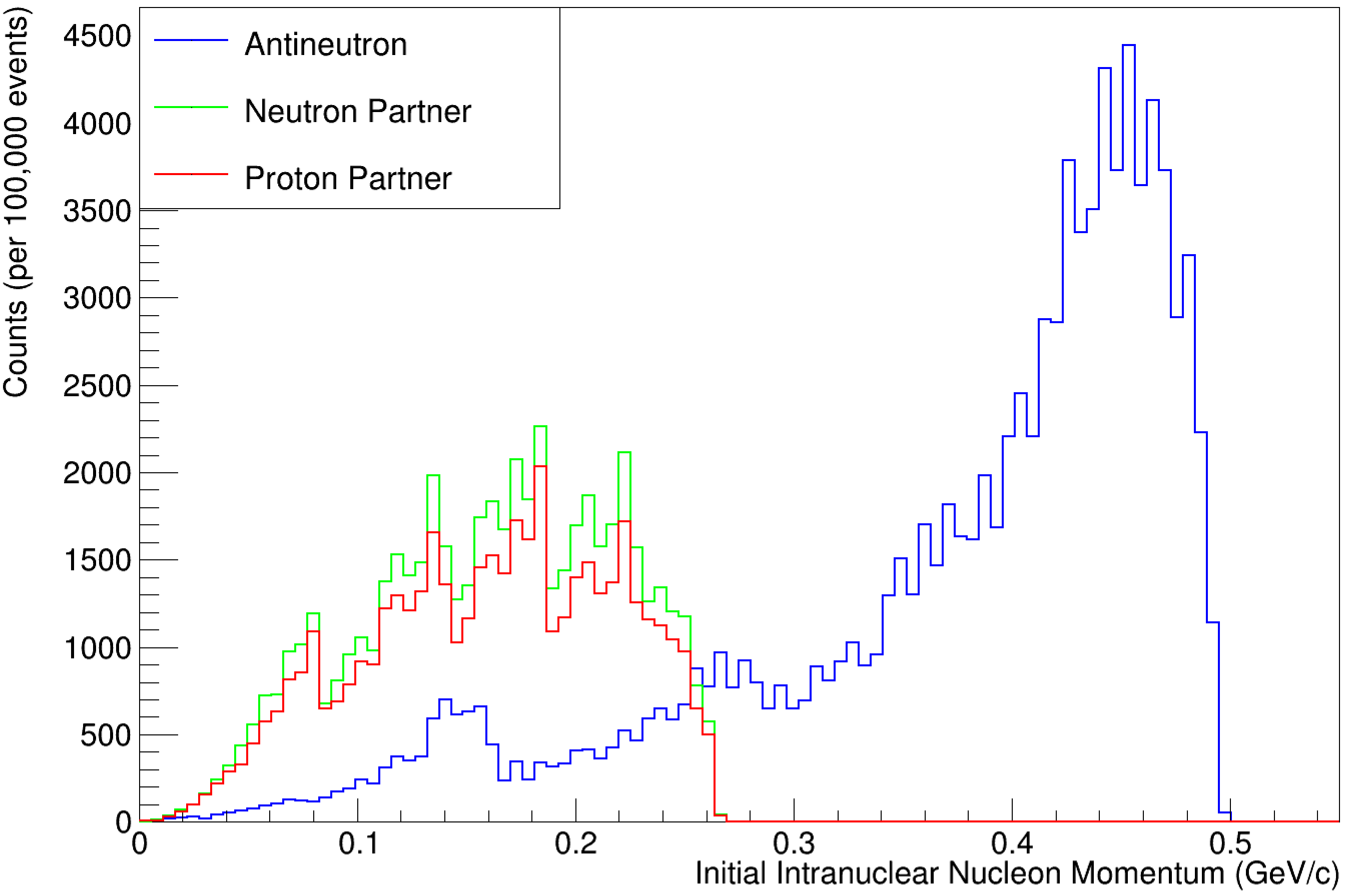}
    \includegraphics[width=0.30\columnwidth]{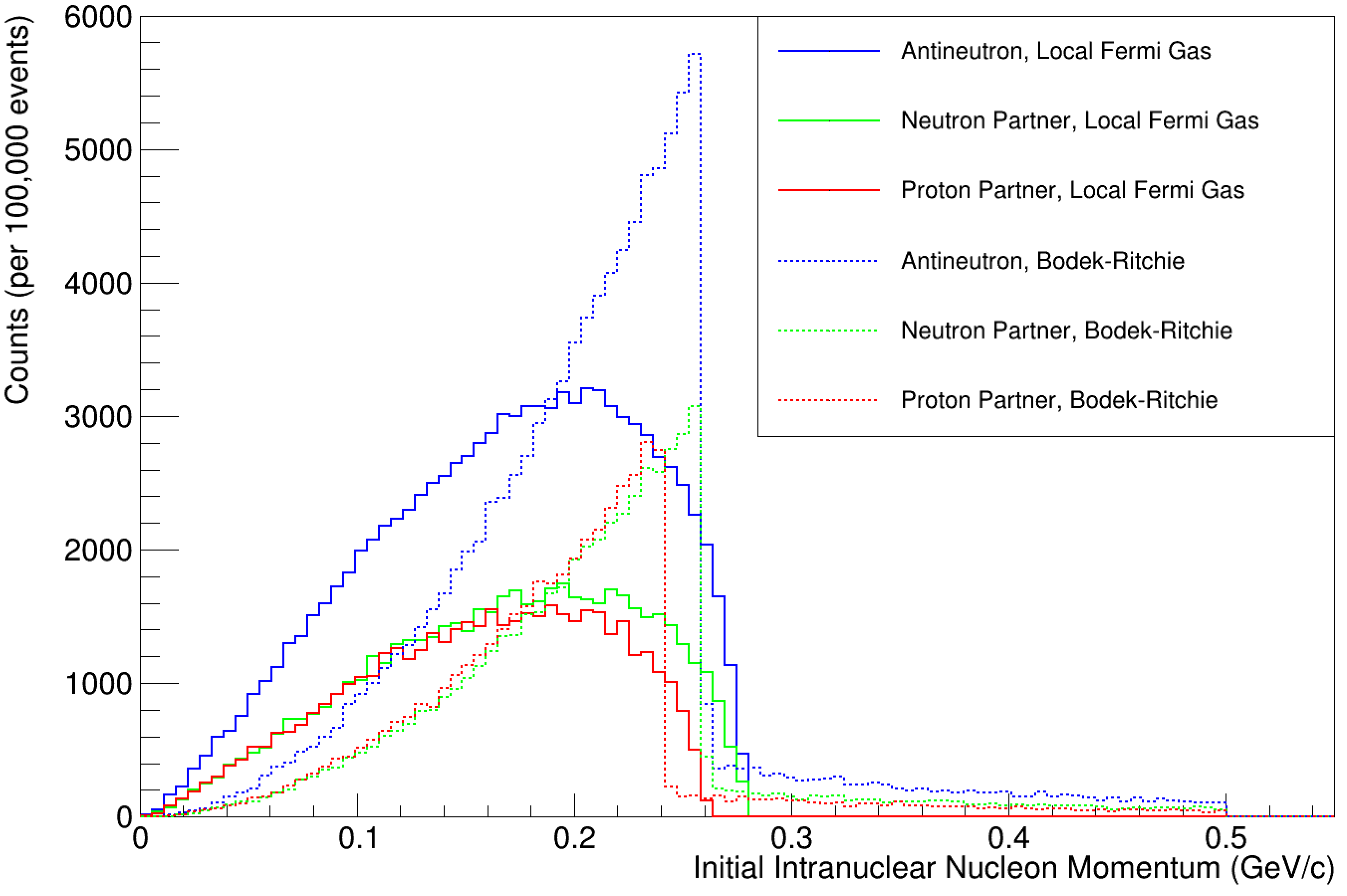}
    \includegraphics[width=0.30\columnwidth]{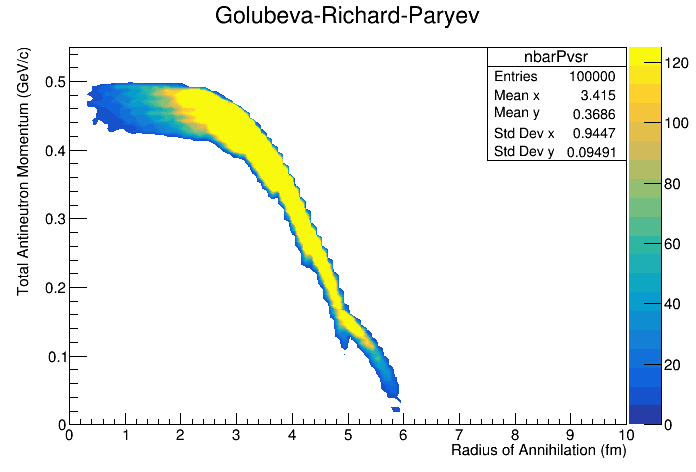}
    \includegraphics[width=0.30\columnwidth]{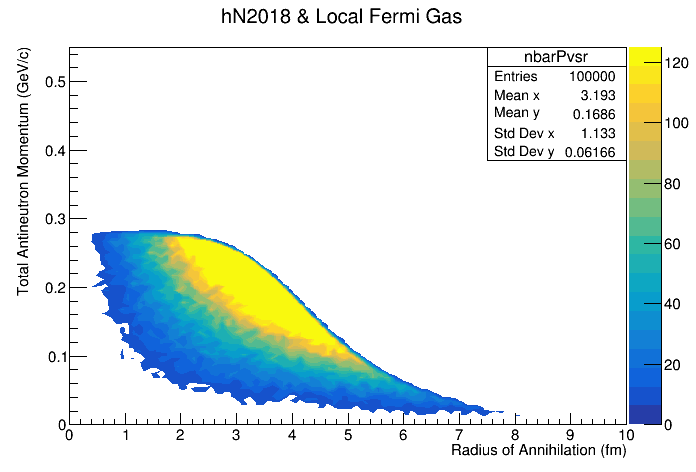}
    \includegraphics[width=0.30\columnwidth]{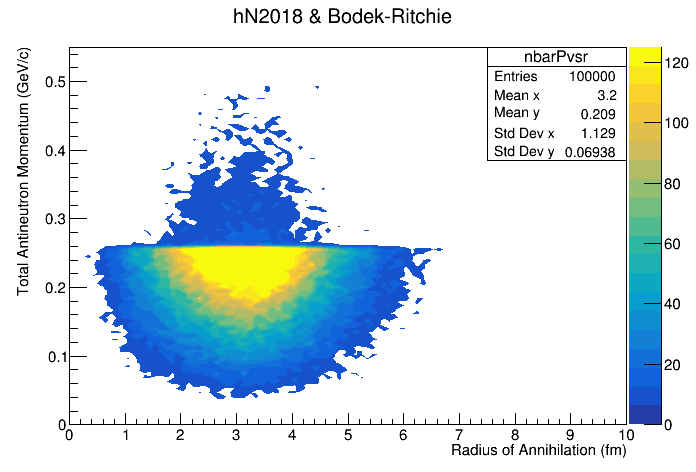}
    \caption{\textbf{Top Left}: Two curves are shown for various generator assumptions. In blue is the naive intranuclear radial position of $\bar{n}$ annihilation, a probability distribution generated by a Woods-Saxon nuclear density as presented in GENIE~\citep{Andreopoulos:2009rq}. In orange is the modern, quantum-mechanically derived intranuclear radial position of annihilation probability distribution as developed in~\citep{Barrow:2019viz}. The vertical scale is arbitrary. 
    \textbf{Top Center}: The initial (anti)nucleon momentum distributions are shown using a local Fermi gas (LFG) model with an additional $\bar{n}$ potential~\citep{Barrow:2019viz}.
    \textbf{Top Right}: The same for the GENIEv3.0.6~\citep{Andreopoulos:2009rq}, showing an LFG model and the default nonlocal Bodek-Ritchie (BR) model.
    \textbf{Bottom Left}: A two dimensional plot of intranuclear $\bar{n}$ momentum-radius correlation is shown using an LFG and the newly-derived annihilation position distribution~\citep{Barrow:2019viz} (top left, orange).
    \textbf{Bottom Center}: The same using GENIEv3.0.6's~\citep{Andreopoulos:2009rq} LFG model of (anti)nucleon momentum and a Woods-Saxon nuclear density (top left, blue), showing good correlation.
    \textbf{Bottom Right}: The same using GENIEv3.0.6's~\citep{Andreopoulos:2009rq} \textit{nonlocal} BR nuclear model of (anti)nucleon momentum and a Woods-Saxon nuclear density (top left, blue), showing no positional correlation, and thus over-selecting high momenta.}
    \label{fig:AnnDist}
\end{figure}

To explain this particular aspect in some detail, let us begin by stating the obvious: there is actual importance in the maintaining of particularly relevant physical correlations in MC simulations which act to inform the detection of unknown (rare) processes. To illustrate this, consider Figs.~\ref{fig:AnnDist}~\citep{Barrow:2019viz}. Some correlations which have gone previously unexplored include the expected position of the intranuclear $\bar{n}$ annihilation following $n\rightarrow\bar{n}$ conversion, as shown at top left. Considering the $\{n,\bar{n}\}$ mass splitting which suppresses $n\rightarrow\bar{n}$ lessens as one hypothetically decreases the binding energy, a radial dependence is expected in the transformation probability beyond the simplistic assumption of the assumed Woods-Saxon nuclear density--indeed, such transformations are expected to occur predominately near the nuclear surface where $n$ binding is low.
Further, and as illustrated in the top central figure, given the strength of the annihilation cross section, the optical potential describing the Fermi motion of a previously converted $\bar{n}$ is expected to be deeper than that of normal nuclear matter, imbuing the converted $\bar{n}$ with higher available momentum. Similarly, due to $n\rightarrow\bar{n}$ occurring predominately on the surface of the nucleus, one may expect a reduction in the available Fermi motion-derived momentum; thus, simulations which are inherently \textit{nonlocal} in their assumptions of Fermi motion (as is the case in the GENIE~\citep{Andreopoulos:2009rq} default nuclear model) can lead to biases in any particular ROI. The (non)locality of a given nuclear model is shown by a decrease in (anti)nucleon momentum as one moves further out toward the nuclear envelope ($r\rightarrow R$), as shown in the bottom figures for two generators; if a hemicircular region is occupied in this parameter space, then no momentum-radius correlations are preserved.

\begin{figure}[h!]
    \centering
    \includegraphics[width=0.40\columnwidth]{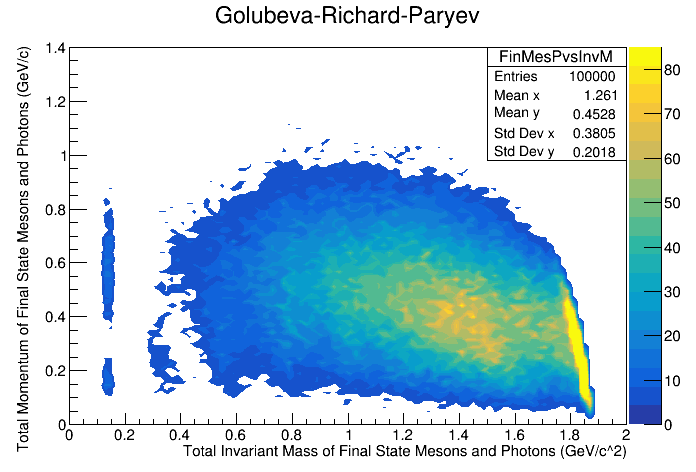}
    \hspace{0.07\columnwidth}
    \includegraphics[width=0.40\columnwidth]{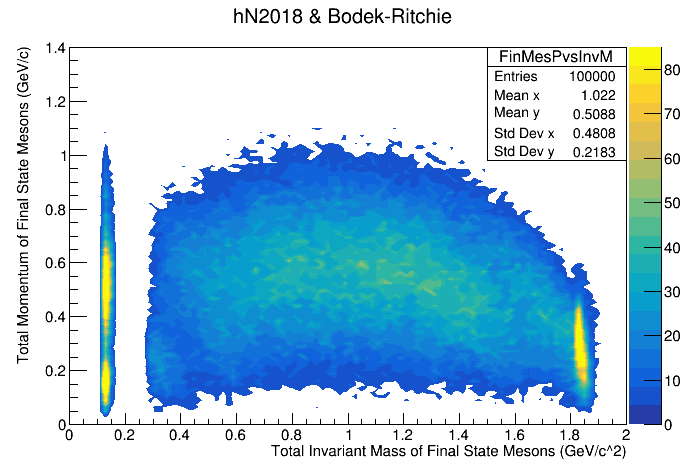}
    \caption{The final state mesonic/pionic parameter space (total momentum versus invariant mass)~\citep{Super-Kamiokande:2011idx} after stochastic intranuclear transport of $\bar{n}$ annihilation generated mesons, compared for a few NMCs, not including detector effects. The ROI is generally considered to the be ``hot-spot" in the lower right hand corner, implying the expected low Fermi momentum and high invariant mass derived from the annihilation of two nucleons creating a topologically spherical $\pi$-star; differences in these may lead to different detector signal efficiencies via automated methods.
    \textbf{Left}: an LFG model with an additional $\bar{n}$ potential and a full intranuclear cascade~\citep{Golubeva:2018mrz,Barrow:2019viz}.
    \textbf{Right}: GENIEv3.0.6~\citep{Andreopoulos:2009rq} using the default nonlocal BR relativistic Fermi gas and a full intranuclear cascade via the 2018 hN Intranuke model (hN).}
    \label{fig:FinMesonTotPvInvMass}
\end{figure}

The investigation in maintaining these physically relevant correlations can go even further, as one may expect fewer FSIs to be experienced by annihilation-generated mesons due to reduced views on intranuclear scattering centers near the nuclear periphery. Also, when evaluating the $n\rightarrow\bar{n}$ signal's ROI (before consideration of further skewing due detector effects), choices across models of these FSIs can have a critical role in determining signal efficiencies through topological selection of high multiplicity events involving knock-out $p$s, which may be overproduced~\citep{Barrow:2019viz}. When considering outgoing mesons only, the nature of the ROI can be seen to remain disparate through the comparison of various NMCs across multiple (and single) generators, as shown in Figs.~\ref{fig:FinMesonTotPvInvMass}.

The importance of these correlations goes beyond the signal simulation; indeed, the same NMCs and correlations should be respected consistently across atmospheric neutrino background simulations. Iterating across the available NMCs, between and within single generators (for instance, GENIE~\citep{Andreopoulos:2009rq}), and intermixing these together allows for the evaluation of uncertainties in a ``universe" style approach, though going beyond simple knob turning; a project of this scale requires massive automated analysis techniques to be successful, especially given the nonreweightable nature of some of the more theoretically well-motivated stochastic FSI models employed. Such uncertainty evaluations will permit a better understanding of the signal to background ratio, informing the final expected sensitivities 
to $n\rightarrow\bar{n}$: if this ratio remains stable across NMCs, then the minutia of certain physical correlations in simulation will be shown to be unnecessary; however, the opposite is the more likely case.
One study showed that a change in the NMC reduced the sensitivity by roughly 40\% all else being equal~\citep{Barrow:2021odz}.
Thus, this greatly encourages not only the evaluation of dependencies of automated analyses' (CNNs, BDTs) responses, but also the necessity of improved reconstruction. 
DUNE has had successes in machine learning being applied to PID~\cite{DUNE:2020fgq}, which was not included in~\citep{Hewes:2017xtr}; implementing a CNN score describing the probability of particular track's PID could better discriminate signal from background considering the unique high-multiplicity $\pi$-star expected to emanate from a nucleus after $\bar{n}$ annihilation.

\begin{figure}[htp]
    \centering
    \includegraphics[width=0.6\columnwidth]{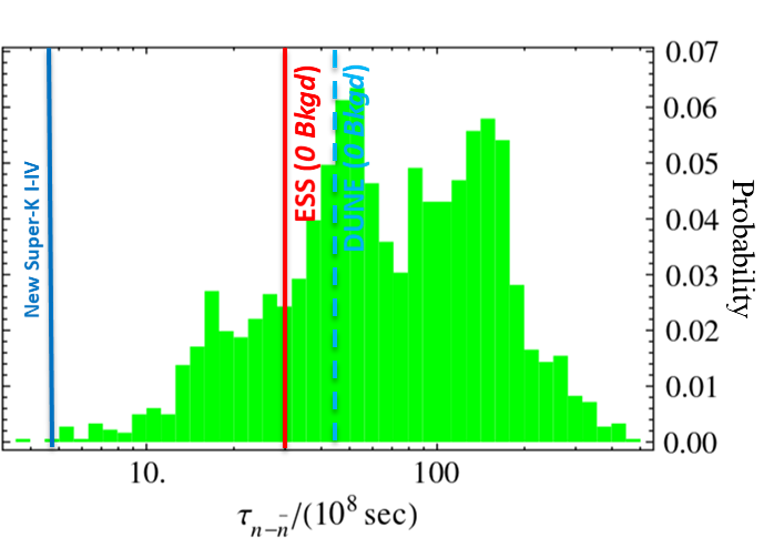}
    \caption{A reproduction of Fig. 9 in~\citep{Babu:2013yca} is shown with sensitivity estimates for DUNE--assuming a 25\% efficiency--and ESS NNBAR--assuming an ILL-like efficiency~\citep{Baldo-Ceolin:1994hzw}--each in a case of zero background compared to the Super-Kamiokande I-IV limit~\citep{Super-Kamiokande:2020bov}.}
    \label{fig:expectedlimits}
\end{figure}

From these discussions,
one may hypothesize as to the regions of $\tau_{n\bar{n}}$ parameter space probed by DUNE~\citep{DUNE:2020fgq}, Super-Kamiokande~\citep{Super-Kamiokande:2020bov}, and ESS NNBAR~\citep{Addazi:2020nlz}. In the context of post-sphaleron baryogenesis~\citep{Babu:2006xc,Babu:2013yca}, one can compare potential lower limits to model predictions, as shown in Figure~\ref{fig:expectedlimits}. A spread in the expected values of $\tau_{n\bar{n}}$ given various nuclear model configurations is \textit{concerning}, especially as the underlying nuclear physics, though well constrained~\citep{Barrow:2021odz}, is not definitively known. 

By considering the above in greater detail into the future, 
sensitivities
to $\mathcal{B-L}$-violating $\Delta\mathcal{B}=2$ processes such as $n\rightarrow\bar{n}$ and related dinucleon decay modes are expected to greatly improve through enhanced physical modeling, reconstruction, and PID. 
Uncertainties due to specific NMC choices remain under investigation, potentially lowering the sensitivities without other improvements. These steps are required in order 
to pursue truly complementary physics goals beyond proton decay and neutrino oscillation studies, enabling the broader particle physics community
to fully utilize future large underground detectors to better search for BSM physics, potentially informing us of our universe's origins.

\subsection{Inclusive nucleon decay searches}

As summarized in Sec.~\ref{sec:theory}, $\mathcal{B}$-violating processes are associated with a broad variety of theories and can manifest through distinct nucleon decay channels. The strongest limits can be set on particular exclusive nucleon decay processes, motivated by specific theoretical models. However, as discussed in Sec.~\ref{sec:BLtheor}, higher order operators leading to complicated processes can readily dominate over typical two-body nucleon decay channels. With increased complexity of interactions, performing exclusive experimental searches for all specific nucleon decay channels becomes unfeasible beyond the simplest realizations.

These considerations highlight the importance of inclusive nucleon decay searches and set limits on classes of baryon number-violating processes simultaneously. Such searches can be classified as follows~\cite{Heeck:2019kgr}:

\begin{itemize}
    \item \textit{model-independent and invisible mode searches}
    
    - model-independent searches probe all nucleon decays simultaneously, unconstrained by specific final states. These searches also constitute the primary approach for studying invisible channels such as $n \rightarrow invisible$. Invisible channels such as $n \rightarrow 3\nu$ could become significant in models based on extra-dimensions~(e.g.~\cite{Mohapatra:2002ug}). Analysis of $p \rightarrow invisible$ will also serve as a test of charge conservation. Prominent searches of this class include 
    isotope fission products as well as signatures of nuclear de-excitation, with current limits being $\tau \gtrsim 10^{30}$~yrs. The latter method is expected to be a promising target for future neutrino experiments such as JUNO, Hyper-Kamiokande and DUNE.
    
    \item \textit{$N \rightarrow X + \text{anything}$} 
    
    - allowing for model-independence while also taking advantage of precise particle identification in current and upcoming large neutrino experiments, such as Super-Kamiokande (see related searches of Ref.~\cite{Super-Kamiokande:2015pys}), motivates $N \rightarrow X + \text{anything}$ searches, where $X$ is a SM particle of unknown energy. These include:

    \subitem A) $N \rightarrow (e^{\pm}, \mu^{\pm}, \pi^{\pm}, K^{\pm}, \rho^{\pm}, K^{
\ast,\pm}$)~+~\textit{anything}: primary decay charged particles are often directly visible with high efficiency in experimental searches. From electric-charge conservation any
proton decay will eventually result in at least one
positron, albeit potentially space-time-delayed if it is originated from the decay of a heavier charged final-state
particle.

    \subitem B) $N \rightarrow (\pi^{0}, K^{0}, \eta, \rho, \omega, K^{\ast,0}$)~+~\textit{anything}: searches involving neutral mesons rely on
the electromagnetically interacting daughter particles for detection (e.g. $\pi^0 \rightarrow \gamma\gamma$).

    \subitem C) $N \rightarrow \gamma$~+~\textit{anything}: assuming multi-body nucleon decays,
the emission of a photon will typically be suppressed compared to the same multi-body channel without a photon. 

    \subitem D) $N \rightarrow \nu$~+~\textit{anything}: similar to other invisible searches, can also be studied through neutrino flux measurements from accumulation of nucleon decays inside the Earth.
    
    \item \textit{$\Delta B > 1$ processes} 
    
    - inclusive searches also offer prospects for $\Delta B > 1$ processes. Note that inclusive $\Delta B = 1$ searches also constrain $\Delta B > 1$. The increased available energy for final state particles in multi-nucleon decay processes modifies the search compared to single-body decays.

\end{itemize}

%% file: Sections/Conclusions.tex
\section{Conclusion and Outlook }\label{sec:conclusions}

We have discussed the current status and future prospects of baryon number violation searches with large underground detectors for neutrino experiments. These detectors used for next-generation neutrino experiments have the capability to improve the existing limits on nucleon lifetime by up to two orders of magnitude. A broad class of GUT models (both non-supersymmetric and supersymmetric) can be probed. Some of them predict an upper limit on the nucleon lifetime, which might be fully within reach of the future experiments. Apart from the $|\Delta B|=1$ nucleon decay, there exist other interesting BNV observables as well, such as neutron-antineutron oscillation ($|\Delta B|=2$) and $B\to LM$ decays ($|\Delta(B-L)|=2$), which also have promising prospects for underground detectors to be used in future neutrino experiments. The discovery of baryon number violation will be an unambiguous signal of new physics, and therefore, it is important to search for as many BNV channels as possible. 

%% file: common/EndMatter.tex
\section*{Acknowledgements}
We thank Borut Bajc, Ilja Dor\v{s}ner, Natsumi Nagata, Jogesh Pati, Stuart Raby, Zurab Tavartkiladze, and Cem Salih Un for useful comments on the manuscript.  
The work of P.S.B.D. is supported in part by the U.S.~Department of Energy under grant No. DE-SC0017987.  This research was supported in part by the National Science Foundation under Grant No. NSF PHY-1748958. S.~Syritsyn is supported by the National Science Foundation under CAREER Award PHY-1847893.  V.T. is supported by
the World Premier International Research Center Initiative (WPI), MEXT, Japan. The work of R.N.M. is supported by the US National Science Foundation grant no. PHY-1914631.
The work of M.M. is supported by the Grant Agency of the
Czech Republic (GA\v{C}R) through contract number~20-17490S and from
the Charles University Research Center UNCE/SCI/013. The work of R.S. is supported in part by NSF Grant PHY-1915093.

\bibliographystyle{JHEP}
\bibliography{refs}